\documentstyle[12pt,psfig]{article}

\newcommand{\cF}{ {\cal F} }
\newcommand{\ra}{\rightarrow}

\input xy
\xyoption{all}

\begin{document}

\hfill ILL-(TH)-03-6

\hfill hep-th/0307245

\vspace{1.0in}

\begin{center}

{\large\bf Lectures on D-branes and Sheaves }

\vspace{0.5in}

Eric Sharpe\\
Department of Mathematics \\
1409 W. Green St., MC-382 \\
University of Illinois \\
Urbana, IL  61801 \\
{\tt ersharpe@uiuc.edu} \\

$\,$

\end{center}

These notes are a writeup of lectures given at the twelfth Oporto 
meeting on ``Geometry, Topology, and Physics,''
and at the Adelaide workshop ``Strings and Mathematics 2003,''
primarily geared towards a physics audience.
We review current work relating boundary states in the open string
B model on Calabi-Yau manifolds
to sheaves.  Such relationships provide us with a mechanism
for counting open string states in situations where the physical spectrum
calculation is essentially intractable -- after translating to
mathematics, such calculations become easy.  We describe several different
approaches to these models, and also describe how these models are
changed by varying physical circumstances -- flat $B$ field backgrounds,
orbifolds, and nonzero Higgs vevs.  We also discuss mathematical 
interpretations of operator products, and how such mathematical interpretations
can be checked physically.  One of the motivations for this work is
to understand the precise physical relationship between boundary
states in the open string B model and derived categories in mathematics,
and we outline what is currently known of the relationship.

\begin{flushleft}
July 2003
\end{flushleft}

\newpage

\tableofcontents

\newpage

\section{Introduction}   \label{intro}

Using sheaves as a mathematical tool to model D-branes on large-radius
Calabi-Yau manifolds was first
suggested many years ago by J.~Harvey and G.~Moore in \cite{hm}.
Since then, it has become popular to assume that such a model
is a reasonable one, and furthermore to assume that sheaves can
be used to calculate physical properties such as, for example:
\begin{enumerate}
\item Open string spectra between D-branes, which are believed to
be counted by Ext groups.
\item Boundary-boundary OPE's, which are believed to be given by 
Yoneda pairings of Ext groups.
\item T-duality, which is believed to be described by a Fourier-Mukai
transformation.
\end{enumerate}
However, although these properties have been checked in
certain special cases,
until relatively recently no one knew how to show any of these properties
hold in any degree of generality.

One motivation for such a mathematical model is that it allows us
to convert physics questions that are often tricky and difficult
into comparatively easy mathematics computations.  Another motivation
comes from understanding mirror symmetry.  Before D-branes were
popularized, Kontsevich \cite{kontsevich} 
proposed an understanding of mirror symmetry
involving mathematical objects known as derived categories
(collections of complexes of sheaves, for the moment).
At the time, the physical meaning of this proposal was far from clear.

Using sheaves as a tool to describe D-branes was progress towards
understanding the physical meaning of Kontsevich's proposal, but only
a first step.  Another important physical step was Sen's work 
(see {\it e.g.} \cite{sen})
on brane/antibrane annihilation, which helped suggest a physical meaning
for the objects in derived categories:  they should be complexes of
alternating branes and antibranes.  This proposal
first appeared in print in \cite{medercat}, which helped motivate
the suggestion by using properties of Fourier-Mukai transforms
modelling T-duality to suggest that such brane/antibrane complexes were
required.  (See also \cite{paulron} for earlier work suggesting a different
physical
meaning of derived categories, in a very different context.)

Derived categories did not become relatively popular in physics,
however, until the appearance of \cite{doug1,paulalb}, which made significant
progress towards clarifying physical details of the realization of
derived categories in physics, by giving vevs to tachyons (as in
\cite{medercat}), checking consequences of such vevs for the resulting
massive worldsheet theories, and suggesting a notion of stability
for objects of derived categories.  However, giving vevs to tachyons
results in massive worldsheet theories, not BCFT's,
and although these theories were believed to flow in the IR to
BCFT's, no general methods for verifying consequences of the derived
category program directly in BCFT were known until the work of
\cite{ks}.

These lectures are an attempt to explain some of this progress to
a more general physics audience.  Hopefully, they will also serve
to educate mathematicians about how some of this mathematics appears
in physics.  We shall concentrate primarily on understanding
sheaves as models of D-branes, rather than
derived categories.  Although we will only outline work on
derived categories and stability, other lecture notes concentrating
on those topics will appear soon \cite{paultoappear}.

In passing, we should also mention K-theory and
topological D-branes, topics which have recently been reviewed in
\cite{gregk}, as one particular lesson learned from K-theory will
be very important in these lectures. 
Specifically, there is an anomaly due to D.~Freed and E.~Witten
\cite{freeded} that will be very important here.
Recall their work had two implications:
\begin{enumerate}
\item 
If the normal bundle to the submanifold on which a D-brane is wrapped
does not admit a Spin$^c$ structure, then there is an anomaly.
This anomaly enters into the physical realization of the Atiyah-Hirzebruch
spectral sequence for K-theory as the first nontrivial differential.
However, complex vector bundles always admit Spin$^c$ lifts, and so
this very interesting story is always trivial for the cases that we
will be concerned with in these lectures.  
\item If the normal bundle to the submanifold on which a D-brane
is wrapped admits a Spin$^c$ structure but not a Spin structure,
then the bundle on the D-brane worldvolume will be twisted.
Understanding this feature will prove essential to correctly
computing open string spectra, and more generally to attempts to
model D-branes with derived categories.
\end{enumerate}
Although the Freed-Witten anomaly will prove important in these
lectures, K-theory {\it per se} will not, and so we will say nothing
more about K-theory here.

Now, how can one check whether sheaves are a good mathematical model
of D-branes?  
There is no known direct systematic map between open string
boundary conditions and sheaves, but rather sheaf models are
ansatzes for descriptions of D-branes.  We have a dictionary
relating some open string boundary conditions and some sheaves,
but how can we check that dictionary?
The answer is that this mathematical model makes
physical predictions, and we can check those predictions.

In particular, the statement that D-branes are described by sheaves
goes hand-in-hand with the statement that open string spectra
between D-branes should be counted by mathematical objects called  
Ext groups between the sheaves, in the sense that the vector space
of open string states should coincide with the vector space formed from
the Ext groups between the sheaves.  We will explain Ext groups later,
but for the moment, suffice it to say, the relationship proposed
is mathematically extremely natural and is more or less necessary
for current pictures of derived categories in physics to make sense.

To a mathematician, statements that D-branes should be well-modelled
by sheaves sound obvious.
However, physically such statements are often far less clear.
For example, physical quantities are usually only counted by
algebraic geometry when one has supersymmetry, but often brane
intersections are not mutually supersymmetric.  Does one only get Ext groups
in the special case of open strings connecting mutually supersymmetric
D-branes?  Another possible objection comes from thinking about
the origin of parts of the open string spectrum.  Descriptions of
D-branes as sheaves appear to omit information about Higgs fields;
they only appear to encode a choice of submanifold and a bundle on that
submanifold.  If the sheaf description does not encode Higgs fields,
then how can one hope to recover the full open string spectrum
(including parts from Higgs fields) from Ext groups between sheaves?
Perhaps sheaf-theoretic models of D-branes are simply too
simplistic to be very useful.

We will answer these questions and many more, by showing how one
can check sheaf-theoretic models of D-branes on the open string
worldsheet.  We will recover Ext groups in a variety of
different ways, always as a BRST cohomology calculation.
Furthermore, we will use such Ext group computations as a tool
for extending sheaf-theoretic descriptions of D-branes,
checking ansatzes for physical meanings of more general sheaves
by comparing open string spectra to Ext groups.
BRST cohomology and Ext group calculations will be a recurring theme
of these lectures.

We begin in section~\ref{sheafrev} with an overview of sheaves and
Ext groups, mathematics that will be used throughout these lectures.
In section~\ref{Bmodelrev} we review the B model topological field theory,
that will be the physical basis of these lectures.  In section~\ref{vertexext}
we outline how to see Ext groups as boundary R-sector spectra
between D-branes, directly in the BCFT of the B model.
This gives us a powerful computational tool, as we describe some
particularly interesting examples where the physical calculation is
very subtle, but the corresponding mathematics is much easier.
In section~\ref{tachyonext} we review the analogous computation in the
massive theory constructed by realizing complexes of locally-free sheaves
(resolving the sheaves corresponding to the D-branes) as tachyon condensation
in a brane-antibrane system.  In sections~\ref{dercatoverview} and
\ref{gencpx} we review derived categories and their physical realization
via the massive theories introduced in section~\ref{tachyonext}.
In section~\ref{orb} we return to sheaves (instead of derived categories
of sheaves), and compute open string spectra in orbifolds.
The result, expressed in terms of Ext groups, gives us a mechanism for
efficiently computing spectra in examples where other techniques
({\it e.g.} guessing gauge theories or quiver representations) are
untenable.  In section~\ref{btwist} we review how nontrivial $B$ field
backgrounds twist the bundles on D-branes, and the effect this has on
our mathematical models:  honest sheaves are replaced by ``twisted sheaves,''
and the boundary R-sector spectra are counted by corresponding twisted
Ext groups, as we see directly in BCFT.  For most of these lectures
we have implicitly assumed that all Higgs vevs vanish, but in 
section~\ref{nonred} we correct that omission and analyze how Higgs vevs
can also be encoded in sheaves.  In general, a Higgs vev deforms the sheaf
to another sheaf inside the total space of the normal bundle to the submanifold
on which the D-brane lives.  Diagonalizable Higgs vevs have the effect
of moving the sheaf; nilpotent Higgs vevs yield more general types of
sheaves than previously considered, {\it e.g.} structures sheaves of
nonreduced schemes.  In section~\ref{algebraic} we discuss algebraic
properties of the open string B model -- mathematical realizations of
boundary-boundary and bulk-boundary OPE's, and how that mathematics
is realized physically.  Finally in section~\ref{stability} we briefly
outline stability issues in derived categories.

%D-branes are not the only context in which sheaves have made
%an appearance in physics in the last ten years.
%Another place is in heterotic compactifications \cite{distgm},
%where it has been argued that certain very special sheaves,
%closely related to bundles, can be used in heterotic compactifications.

In these notes, we shall assume the reader has had prior exposure
to vector bundles, but little else.  We have attempted to include
enough basic material on sheaf theory to make these notes
self-contained.

\section{Overview of mathematics of sheaves and Ext groups}
\label{sheafrev}

In this section, we shall review some mathematics that will
play an important role later:  sheaves, sheaf cohomology, and
Ext groups.  We begin with some even more basic mathematics,
that of exact sequences, which will be used throughout the text.

\subsection{Complexes and exact sequences}

The language of complexes and exact sequences, standard in
algebraic topology, will be used throughout these lectures.
However, many physicists do not know this language, so in this
introductory section we shall review these concepts.

A {\it complex} of groups, rings, modules, sheaves, {\it etc} is
a collection ${\cal A}_n$ of groups, rings, {\it etc}, with
maps $\phi_n:  {\cal A}_n \rightarrow {\cal A}_{n+1}$
satisfying the important property that $\phi_{n+1} \circ \phi_n = 0$.
Complexes are typically denoted as follows:
\begin{displaymath}
\cdots \:
\stackrel{ \phi_{n-1} }{ \longrightarrow } \:
{\cal A}_n \: 
\stackrel{ \phi_n }{ \longrightarrow } \:
{\cal A}_{n+1} \:
\stackrel{ \phi_{n+1} }{ \longrightarrow } \:
{\cal A}_{n+2} \:
\stackrel{ \phi_{n+2} }{ \longrightarrow } \: \cdots
\end{displaymath}
The defining property of a complex is the relation
$\phi_{n+1} \circ \phi_n = 0$, which tells us that the image
of any map is a subset of the kernel\footnote{The set of elements mapped
to zero.} of the next map in the sequence. 

An {\it exact} sequence is a special kind of complex,
namely one in which the image of each map is the same as the
kernel of the next map, not just a subset.
This is a stronger statement than merely 
$\phi_{n+1} \circ \phi_n = 0$.
For example, for the complex
\begin{displaymath}
{\cal A} \:
\stackrel{ \phi }{ \longrightarrow } \:
{\cal B} \: \longrightarrow \: 0
\end{displaymath}
to be exact implies that $\phi$ is surjective (onto):  
the kernel of the right map
is all of ${\cal B}$, since the right map sends all of ${\cal B}$ to zero,
yet since the complex is exact, the kernel of each map equals the image
of the previous map, so the image of $\phi$ is all of ${\cal B}$,
hence $\phi$ is surjective.
Similarly, for the complex
\begin{displaymath}
0 \: \longrightarrow \:
{\cal A} \:
\stackrel{ \phi }{ \longrightarrow } \:
{\cal B} 
\end{displaymath}
to be exact implies that $\phi$ is injective (one-to-one):
the image of the left map is zero, but since the complex is exact,
the image of each map equals the kernel of the next, so the kernel
of $\phi$ is zero, hence $\phi$ is injective.

{\it Short exact sequences} are three-element exact sequences of the form
\begin{displaymath}
0 \: \longrightarrow \:
{\cal A} \:
\stackrel{ \phi_1 }{ \longrightarrow } \:
{\cal B} \:
\stackrel{ \phi_2 }{\longrightarrow } \:
{\cal C} \:
\longrightarrow \: 0
\end{displaymath}
From the discussion above, we see that $\phi_1$ is injective and
$\phi_2$ is surjective, and the image of $\phi_1$ equals the kernel
of $\phi_2$.

\subsection{Sheaves}

Nowadays most physicists are familiar with bundles, and the important role they
have played in gauge theories.  But, what is a sheaf?
One motivation for sheaves is as the mathematical machinery
needed to make sense of, for example,
a vector bundle living only over a submanifold
(a notion with obvious applications to modelling D-branes),
and other more abstract settings where bundles are no longer a sensible
concept.

One definition of a sheaf on a space $X$ is as a mechanism for
associating a set, or group, or ring, or module, or even a category, 
to every open set on 
$X$.  For simplicity, we will focus on sheaves of
abelian groups, but the definitions extend in a straightforward
way to other cases.  Now, let ${\cal S}$ be such a sheaf.
The abelian group ${\cal S}(U)$ assigned to an open set $U$ is known
as the (set of) {\it sections} of the sheaf over $U$.
Now, not any collection of sections over open sets will do:
in order to be a sheaf, a number of properties must be satisfied.
First, for every inclusion $V \subseteq U$, we need to specify
a {\it restriction map} $\rho_{U,V}: {\cal S}(U) \rightarrow {\cal S}(V)$,
such that for any triple inclusion $W \subseteq V \subseteq U$,
the restriction from $U$ to $W$ is the same as the composition of the
restrictions from $U$ to $V$ and then from $V$ to $W$:
\begin{displaymath}
\rho_{U,W} \: = \: \rho_{V,W} \circ \rho_{U,V}
\end{displaymath}
and such that the restriction map associated to the identity $U \subseteq U$
is the identity map ${\cal S}(U) \rightarrow {\cal S}(U)$.
We shall usually denote the restriction map with a vertical bar,
as $|_V$, in the usual way.
A collection of sets $\{ {\cal S}(U) \}$ together with restriction
maps gives us a {\it presheaf} of sets.
To get a {\it sheaf} of sets, we must impose what are called the
``gluing'' conditions:
\begin{enumerate}
\item For any pair of intersecting open sets $U$, $V$, and sections
$\sigma \in {\cal S}(U)$, $\tau \in {\cal S}(V)$ such that
\begin{displaymath}
\sigma|_{U \cap V} \: = \: \tau_{U \cap V}
\end{displaymath}
there exists a section $\rho \in {\cal S}(U \cup V)$ such that
$\rho|_U = \sigma$ and $\rho|_V = \tau$.
In other words, sections glue together in the obvious way.
\item If $\sigma \in {\cal S}(U \cup V)$ and
\begin{displaymath}
\sigma|_U \: = \: \sigma|_V \: = \: 0
\end{displaymath}
then $\sigma = 0$.
\end{enumerate}

One of the most basic examples of a sheaf is the sheaf of sections of
a vector bundle.  Given a vector bundle on a space, we can associate to
any open set $U$ the group of sections of the bundle over that open set.
In fact, technically there are several ways to get a sheaf from a vector
bundle -- we could associate smooth sections, or we could associate 
holomorphic sections if the vector bundle has a complex structure,
or instead of creating a sheaf of sets, we could create a sheaf of modules,
which is the more usual construction.
For the purposes of these physics lectures, these distinctions will
largely be irrelevant.  Technically, we will almost always be interested
in sheaves of modules of holomorphic sections, but will speak loosely
of other cases.

Sheaves have a property known as being ``locally free'' if they come
from holomorphic vector bundles, in the fashion above.  For most of these notes,
we shall ignore the distinction between locally-free sheaves and
holomorphic vector bundles, and will use the terms interchangeably.

We should also mention some notation that will be used throughout
these notes.  A holomorphic line bundle with first Chern class $c_1$
on a given space will typically be denoted ${\cal O}(c_1)$.
This notation makes most sense on projective spaces, where the
first Chern class is simply an integer, so that ${\cal O}(n)$ denotes
a holomorphic line bundle of first Chern class $n$.

Another easy example of a sheaf is the {\it skyscraper sheaf}.
Let $X$ be a space, and $p$ be a point on that space.
Now, associate sections ${\cal S}(U)$ to open sets as follows:
\begin{itemize}
\item If $U$ contains $p$, then ${\cal S}(U) = {\bf C}$.
\item If $U$ does not contain $p$, then ${\cal S}(U) = \{ 0 \}$.
\end{itemize}
It is easy to check that this is a sheaf.
Moreover, the support of the sheaf ${\cal S}$ (meaning, the subset of $X$
over which the sheaf is nonzero) is only the point $p$.

Skyscraper sheaves are the simplest examples of ``vector bundles on
submanifolds'' alluded to earlier.

Given a continuous map $i: Y \rightarrow X$ and a sheaf ${\cal S}$ on $Y$,
we can form a sheaf denoted $i_* {\cal S}$ on $X$, defined as follows:
$i_* {\cal S}(U) \equiv {\cal S}(i^{-1}(U))$.

For example, given a vector bundle ${\cal E}$ (or rather, the associated sheaf)
on a submanifold $S$ of a manifold $X$, with inclusion map $i: S
\hookrightarrow X$, we can form the sheaf $i_* {\cal E}$ on $X$.
It is easy to check that $i_* {\cal E}$ only has support on $S$ -- if an
open set in $X$ does not intersect $S$, then there are no sections associated
to that open set.  The sheaf $i_* {\cal E}$ is the more precise meaning
of the phrase ``vector bundle on a submanifold,'' and is an example
of what is known technically as a {\it torsion sheaf}.
Notice that a skyscraper sheaf can also be put in the form
$i_* {\cal E}$, where $i: p \hookrightarrow X$ and ${\cal E}$ is the rank
one line bundle on the point $p$, and so skyscraper sheaves
are also examples of torsion sheaves.

In particular, later we shall study the extent to which we can
describe a D-brane on a complex submanifold $S$ with holomorphic
vector bundle ${\cal E}$ by the sheaf $i_* {\cal E}$, for example by
comparing open string spectra to Ext groups (which will be discussed
momentarily).  

Another example of a sheaf is the sheaf of maps into ${\bf Z}$,
that assigns, to every open set $U$, the set of continuous maps
$U \rightarrow {\bf Z}$.
This sheaf comes up in considerations of cohomology, but is less useful
for modelling D-branes.

There are many more kinds of sheaves, more than we shall be able to 
discuss here.  Another way to describe sheaves is in terms of
modules over coordinate rings (\cite[section 5.3]{gh},
\cite[section II.5]{hart}), which
often gives results not of forms previously discussed.
For example, sheaves on ${\bf C}^2$ can be built by specifying
${\bf C}[x,y]$-modules.  The trivial line bundle on ${\bf C}^2$
is described by the ${\bf C}[x,y]$ module given by
${\bf C}[x,y]$ itself.
The skyscraper sheaf supported at the origin
corresponds to the ${\bf C}[x,y]$-module
\begin{displaymath}
{\bf C}[x,y]/(x,y)
\end{displaymath}
where $(x,y)$ denotes the ideal of the ring ${\bf C}[x,y]$ generated
by $x$ and $y$.  The sheaf corresponding to the trivial line bundle
over the subvariety $x=0$ is described by the ${\bf C}[x,y]$-module
\begin{displaymath}
{\bf C}[x,y]/(x).
\end{displaymath}
A more interesting family of sheaves supported over the origin of
${\bf C}^2$ is described by the one-parameter-family of ${\bf C}[x,y]$-modules
given by
\begin{displaymath}
{\bf C}[x,y]/(x^2, y- \alpha x)
\end{displaymath}
for some complex number $\alpha$.  These are {\it not} vector bundles
over their supports in any sense, making their physical interpretation
somewhat confusing.  We shall see in section~\ref{nonred} 
that these
sheaves model pairs of D0 branes, sitting at the origin of ${\bf C}^2$,
with nilpotent Higgs vevs.

A word on notation.  Sections of a sheaf are often denoted with $\Gamma$,
{\it i.e.} the sections of a sheaf ${\cal S}$ over $U$ are
sometimes denoted $\Gamma(U, {\cal S})$.

Some basic references on sheaves are \cite[section 0.3]{gh}, 
\cite[section 2]{gunning}, and \cite[section II.10]{botttu}.

\subsection{Sheaf cohomology}

One can define cohomology groups with coefficients in sheaves,
just as one can define cohomology with coefficients in
${\bf R}$ and ${\bf Z}$.  Such cohomology groups are called sheaf
cohomology.  Sheaf cohomology groups commonly arise in physics
when counting vertex operators in string compactifications.
See for example \cite{dg} for a discussion of how sheaf cohomology
counts spectra in heterotic compactifications, and
\cite{edtft} for a discussion of how sheaf cohomology counts
spectra in the B model topological field theory, for just two examples.
Sheaf cohomology will play an important role in understanding open string
spectra also, so we shall take a few minutes to review it.

In principle sheaf cohomology of degree $n$ is defined as
$n$-cochains, closed with respect to a coboundary operator, modulo
exact cochains.  Let us describe what that means more precisely,
then work through some applications.

Let $\{ U_{\alpha} \}$ be a good open cover of our space $X$, and
${\cal S}$ a sheaf on $X$.  Define cochains of degree $n$,
denoted $C^n({\cal S})$, as follows:
\begin{eqnarray*}
C^0({\cal S}) & = & \coprod_{\alpha} {\cal S}(U_{\alpha}) \\
C^1({\cal S}) & = & \coprod_{\alpha \neq \beta} {\cal S}(
U_{\alpha} \cap U_{\beta} ) \\
& \cdots & \\
C^n({\cal S}) & = & \coprod_{\alpha_0 \neq \cdots \neq \alpha_n}
{\cal S}( U_{\alpha_0} \cap \cdots \cap U_{\alpha_n} )
\end{eqnarray*}
We define a coboundary operator $\delta: C^p({\cal S}) \rightarrow
C^{p+1}({\cal S})$ by,
\begin{displaymath}
( \delta \sigma )_{ i_0, \cdots, i_p } \: = \:
\sum_{j=0}^{p+1} (-)^j \sigma_{i_0, \cdots, \widehat{ i_j }, \cdots,
i_{p+1} }|_{ U_{i_0} \cap \cdots \cap U_{i_p} }.
\end{displaymath}
Then, sheaf cohomology is defined as $\delta$-closed cochains,
modulo $\delta$-exact cochains.  (Strictly speaking, we should take
a direct limit over open covers in order to obtain sheaf cohomology,
but for our purposes that technicality will be irrelevant.)
The resulting cohomology groups are denoted
$H^n(X, {\cal S})$, where $n$ is called the degree.

To help clarify the meaning of sheaf cohomology, let us take a moment
to verify the following claim:  line bundles are classified by degree 1
sheaf cohomology.
Consider the sheaf $C^{\infty}(U(1))$ of smooth maps into $U(1)$,
or, alternatively, the sheaf ${\cal O}^*$ of nowhere-zero holomorphic functions.
A degree one element of sheaf cohomology is a collection of 
either smooth $U(1)$-valued maps or nowhere-zero holomorphic functions,
defined on overlaps of elements of a good open cover.
Let such a collection of functions on overlaps be denoted $g_{\alpha \beta}$.
Then, the condition of $\delta$-closedness implies that the $g_{\alpha \beta}$
close on triple overlaps:
\begin{displaymath}
g_{\alpha \beta} g_{\beta \gamma} g_{\gamma \alpha} \: = \: 1.
\end{displaymath}
Thus, closed degree one cochains, valued in such sheaves, define transition
functions for a smooth principal $U(1)$ bundle, or, with the alternative
sheaf, a holomorphic line bundle.
Moreover, for two such cochains $\{ g_{\alpha \beta} \}$,
$\{ g'_{\alpha \beta} \}$ to be related by $\delta$-exact cochains
means that
\begin{displaymath}
g'_{\alpha \beta} \: = \:
\phi_{\alpha} g_{\alpha \beta} \phi_{\beta}^{-1}
\end{displaymath}
which is precisely the condition for the two sets of transition functions
to define isomorphic bundles.
Thus, we see explicitly that $H^1(X, C^{\infty}(U(1)) )$ classifies
isomorphism classes of principal $U(1)$ bundles,
and that $H^1(X, {\cal O}^*)$ classifies isomorphism classes of holomorphic
line bundles -- in both cases, the cochains are the transition functions.

Degree zero sheaf cohomology also has an easy interpretation.
From the definition, degree zero sheaf cohomology consists of
one section $\sigma_{\alpha} \in {\cal S}(U_{\alpha})$ for each $\alpha$,
and closure under $\delta$ implies that 
\begin{displaymath}
\sigma_{\alpha}|_{U_{\alpha} \cap U_{\beta} } \: = \:
\sigma_{\beta}|_{ U_{\alpha} \cap U_{\beta} }.
\end{displaymath}
In other words, degree zero sheaf cohomology consists of a section
over each open set, such that the sections agree on overlaps.
It should be clear that such a set of data is precisely a global
section of the sheaf:  
\begin{displaymath}
H^0(X, {\cal S}) \: = \: \Gamma(X, {\cal S}) \: = \: {\cal S}(X).
\end{displaymath}

When the sections of the coefficient sheaf are {\it constant} maps,
as for the sheaf ${\bf Z}$ described earlier, then sheaf cohomology
is identical to ordinary cohomology:
\begin{displaymath}
H^n_{sheaf}(X, {\bf Z}) \: = \: H^n_{std}(X, {\bf Z} ).
\end{displaymath}

For most of this paper, we shall only be interested in sheaf
cohomology on complex manifolds with coefficients in holomorphic sheaves.
However, there are some amusing games one can play with more
general sheaves, which we wish to take a moment to describe.
In general, any map of sheaves ${\cal S} \rightarrow {\cal T}$
induces a map between sheaf cohomology groups
$H^n(X, {\cal S}) \rightarrow H^n(X, {\cal T})$,
and also, a short exact sequence of sheaves
\begin{displaymath}
0 \: \longrightarrow \: {\cal S} \: \longrightarrow \: {\cal T} \:
\longrightarrow \: {\cal U} \: \longrightarrow \: 0
\end{displaymath}
induces a long exact sequence of sheaf cohomology groups
\begin{displaymath}
\cdots \: \longrightarrow \:
H^n(X, {\cal S} ) \: \longrightarrow \: H^n(X, {\cal T}) \: \longrightarrow
\: H^n(X, {\cal U} ) \: \longrightarrow \:
H^{n+1}(X, {\cal S} ) \: \longrightarrow \: \cdots
\end{displaymath}
Using this fact we can quickly derive the topological classification
of principal $U(1)$ bundles by first Chern classes, as follows.
There is a short exact sequence of sheaves
\begin{displaymath}
0 \: \longrightarrow \: {\bf Z} \: \longrightarrow \:
C^{\infty}({\bf R}) \: \longrightarrow \:
C^{\infty}(U(1)) \: \longrightarrow \: 0
\end{displaymath}
which induces a long exact sequence as above.  However, 
$H^n(X, C^{\infty}({\bf R})) = 0$ for $n>0$, so, for example, we
find immediately that
\begin{displaymath}
H^1(X, C^{\infty}(U(1))) \: \cong \: H^2(X, {\bf Z}).
\end{displaymath}
As we have already seen, the group on the left classifies principal
$U(1)$ bundles, and their images in $H^2(X, {\bf Z})$ under the induced
map in the long exact sequence are just their first Chern classes.
Another relationship that can be immediately derived from the
associated long exact sequence is that
\begin{displaymath}
H^2(X, C^{\infty}(U(1))) \: \cong \: H^3(X, {\bf Z})
\end{displaymath}
which will play an important role when we discuss $B$ fields.

Just as ordinary cohomology of a space can be realized by differential
forms for special coeffients ({\it i.e.}, ${\bf R}$ coefficients),
sheaf cohomology can be realized in terms of differential forms when the
coefficients are locally-free sheaves -- holomorphic vector bundles.
In particular, if ${\cal E}$ is a holomorphic vector bundle,
then $H^n(X, {\cal E})$ is the same as $\overline{\partial}$-closed
$(0,n)$-differential forms valued in the bundle ${\cal E}$,
modulo $\overline{\partial}$-exact differential forms.

Here are some useful facts for calculating sheaf cohomology on complex
manifolds,
for coefficients in holomorphic sheaves:
\begin{itemize}
\item The trivial rank one line bundle ${\cal O}$ has only one
(constant) section on a compact manifold,
up to scale:  $H^0(X, {\cal O}) = {\bf C}$.
\item A line bundle of negative degree has no sections:
$H^0(X, {\cal O}(-n) ) = 0$.
\item A skyscraper sheaf ${\cal O}_p$ has one section and no
higher cohomology:
\begin{displaymath}
H^n(X, {\cal O}_p ) \: = \: \left\{ \begin{array}{cl}
                                    {\bf C} & n=0 \\
                                     0 & n>0
                                     \end{array} \right.
\end{displaymath}
\item There is a symmetry known as Serre duality 
which can be used to relate sheaf cohomology groups
of different degrees.  Specifically, on a complex manifold of
(complex) dimension $n$, for a bundle ${\cal E}$, Serre duality
is the statement that
\begin{displaymath}
\mbox{dim }H^p(X, {\cal E}) \: = \:
\mbox{dim }H^{n-p}(X, {\cal E}^{\vee} \otimes K_X)
\end{displaymath}
where $K_X$ is the canonical line bundle (the top exterior power
of the holomorphic part of the cotangent bundle), and 
${\cal E}^{\vee}$ denotes the dual bundle.
\item The alternating sum of the dimensions of sheaf cohomology groups
with coefficients in a bundle ${\cal E}$ can be computed
with the Hirzebruch-Riemann-Roch formula \cite{hirzebruch}:
\begin{displaymath}
\sum_i (-)^i \mbox{dim }H^i(X, {\cal E}) \: = \:
\int_X  \mbox{ch}({\cal E}) \wedge \mbox{td}(TX) 
\end{displaymath}
\end{itemize}

Elementary discussions of sheaf cohomology can be found in
\cite[section 0.3]{gh}, \cite[section 3]{gunning}, and 
\cite[section II.10]{botttu}.

\subsection{Ext groups}   \label{extrev}

Why are sheaves a useful tool for describing D-branes?
How can one extract physics from such a description?
One of the standard claims is that spectra of open strings
between D-branes described by sheaves, should be counted by
mathematical objects called Ext groups, namely
integrally-graded sets of vector spaces that are computed from
a pair of sheaves.  Sheaf cohomology, by contrast, was
integrally-graded sets of vector spaces computed from a single sheaf.
In this section, we shall
describe Ext groups, and check that they count D-brane spectra
in a few examples.  Later we will show several general arguments for
the correspondence.

First, some generalities.
Ext groups are often described between modules over rings.
%Degree one Ext groups can be interpreted as classifying equivalence
%classes of extensions.  However, Ext groups can be defined in more
%general degrees, not just degree one.
We outlined earlier that one way to think of certain kinds of 
sheaves is as modules over rings of algebraic functions,
so the reader should not be surprised to learn that
the notion of Ext groups between modules can be extended to 
certain kinds of sheaves.
In fact,
Ext groups $\mbox{Ext}^n_A(M,N)$ between two $A$-modules $M$, $N$
can be used to define two notions of Ext between sheaves ${\cal S}$,
${\cal T}$:
\begin{itemize}
\item Global Ext:  $\mbox{Ext}^n_X\left( {\cal S}, {\cal T}\right)$
are groups, one for each integer $n$, defined for pairs of sheaves.
Physically these global Ext groups count open string spectra, as we shall
check in some examples shortly.
\item Local Ext:  $\underline{\mbox{Ext}}^n_{ {\cal O}_X }\left(
{\cal S}, {\cal T}\right)$.  These are not groups, but rather
are sheaves, obtained by applying Ext's for modules on a 
``fiber-by-fiber'' basis.
\end{itemize}
In each case, the integer $n$ is known as the degree.
Local and global Ext's between two sheaves are closely related.
In fact, we shall soon describe a method for computing global Ext's
that relies on computing local Ext's as an intermediate step.

We mentioned that local and global Ext's can only be defined
for some kinds of sheaves.  For the purposes of these lectures,
we shall only define Ext's between {\it coherent} sheaves.
A coherent sheaf is, for our purposes,
a finite-rank (holomorphic) sheaf with the property that there exists
an exact sequence whose elements are all vector bundles,
except for the last object in the sequence, which is the
sheaf in question.  In other words, we say ${\cal S}$ is a coherent
sheaf if there exist bundles (locally-free sheaves) ${\cal E}_0, \cdots,
{\cal E}_n$ and maps such that
\begin{displaymath}
0 \: \longrightarrow \: {\cal E}_n \: \longrightarrow \:
{\cal E}_{n-1} \: \longrightarrow \: \cdots \: 
\longrightarrow \: {\cal E}_0 \: \longrightarrow \: {\cal S}
\: \longrightarrow \: 0
\end{displaymath}
is exact, for some $n$.
Such an exact sequence is known as a locally-free {\it resolution}
of the (coherent) sheaf ${\cal S}$, and is also an example
of a projective resolution (for local Ext, in any event).

Intuitively, a coherent sheaf is a sheaf that can be sensibly acted upon
by the ring of algebraic functions on a space.
Holomorphic vector bundles define coherent sheaves, as do skyscraper sheaves,
and pushforwards of vector bundles.  Non-holomorphic sheaves such as the
sheaf $C^{\infty}(U(1))$ are not coherent.  Another non-coherent sheaf
is the sheaf ${\bf Z}$ -- multiplying a local section by an algebraic
function would give a non-constant map, which could no longer be
a section of that sheaf.  In this language, D-branes will always be
modelled by coherent sheaves.

Now, how do we calculate global Ext groups between two coherent sheaves?
Later, while learning how Ext groups are realized physically,
we shall learn more efficient computational methods, but for the moment,
we shall describe a more nearly first-principles approach.
Let ${\cal S}$ and ${\cal T}$ be two coherent sheaves.
For a first-principles calculation, we shall need to know
a locally-free resolution of ${\cal S}$:
\begin{displaymath}
0 \: \longrightarrow \: {\cal E}_n \: \longrightarrow \: 
{\cal E}_{n-1} \: \longrightarrow \: \cdots \:
\longrightarrow \: {\cal E}_0 \: \longrightarrow \:
{\cal S} \: \longrightarrow \: 0
\end{displaymath}

For our first step to computing the global Ext groups
$\mbox{Ext}^n_X\left( {\cal S}, {\cal T} \right)$, we need to compute
the local Ext sheaves $\underline{\mbox{Ext}}^n_{ {\cal O}_X }\left(
{\cal S}, {\cal T} \right)$.  These sheaves are defined as the
cohomology of the complex
\begin{displaymath}
0\: \longrightarrow \: \underline{\mbox{Hom}}\left( {\cal E}_0, {\cal T} \right) 
\: \longrightarrow \:
\underline{\mbox{Hom}}\left( {\cal E}_1, {\cal T} \right)
\: \longrightarrow \: \cdots \: \longrightarrow \:
\underline{\mbox{Hom}}\left( {\cal E}_n, {\cal T} \right) 
\: \longrightarrow \: 0
\end{displaymath}
Although computing the cohomology of this complex looks very messy
in general, in our first examples it will actually be much easier,
and soon we shall learn alternative methods for computing global Ext groups.
Locally-free resolutions of a coherent sheaf are not unique in
general; however, the local Ext sheaves obtained from this procedure
{\it are} uniquely determined.

Given local Ext sheaves, we can now obtain global Ext groups
using the famous ``local-to-global'' spectral sequence:
\begin{displaymath}
E_2^{p,q} \: = \:
H^p\left( X, \underline{\mbox{Ext}}^q_{ {\cal O}_X } \left(
{\cal S}, {\cal T} \right) \right) \:
\Longrightarrow \:
\mbox{Ext}^{p+q}_X\left( {\cal S}, {\cal T} \right).
\end{displaymath}

For those readers who have not seen spectral sequences before,
a spectral sequence is a way of constructing a group through a series
of successive approximations, much like a sculptor creating a statue
by chipping away at a block of marble.  At stage $r$, the next approximation
is created by taking kernels and modding out images of
differentials $d_r: E_r^{p,q} \rightarrow E_r^{p+r,q-r+1}$.
That set of kernels, modulo images, defines the groups $E_{r+1}^{p,q}$,
to which one applies the differentials again, and so forth, until
the process terminates.

Now, in general computing spectral sequences can be quite tricky,
but in practice, it is often the case that most of the groups $E_2^{p,q}$
vanish, so most of the differentials are necessarily trivial, making
the computation much easier.  We shall not discuss the differentials
of the local-to-global spectral sequence here (indeed, one usually
tries to avoid having to explicitly describe differentials),
but later in discussing vertex operator representations of Ext groups,
we shall explicitly write out some differentials in another spectral
sequence.

Another important property of Ext groups is Serre duality.
We previously mentioned Serre duality in connection with sheaf
cohomology of bundles.  There is also a generalization involving
Ext groups between sheaves.  The generalization states that for
$X$ an $n$-fold,
\begin{displaymath}
\mbox{dim }\mbox{Ext}^p_X\left( {\cal S}, {\cal T}\right) \: = \:
\mbox{dim }\mbox{Ext}^{n-p}_X\left( {\cal T}, {\cal S} \otimes K_X \right)
\end{displaymath}
where $K_X$ is the canonical (line) bundle on $X$.
For D-branes on a Calabi-Yau, Serre duality encodes the relation between
open string spectra for two open strings with opposite orientations
but the same boundary D-branes.

As an easy example of this technology in action, let us compute
$\mbox{Ext}^n_X\left( {\cal E}, {\cal F}\right)$, where ${\cal E}$ and
${\cal F}$ are both bundles on $X$.
Since ${\cal E}$ is a bundle, its locally-free resolution is trivial:
\begin{displaymath}
0 \: \longrightarrow \: {\cal E} \:
\stackrel{ = }{\longrightarrow} \: {\cal E} \:
\longrightarrow \: 0
\end{displaymath}
Local Ext's are then the cohomology of the complex
\begin{displaymath}
0 \: \longrightarrow \:
\underline{\mbox{Hom}}\left( {\cal E}, {\cal F} \right)
\: \longrightarrow \: 0
\end{displaymath}
from which we read off that
\begin{displaymath}
\underline{\mbox{Ext}}^n_{ {\cal O}_X } \left( {\cal E}, {\cal F} \right)
\: = \: \left\{ \begin{array}{cl}
                \underline{\mbox{Hom}}( {\cal E}, {\cal F} ) = {\cal E}^{\vee}
\otimes {\cal F} & n=0 \\
                0 & n>0
                \end{array} \right.
\end{displaymath}
Global Ext's are then computed from the spectral sequence
\begin{displaymath}
E_2^{p,q} \: = \:
H^p\left( X, \underline{\mbox{Ext}}^q\left( {\cal E}, {\cal F} \right) \right)
\: \Longrightarrow \: \mbox{Ext}^{p+q}_X\left( {\cal E}, {\cal F} \right).
\end{displaymath}
In the present case, local Ext vanishes for $q>0$,
so $E_2^{p,q} = 0$ for $q>0$, and since all the $d_r$ map between
different $q$ for $r \geq 2$, we see that the spectral sequence is
necessarily trivial, and
\begin{displaymath}
\mbox{Ext}^n_X\left( {\cal E}, {\cal F} \right) \: = \:
H^n\left( X, {\cal E}^{\vee} \otimes {\cal F} \right).
\end{displaymath}

Physically, the Ext groups above are supposed to count open string
states between two D-branes wrapped on all of $X$, with gauge bundles
${\cal E}$ and ${\cal F}$, in the sense that the vector space of
open string states should coincide with the vector space
\begin{displaymath}
\bigoplus_n \mbox{Ext}^n_X\left( {\cal E}, {\cal F} \right).
\end{displaymath}
Now, the open string spectrum in precisely
this case was computed previously in \cite{edcs}, and indeed, there it
was shown that the open string spectrum is counted by
$H^n(X, {\cal E}^{\vee} \otimes {\cal F} )$, in perfect agreement
with $\mbox{Ext}^n_X\left( {\cal E}, {\cal F}\right)$.

For another example, let us consider open strings stretched between
two D0 branes both supported at a point on ${\bf C}^3$.
Let us check the claim that the open string spectrum is counted by
Ext groups, by calculating the Ext groups in this case, and then
comparing to the (known) open string spectrum.
In particular, let us calculate $\mbox{Ext}^n_{ {\bf C}^3 }\left(
{\cal O}_p, {\cal O}_p \right)$, where ${\cal O}_p$ is a skyscraper
sheaf in ${\bf C}^3$ supported at the point $p$.
Without loss of generality, assume $p$ is the origin.
Then, a locally-free resolution of ${\cal O}_p$ is given\footnote{
In special cases, such as the present case, there are methods to
determine locally-free resolutions of sheaves ({\it e.g.} Koszul resolutions).  
Explaining such methods would take us rather far afield.  Rather than
try to explain how to find such resolutions in general, we shall simply
state them wherever needed, and refer the reader to the references
for information on their derivation.
} by
\begin{displaymath}
0 \: \longrightarrow \:
{\cal O} \: \stackrel{{\scriptsize \left[ \begin{array}{c}
                       -x \\ y \\ -z \end{array} \right] }}{\longrightarrow}
\: {\cal O}^3 \:
\stackrel{{\scriptsize \left[ \begin{array}{ccc}
                  0 & -z & -y \\
                  -z & 0 & x \\
                  y & x & 0 \end{array} \right] }}{ \longrightarrow } \:
{\cal O}^3 \: 
\stackrel{ [ x,y,z] }{ \longrightarrow } \:
{\cal O} \: \longrightarrow \: {\cal O}_p \: \longrightarrow \: 0
\end{displaymath}
Thus, local Ext's are the cohomology of the complex
\begin{displaymath}
0 \: \longrightarrow \:
\underline{\mbox{Hom}}\left( {\cal O}, {\cal O}_p \right) 
\: \longrightarrow \:
\underline{\mbox{Hom}}\left( {\cal O}^3, {\cal O}_p \right) 
\: \longrightarrow \:
\underline{\mbox{Hom}}\left( {\cal O}^3, {\cal O}_p \right) 
\: \longrightarrow \:
\underline{\mbox{Hom}}\left( {\cal O}, {\cal O}_p \right) 
\: \longrightarrow \: 0
\end{displaymath}
Now,
\begin{eqnarray*}
\underline{\mbox{Hom}}\left( {\cal O}, {\cal O}_p \right) & = & {\cal O}_p, \\
\underline{\mbox{Hom}}\left( {\cal O}^3, {\cal O}_p \right) & = & {\cal O}_p^3, 
\end{eqnarray*}
and the maps in the complex all vanish\footnote{
Each map in the complex is obtained from a map in the resolution,
in the following fashion.  Given a map $f: A \rightarrow B$,
we can define a map $f^{\vee}: \mbox{Hom}(B,C) \rightarrow 
\mbox{Hom}(A,C)$ by composing any element of $\mbox{Hom}(B,C)$
with $f$ to get an element of $\mbox{Hom}(A,C)$.
In the present case, each of the maps in the locally-free resolution
of ${\cal O}_p$
vanishes where $x=y=z=0$.  Since the `dual' maps are obtained by
composing with the original maps, and the original maps all vanish
at the only places where the sheaves are nonzero, we see that the
maps in the complex defining local $\underline{\mbox{Ext}}$'s necessarily
all vanish.
} where ${\cal O}_p$ has
support, so computing the cohomology is trivial -- the cohomology at
degree $n$ is the same as the sheaf at position $n$.
In other words, the local Ext's are given by
\begin{displaymath}
\underline{\mbox{Ext}}^n_{ {\cal O}_{ {\bf C}^3 } } \left(
{\cal O}_p, {\cal O}_p \right) \: = \:
\left\{ \begin{array}{cl}
        \underline{\mbox{Hom}}( {\cal O}, {\cal O}_p ) = {\cal O}_p & n=0 \\
     \underline{\mbox{Hom}}( {\cal O}^3, {\cal O}_p ) = {\cal O}_p^3 & n=1 \\
     \underline{\mbox{Hom}}( {\cal O}^3, {\cal O}_p ) = {\cal O}_p^3 & n=2 \\
     \underline{\mbox{Hom}}( {\cal O}, {\cal O}_p ) = {\cal O}_p & n=3
     \end{array} \right.
\end{displaymath}
Since the support of the local Ext's is in dimension zero, the local-to-global
spectral sequence is trivial -- only degree zero sheaf cohomology can be
nonzero, so $E_2^{p,q} = 0$ if $p>0$, and $d_r$ maps between different $p$'s
for $r > 0$ -- hence we can read off
\begin{displaymath}
\mbox{Ext}^n_{ {\bf C}^3 } \left( {\cal O}_p, {\cal O}_p \right) \: = \:
\left\{ \begin{array}{cl}
        {\bf C} & n=0 \\
        {\bf C}^3 & n=1 \\
        {\bf C}^3 & n=2 \\
        {\bf C} & n=3
        \end{array} \right.
\end{displaymath}

Physically the Ext groups above count open string spectra for 
open strings connecting a D-brane at a point on ${\bf C}^3$ back
to itself, or to another D-brane at the same point, in that the 
vector space of open string states should coincide with the
vector space
\begin{displaymath}
\bigoplus_n \mbox{Ext}^n_{ {\bf C}^3 }\left( {\cal O}_p, {\cal O}_p \right).
\end{displaymath}
The spectrum of such open strings should contain a single vector
 -- the abelian gauge field on the brane -- plus three scalars --
Higgs fields corresponding to the normal bundle, and indeed in degree zero
we have one element, and in degree one we have three elements.

Now, as described in more detail in \cite{dg}, string spectrum calculations
count both antichiral superfields as well as chiral superfields, for example.
The degree two and three states are Serre dual to the degree one and two states
above, and are interpreted essentially as antiparticles, coming from open
strings with the opposite orientation.

In this simple example, the fact that the gauge field comes from
Ext groups of degree zero, and Higgs fields from Ext groups of degree one,
is not precisely a coincidence (nor, unfortunately, will it always be the case).
This can be viewed as a direct effect of the physical constraint
known as the GSO projection,
which in general can be used to determine the spacetime properties of 
vertex operators
from the charges of factors in the internal CFT.  
GSO selects out charge one states
 -- so if we start with a charge zero part from the internal CFT,
then we must multiply by a worldsheet fermion $\psi^{\mu}$, creating
a charge one state which describes a vector in spacetime.
Similarly, starting with a charge one state from the internal CFT
yields spacetime scalars.

Unfortunately, unlike many other cases where cohomology can be used
to count spectra, GSO arguments as the above tell us that
the full $U(1)_R$ charge of a boundary state in the boundary ${\cal N}=2$
algebra
is {\it not} the same as the degree of the Ext group, when the boundary
conditions on the two sides of the open string are not the same,
due to contributions to the charge from the vacuum.  We shall
see this more explicitly in our next example.

For our next example, let us compute the open string spectrum
corresponding to the ADHM construction.  
In particular, it is well-known that
the moduli space of $k$ instantons in $U(N)$ is equivalent
to the classical Higgs moduli space of a six-dimensional $U(k)$ gauge
theory with hypermultiplets in the $(k,N)$ of $U(k) \times U(N)$ and
a single hypermultiplet in the adjoint of $U(k)$.  In other words,
the ADHM construction of $k$ instantons in $U(N)$ can be recovered from,
say, $k$ D5-branes on $N$ D9-branes.
As an easy special case, let us look at open strings connecting
a D-brane at a point on ${\bf C}^2$ to a D-brane supported on all
of ${\bf C}^2$.  The spectrum of such open strings gives the
hypermultiplets valued in the $(k,N)$ of $U(k) \times U(N)$.
We will check our claims that Ext groups compute open string spectra
by reproducing that result from Ext groups between a skyscraper sheaf
on ${\bf C}^2$, and the trivial line bundle on ${\bf C}^2$.

Let us calculate $\mbox{Ext}^n_{ {\bf C}^2 }\left(
{\cal O}_p, {\cal O} \right)$.
A locally-free resolution of the skyscraper sheaf ${\cal O}_p$ on ${\bf C}^2$,
supported at the origin, say, is given by
\begin{displaymath}
0 \: \longrightarrow \: {\cal O} \:
\stackrel{{\scriptsize \left[ \begin{array}{c}  -y \\ x \end{array} \right] 
}}{\longrightarrow} \: {\cal O}^2 \:
\stackrel{ [ x,y] }{\longrightarrow} \: {\cal O} \:
\longrightarrow \: {\cal O}_p \: \longrightarrow \: 0
\end{displaymath}
so the local $\underline{\mbox{Ext}}$ sheaves are the cohomology sheaves
of the complex
\begin{displaymath}
\underline{\mbox{Hom}}\left( {\cal O}, {\cal O} \right) \:
\stackrel{ {\scriptsize [x,y]^{\vee} } }{ \longrightarrow } \:
\underline{\mbox{Hom}}\left( {\cal O}^2, {\cal O} \right) \:
\stackrel{ {\scriptsize \left[ \begin{array}{c}
                                -y \\ x \end{array} \right]^{\vee} }}{
\longrightarrow } \:
\underline{\mbox{Hom}}\left( {\cal O}, {\cal O} \right)
\end{displaymath}
For essentially the same reasons that the exact sequence defining
the resolution of ${\cal O}_p$ above is exact, the left-most map is
injective, hence zero kernel, and the kernel of the right map is
the same as the image of the left map, so the only nonzero cohomology
sheaf is in degree two, and is the skyscraper sheaf ${\cal O}_p$:
\begin{displaymath}
\underline{\mbox{Ext}}^n_{ {\cal O}_{ {\bf C}^2 } } \left(
{\cal O}_p, {\cal O} \right) \: = \:
\left\{ \begin{array}{cl}
        0 & n=0,1 \\
        {\cal O}_p & n=2
        \end{array} \right.
\end{displaymath}
Since the local $\underline{\mbox{Ext}}$'s are supported only over points,
the local-to-global spectral sequence is again trivial
(the only nonzero sheaf cohomology groups are in degree zero,
and the differentials $d_r$ move between different degrees whenever
$r \geq 0$), so we can read off that
\begin{displaymath}
\mbox{Ext}^n_X\left( {\cal O}_p, {\cal O} \right) \: = \:
\left\{ \begin{array}{cl}
        0 & n=0,1 \\
        H^0(X, {\cal O}_p) = {\bf C} & n=2
       \end{array} \right.
\end{displaymath}
More generally, it is straightforward to check that
\begin{displaymath}
\mbox{Ext}^n_X\left( {\cal O}_p^k, {\cal O}^N \right) \: = \:
\left\{ \begin{array}{cl}
        0 & n=0,1 \\
        {\bf C}^{kN} & n=2
        \end{array} \right.
\end{displaymath}
exactly right to match one set of open strings in the ADHM construction,
giving a $(k,N)$-valued hypermultiplet.

We can use Serre duality to calculate the spectrum of open strings with
the opposite orientation:
\begin{displaymath}
\mbox{Ext}^n_X\left( {\cal O}^N, {\cal O}_p^k \right) \: = \:
\left\{ \begin{array}{cl}
        {\bf C}^{kN} & n=0 \\
        0 & n=1,2
        \end{array} \right.
\end{displaymath}
exactly right to match ADHM.
As a useful exercise, let us also check this result directly,
by computing local $\underline{\mbox{Ext}}$ sheaves and taking cohomology.
A locally-free resolution of ${\cal O}$ is just ${\cal O}$ itself:
\begin{displaymath}
0 \: \longrightarrow \: {\cal O} \:
\stackrel{ = }{\longrightarrow } \:
{\cal O} \: \longrightarrow \: 0
\end{displaymath}
so local  $\underline{\mbox{Ext}}$ sheaves are the cohomology sheaves
of the one-element complex
\begin{displaymath}
0 \: \longrightarrow \: \underline{\mbox{Hom}}( {\cal O}, {\cal O}_p )
\: \longrightarrow \: 0
\end{displaymath}
so
\begin{displaymath}
\underline{\mbox{Ext}}^n_{ {\cal O}_{ {\bf C}^2 } } \left( {\cal O},
{\cal O}_p \right) \: = \:
\left\{ \begin{array}{cl}
        \underline{\mbox{Hom}}( {\cal O}, {\cal O}_p ) = {\cal O}_p & n=0\\
        0 & n > 0
        \end{array} \right.
\end{displaymath}
The local-to-global spectral sequence trivializes, again, so we calculate
\begin{displaymath}
\mbox{Ext}^n_{ {\bf C}^2 }\left( {\cal O}, {\cal O}_p \right)
\: = \:
\left\{ \begin{array}{cl}
        H^0\left({\bf C}^2, {\cal O}_p \right) = {\bf C} & n=0 \\
        0 & n>0
        \end{array} \right.
\end{displaymath}
from which the advertised result follows. 

Before proceeding, we should mention that there is an issue involving
$U(1)_R$ charges.  Usually cohomology degrees turn out to be the
same as $U(1)_R$ charges, but this does not seem to apply in the
present case.  We can see this issue via the GSO projection,
which can be used to determine the spacetime Lorentz properties
of a state from its charge.  If, for example, a state has charge zero,
then in order to construct a state satisfying the GSO projection,
one must add a worldsheet fermion $\psi^{\mu}$ from an uncompactified
direction, so that the state describes a vector in the low-energy theory.
Similarly, charge one states correspond to low-energy scalars.

This association works perfectly in the case of open strings connecting
a D0-brane back to itself, where the low-energy vector did indeed
correspond to a degree zero Ext group, and the low-energy scalars
corresponded to elements of a degree one Ext group.
For the ADHM construction, on the other hand, we have a problem
-- the states we have just counted are all supposed to be low-energy
scalars, yet there are no degree one Ext group elements.  All of these
states come from either degree zero or its Serre dual, degree two.

This apparent mismatch tells us that the $U(1)_R$ charge of the state
cannot be the same as the Ext degree.  We shall see later in explicit
computations of vertex operators that the $U(1)_R$ charge of the vacuum
can create a mismatch between Ext degrees and state charges.

So far we have reviewed the mathematical construction of Ext groups,
and checked in a few examples that Ext groups do indeed calculate
open string spectra.  However, just because a result holds true in
a few special cases hardly implies that it is true in general.
In much of the rest of these lectures,
we shall give more general reasons why Ext groups calculate spectra.

A basic discussion of Ext groups of coherent sheaves can be
found in \cite[section 5.4]{gh}.

\section{Overview of the B model topological field theory}  \label{Bmodelrev}

\subsection{Basics}

Sheaves (and derived categories) are most relevant to describing
the boundary states of the open string B model.
Let us take a few minutes to review the B model.
(See \cite{edtft} for more information on 
the closed string A and B models.  Our conventions will
closely follow \cite{edtft}.)

The B model is a twist of the ordinary nonlinear sigma model in two
dimensions, with (bulk) lagrangian
\begin{displaymath}
g_{\mu \nu} \partial \phi^{\mu} \overline{\partial}
\phi^{\nu} \: + \:
i g_{i \overline{\jmath} } \psi_-^{\overline{\jmath}} D_z \psi_-^{i} \: + \:
i g_{i \overline{\jmath}} \psi_+^{\overline{\jmath}} D_{ \overline{z} } \psi_+^{i}
\: + \:
R_{i \overline{\imath} j \overline{\jmath} } \psi_+^{i} \psi_+^{\overline{\imath}}
\psi_-^{j} \psi_-^{\overline{\jmath}}
\end{displaymath}
We have implicitly assumed the target is a K\"ahler manifold.
This theory has $(2,2)$ supersymmetry, with transformation laws
\begin{eqnarray*}
\delta \phi^i & = & i \alpha_- \psi_+^i \: + \:
i \alpha_+ \psi_-^i \\
\delta \phi^{\overline{\imath}} & = & i \tilde{\alpha}_- 
\psi_+^{\overline{\imath}} \: + \:
i \tilde{\alpha}_+ \psi_-^{\overline{\imath}} \\
\delta \psi_+^i & = & - \tilde{\alpha}_- \partial \phi^i 
\: - \: i \alpha_+ \psi_-^j \Gamma^i_{jm} \psi_+^m \\
\delta \psi_+^{\overline{\imath}} & = & - \alpha_- \partial
\phi^{\overline{\imath}} \: - \: i \tilde{\alpha}_+
\psi_-^{\overline{\jmath}} \Gamma^{\overline{\imath}}_{
\overline{\jmath} \overline{m} } \psi_+^{\overline{m}} \\
\delta \psi_-^i & = & - \tilde{\alpha}_+ \overline{\partial} \phi^i 
\: - \: i \alpha_- \psi_+^j \Gamma^i_{jm} \psi_-^m \\
\delta \psi_-^{\overline{\imath}} & = & - \alpha_+ 
\overline{\partial} \phi^{\overline{\imath}} \: - \:
i \tilde{\alpha}_- \psi_+^{\overline{\jmath}} 
\Gamma^{\overline{\imath}}_{\overline{\jmath} \overline{m} }
\psi_-^{\overline{m}}
\end{eqnarray*}

In the B model, the worldsheet fermions are twisted to be sections
of different bundles than usual:
\begin{eqnarray*}
\psi_{\pm}^{\overline{\imath}} & \in & \Gamma\left( \phi^* T^{0,1}X \right) \\
\psi_+^i & \in & \Gamma\left( K \otimes \phi^* T^{1,0}X \right) \\
\psi_-^i & \in & \Gamma\left( \overline{K} \otimes \phi^* T^{1,0}X \right)
\end{eqnarray*}
Following \cite{edtft}, we define
\begin{eqnarray*}
\eta^{\overline{\imath}} & = & \psi_+^{\overline{\imath}} \: + \:
\psi_-^{\overline{\imath}} \\
\theta_i & = & g_{i \overline{\jmath}} \left( \,
\psi_+^{\overline{\jmath}} \: + \: \psi_-^{\overline{\jmath}} \, \right) \\
\rho_z^i & = & \psi_+^i \\
\rho_{\overline{z}}^i & = & \psi_-^i 
\end{eqnarray*}

The $(2,2)$ supercharges are now scalars and vectors, rather than
worldsheet spinors.
The scalar part is known as the BRST transformation, with Noether
charge denoted $Q$, and acts on the
fields as
\begin{eqnarray*}
\delta_Q \phi^i & = & 0 \\
\delta_Q \phi^{\overline{\imath}} & = & i \alpha \eta^{\overline{\imath}} \\
\delta_Q \eta^{\overline{\imath}} & = & 0 \\
\delta_Q \theta_i & = & 0 \\
\delta_Q \rho^i & = & - \alpha d \phi^i
\end{eqnarray*}

The vector part of the original $(2,2)$ algebra has Noether charge $G$,
and acts as
\begin{eqnarray*}
\delta_G \phi^i & = & i \alpha \left( \, \rho_z^i \: + \:
\rho_{\overline{z}}^i \, \right) \\
\delta_G \phi^{\overline{\imath}} & = & 0 \\
\delta_G \eta^{\overline{\imath}} & = & - \alpha d \phi^{\overline{\imath}} \\
\delta_G \theta_i & = & i \alpha \Gamma^m_{ik} \left(
\, \rho_z^k \: + \: \rho_{\overline{z}}^k \, \right)\theta_m \: - \:
\alpha g_{i \overline{\jmath}} \left( \, \partial \: - \:
\overline{\partial} \, \right) \phi^{\overline{\jmath}} \\
\delta_G \rho_z^i & = & -i \alpha \rho_{\overline{z}}^j \Gamma^i_{jm}
\rho_z^m \\
\delta_G \rho_{\overline{z}}^i & = & -i \alpha \rho_z^j
\Gamma^i_{jm} \rho_{\overline{z}}^m 
\end{eqnarray*}
(The definitions of $\eta$, $\theta$ were made so as to simplify
the action of the BRST operator $Q$, not $G$.) 

In the present variables, the bulk lagrangian can be rewritten in
the form
\begin{displaymath}
g_{\mu \nu} \partial \phi^{\mu} \overline{\partial} \phi^{\nu} \: + \:
i g_{i \overline{\imath}} 
\eta^{\overline{\imath}}\left( \, D_z \rho_{\overline{z}}^i \: + \:
D_{\overline{z}} \rho_z^i \, \right)  \: + \:
i \theta_i \left( \, D_{\overline{z}} \rho_z^i \: - \: D_z
\rho_{\overline{z}}^i \, \right) \: + \:
R_{i \overline{\imath} j \overline{\jmath} }
\rho_z^i \rho_{\overline{z}}^j \eta^{\overline{\imath}} g^{k \overline{\jmath}}
\theta_k 
\end{displaymath}

Two more facts about the B model will be important for
what follows.  First, the B model localizes on constant maps,
so we will never have to consider rational curve corrections.
Second, the B model depends only upon complex structure moduli,
not K\"ahler moduli.

So far we have only discussed closed strings.
In order to define the open string B model,
we must specify boundary conditions on the bulk fields,
and a boundary action.

As described in \cite[section 3.2]{edcs},
the boundary action for the B model for a single D-brane can be written as
\begin{displaymath}
\int_{\partial \Sigma} \phi^* A \: + \: i F_{i \overline{\jmath}}
\eta^{\overline{\jmath}} \rho^i
\end{displaymath}
(For multiple D-branes, the gauge fields are nonabelian;
correlation functions involve path-ordered exponentials of actions of the
form above, describing Wilson loops around the boundary components of
the open string.)

The boundary conditions on the worldsheet fermions can be expressed as follows:
\begin{itemize}
\item For Dirichlet directions, $\eta = 0$.
\item For Neumann directions, for a single D-brane,
$\theta_i = F_{i \overline{\jmath}} \eta^{\overline{\jmath}}$
\cite{abooetal}.
\end{itemize}
In addition, for Neumann directions there is also a condition on
derivatives of the worldsheet bosons \cite{abooetal}, 
closely analogous to the stated
condition for worldsheet fermions, but as we shall only be interested
in constant maps, conditions on derivatives of worldsheet bosons are
irrelevant.

There is one additional subtlety involved in the open string B model
boundary conditions, that will play an important role later.
Specifically, in writing the B model boundary conditions in local
coordinates in the form above, we appear to be assuming that $TX|_S$
splits holomorphically as $TS \oplus {\cal N}_{S/X}$.
Although it is true that $TX|_S = TS \oplus {\cal N}_{S/X}$ as
smooth bundles, that statement is {\it not} true holomorphically
\footnote{For example, consider conic (degree 2) curves $C$ in the projective
plane ${\bf P}^2$.  These curves are topologically ${\bf P}^1$'s,
but are embedded in the projective plane nontrivially.
Here, $TC = {\cal O}(2)$, $T {\bf P}^2|_C = {\cal O}(3) \oplus {\cal O}(3)$,
and ${\cal N}_{C/{\bf P}^2} = {\cal O}(4)$, so clearly
$T {\bf P}^2 |_C \neq TC \oplus {\cal N}_{C/{\bf P}^2}$ holomorphically.
Note that as $C^{\infty}$ bundles, ${\cal O}(3) \oplus {\cal O}(3)
\cong {\cal O}(2) \oplus {\cal O}(4)$, so $TX|_S$ {\it does} split as a
$C^{\infty}$ bundle, but not as a {\it holomorphic} bundle.}.
Holomorphically, we only have a weaker statement that $TX|_S$ is an
extension of ${\cal N}_{S/X}$ by $TS$:
\begin{displaymath}
0 \: \longrightarrow \: TS \: \longrightarrow \: TX|_S \: \longrightarrow
\: {\cal N}_{S/X} \: \longrightarrow \: 0
\end{displaymath}
Even when $TX|_S$ splits holomorphically, the splitting need not be unique.
In such cases,
the choice of B model boundary conditions picks out a particular splitting.
Correctly phrasing the open string B model boundary 
conditions in a case in which $TX|_S$ does not split holomorphically,
is not something we shall attempt to describe here.

\subsection{Two anomalies}   \label{twoanom}

There are two anomalies in the open string B model that will play an important
role in later analysis.

\subsubsection{Open string analogue of the Calabi-Yau condition}

The first anomaly is an analogue for open strings of the statement
that the closed string B model is only well-defined for Calabi-Yau
target spaces \cite{edtft}, unlike the A model.
Recall the reason for this is well-definedness of the integral over fermion
zero modes.  For example, when the worldsheet is a ${\bf P}^1$ and the target
is a three-fold,
there are three $\psi_+^{\overline{\imath}}$ and
three $\psi_-^{\overline{\imath}}$ zero modes, and to make sense of
the integration
\begin{displaymath}
\epsilon_{ \overline{\imath} \overline{\jmath} \overline{k} }
\int d \psi^{\overline{\imath}} d \psi^{\overline{\jmath}}
d \psi^{\overline{k}}
\end{displaymath}
implicitly assumes the existence of a trivialization 
$\epsilon_{\overline{\imath} \overline{\jmath} \overline{k} }$.
But,such a trivialization is a nowhere-zero (anti)holomorphic
top-form, which exists if and only if the target is a Calabi-Yau.
 
There is an analogous issue in the open string B model, though the
form of the anomaly varies depending upon the D-branes.
If the worldsheet is an infinite strip, the gauge bundle on the D-brane
on each side of the strip is trivial (simplifying the boundary conditions),
and $TX|_S$ splits holomorphically as $TS \oplus {\cal N}_{S/X}$ for
each D-brane, then in order for theory on the strip to be well-defined,
the line bundle
\begin{displaymath}
\Lambda^{top} {\cal N}_{S \cap T / S } \otimes
\Lambda^{top} {\cal N}_{S \cap T / T }
\end{displaymath}
(where $S$ and $T$ are the submanifolds that the two D-branes on either
side of the strip are wrapped on)
must also be trivial \cite{ks}, 
so that the fermion zero mode integral is well-defined.

\subsubsection{The Freed-Witten anomaly}

The second anomaly that is relevant here is the Freed-Witten anomaly
\cite{freeded}.  In their analysis\footnote{Although their paper
was originally written for physical untwisted open string theories,
the results also apply to the open string B model \cite{freedpriv}.}
of open string theories, they found two interesting physical effects:
\begin{enumerate}
\item First, they found that a D-brane can only consistently wrap
submanifolds $S$ with the property that the normal bundle $N_{S/X}$
admits a Spin$^c$ structure.
\item Second, if the normal bundle $N_{S/X}$ admits a Spin$^c$ structure,
but not a Spin structure, then the gauge bundle on the D-brane worldvolume
must be twisted.
\end{enumerate}
All complex vector bundles admit Spin$^c$ structures, so the first effect
is irrelevant for our purposes.  The second effect is much more relevant,
as not all complex vector bundles\footnote{For example, the tangent bundle
to the projective plane ${\bf P}^2$ does not admit a Spin structure.  More
generally, any complex vector bundle with $c_1$ odd does not admit a 
Spin structure.} admit Spin structures.  It means that the gauge bundle
on the D-brane worldvolume is not always an honest bundle.

This second effect might seem rather confusing, in light of the fact
that it is very appealing to identify a sheaf $i_* {\cal E}$ with a 
D-brane whose worldvolume gauge bundle is ${\cal E}$.  What this effect
tells us is that the precise relationship between sheaves and D-branes
is slightly more subtle.

There is an easy way to take the second effect into
account, using the fact that for a complex submanifold $S$ of a Calabi-Yau
$X$, the line `bundle' $\sqrt{ K_S^{\vee} }$ is an honest bundle if and only
if the normal bundle to $S$ admits a Spin structure.
Then, for a D-brane wrapped on a submanifold $S$,
with inclusion map $i: S \hookrightarrow X$, we take the sheaf $i_* {\cal E}$
to correspond to a D-brane with worldvolume gauge bundle
${\cal E} \otimes \sqrt{ K_S^{\vee} }$, and not merely ${\cal E}$.

This prescription no doubt sounds somewhat ad-hoc, but as we shall
see later, it is uniquely determined by physics issues,
and can also be motivated by K-theoretic arguments which we shall not review
here.

For most of these lectures, we shall be able to ignore this
twisting by $\sqrt{K_S^{\vee}}$.  For example, when computing spectra
between D-branes on the same submanifold $S$, each Chan-Paton factor
will come with a $\sqrt{K_S^{\vee}}$ factor, and so the factors will
cancel out.  When considering D-branes wrapped on distinct submanifolds,
however, these factors will become extremely important, and we shall
see that their presence is absolutely required in order for Serre duality
to close the spectra back into themselves, and in fact to recover
Ext groups at all.

\subsection{Closed string B model spectra}   \label{bulkstates}

Since we shall shortly be discussing boundary vertex operators in the
B model, let us begin with a discussion of bulk states,
following \cite{edtft}.

The (local) bulk states are required to be BRST-closed states,
modulo BRST-exact states.
It is straightforward to check that the BRST-closed states are
of the form
\begin{equation}   \label{bulkbasic}
b^{j_1 \cdots j_m}_{ \overline{\imath}_1 \cdots \overline{\imath}_n }(\phi)
\eta^{\overline{\imath}_1} \cdots \eta^{\overline{\imath}_n }
\theta_{j_1} \cdots \theta_{j_m}
\end{equation}
However, counting BRST-closed states modulo BRST-exact states
is difficult in this particular language.

To make the counting more clear, we can translate the counting
problem into a mathematics question, where it becomes easy.
Since $Q \cdot \phi^i = 0$ and $Q \cdot \phi^{\overline{\imath}} 
\propto \eta^{\overline{\imath}}$, we can interpret
\begin{eqnarray*}
\eta^{\overline{\imath}} & \sim & d \overline{z}^{ \overline{\imath}} \\
\theta_i & \sim & \frac{ \partial }{ \partial z^i } \\
Q & \sim & \overline{\partial}
\end{eqnarray*}

Thus, the states~(\ref{bulkbasic}) are in one-to-one correspondence
with bundle-valued differential forms
\begin{displaymath}
b^{j_1 \cdots j_m}_{ \overline{\imath}_1 \cdots \overline{\imath}_n }
(\phi)
d \overline{z}^{\overline{\imath}_1} \wedge \cdots \wedge
d \overline{z}^{\overline{\imath}_n} \wedge
\frac{\partial}{\partial z^{j_1} } \wedge \cdots \wedge
\frac{\partial}{\partial z^{j_m} }
\end{displaymath}
and computing $Q$-closed states modulo $Q$-exact states is the
same as computing $\overline{\partial}$-closed states modulo
$\overline{\partial}$-exact states.
The reader should recognize from our earlier discussion
that such $\overline{\partial}$-cohomology of bundle-valued differential
forms provides a de-Rham-like realization of sheaf cohomology;
thus, the B model bulk states are counted by the sheaf cohomology
groups
\begin{displaymath}
H^n\left(X, \Lambda^m T^{1,0}X \right)
\end{displaymath}
which sometimes go by the fancier name ``Hochschild cohomology.''

The form of this result, {\it i.e.} counting vertex operators
via sheaf cohomology, is actually a relatively standard story in
the string compactifications literature.  Vertex operators usually
can be expressed as differential forms, thus the result.
For another example, \cite{dg} contains a derivation of massless states
in heterotic compactifications; the results are also expressed in
terms of sheaf cohomology, mostly in sheaf cohomology valued in exterior
powers of the gauge bundle.

The OPE algebra obeyed by the B model bulk states is relatively
simple:  the OPE of two differential forms is determined by the
wedge product of the differential forms \cite{edtft}.
We shall see in section~\ref{yoncheck} that for open strings,
the OPE algebra can be much more complicated.

\section{Vertex operators for Ext groups}   \label{vertexext}

We just saw how to explicitly construct the B model bulk states.
Next, let us consider the open string B model boundary R-sector states.
(Note that the phrase ``boundary state'' is ambiguous, and in these
notes, we will mostly use the phrase to mean excitations of a
boundary-condition-changing twist field, as opposed to a set of open
string boundary conditions seen by a closed string.)
As outlined previously, it has been believed for some time that
such open string boundary R-sector states should be counted by
Ext groups, but a direct check of that statement has been lacking.

In particular, we saw in the previous section how B model bulk states are
counted by sheaf cohomology, valued in exterior powers of the
tangent bundle.  We would like to repeat that closed string calculation,
directly in open strings, and recover Ext groups.

What we will find is that properly taking into account open string
boundary conditions can be both subtle and difficult,
making direct physical calculations painful in nontrivial examples.
In such cases, the description of the spectrum in terms of
Ext groups gives a far more efficient computational method.

\subsection{Basic analysis}

The form of the closed-string result is very typical -- vertex
operators almost always look like bundle-valued forms, whose BRST cohomology
is some sheaf cohomology computation.
Already we begin to see some potential problems with our claims
that boundary states should be counted by Ext groups:
\begin{enumerate}
\item Unlike sheaf cohomology groups, Ext groups do not have a 
representation in terms of bundle-valued differential forms on the
ambient space, making their realization in terms of vertex operators,
somewhat puzzling.  We will eventually see that open string boundary
conditions have the effect of physically realizing spectral sequences,
thus implicitly solving this problem albeit in a physically nonobvious fashion.
\item Ordinarily we only see nice relations to algebraic geometry
in supersymmetric situations.  However, a generic pair of
D-branes (even those wrapped on complex submanifolds, with holomorphic
gauge fields) will {\it not} be mutually supersymmetric.
Will we still get Ext groups even for such non-mutually-supersymmetric
cases, or are Ext groups going to count states only for
mutually supersymmetric boundaries?
We will see after computing R-sector spectra 
that we get Ext groups in all cases, not just mutually
supersymmetric ones, but as should be clear, such a result is not
{\it a priori} obvious, and is even somewhat surprising.
\item At a looser level, the reader might object that much of the open
string spectrum should be derived from zero modes of Higgs fields on
the worldvolume, which are {\it not} obviously encoded in the sheaf 
description of D-branes.  If Higgs fields are not present, how can
one recover the spectrum?  Nevertheless, Ext's of sheaves do reproduce
the complete R-sector spectrum, and later in section~\ref{nonred} we shall
see how Higgs fields are encoded in sheaves, in a nonobvious fashion.
\end{enumerate}

Let us now turn to computing open string boundary states,
following the same pattern as for closed string bulk states.
For simplicity, in these lectures we shall only consider the special
case of coincident D-branes, wrapped on a submanifold $S$ of the
Calabi-Yau $X$; see \cite{ks} for a much more general
treatment.

Also, to begin, let us assume that the gauge bundle on each D-brane
has no curvature, and that $TX|_S$ splits holomorphically as
$TS \oplus {\cal N}_{S/X}$, where $S$ is the submanifold on which the
D-brane is wrapped.  In this case, the dictionary between 
worldsheet fermions and target-space quantities tells us that
the $\eta$'s, restricted to the boundary, should be interpreted
as antiholomorphic one-forms on $S$, not $X$, since only $\eta$'s
``parallel'' to $S$ can be nonzero.  Furthermore, the $\theta$'s
should couple to the normal bundle ${\cal N}_{S/X}$, and not $TX$,
since they vanish for directions parallel to $S$.

From our previous claims about the physical relevance of Ext groups,
if the gauge bundle on one D-brane is ${\cal E}$ and the gauge bundle
on the other D-brane is ${\cal F}$, then the open string spectrum
must be counted by the Ext groups
\begin{displaymath}
\mbox{Ext}^*_X\left( i_* {\cal E}, i_* {\cal F} \right)
\end{displaymath}
where $i: S \hookrightarrow X$ is the inclusion map,
so that the sheaves $i_* {\cal E}$, $i_* {\cal F}$ naturally
represent the D-branes.

In the special case of D-branes wrapped on the entire space,
the counting was previously worked out in \cite{edcs},
where it was shown that the states are counted by
sheaf cohomology:  $H^n(X, {\cal E}^{\vee} \otimes {\cal F})$,
which happen to agree with Ext groups when the sheaves are actually
bundles on $X$.  

More generally, following the same methods as for the closed string
bulk states and for the special case discussed in \cite{edcs},
the boundary R-sector states are of the form
\begin{displaymath}
b^{\alpha \beta j_1 \cdots j_m}_{ \overline{\imath}_1 \cdots \overline{\imath}_n }(\phi)
\eta^{\overline{\imath}_1} \cdots \eta^{\overline{\imath}_n }
\theta_{j_1} \cdots \theta_{j_m}
\end{displaymath}
where the $\alpha$, $\beta$ are Chan-Paton indices, coupling to the
bundles on either D-brane.  We can now apply the same dictionary
as for the bulk states, with the modifications mentioned above
($\eta$'s are antiholomorphic 1-forms on $S$, $\theta$'s couple
to ${\cal N}_{S/X}$, and the $\phi$ zero modes are restricted to $S$).
We immediately find that the states above are classified by
the sheaf cohomology groups
\begin{displaymath}
H^n\left(S, {\cal E}^{\vee} \otimes {\cal F} \otimes
\Lambda^m {\cal N}_{S/X} \right).
\end{displaymath}

On the one hand, an expression in the form of sheaf cohomology certainly
is consistent with other calculations of massless states.
On the other hand, we promised that the result would be Ext groups,
and these sheaf cohomology groups are {\it not} Ext groups!

To make the distinction between Ext groups and the sheaf cohomology
groups above more precise, let us outline a more precise reason
why they differ.
The sheaf cohomology groups above treat
deformations of the submanifold ( $\sim H^0({\cal N}_{S/X})$)
differently from deformations of the gauge bundle on a submanifold
($H^1( {\cal E}^{\vee} \otimes {\cal E})$).  By contrast, degree one
Ext groups mix the two, when possible -- sometimes, the holomorphic
structure on a bundle can not be consistently dragged along if the
submanifold $S$ is moved, in which case, there are fewer elements in
$\mbox{Ext}^1(i_*{\cal E}, i_*{\cal E})$ than there are in
$H^0( {\cal E}^{\vee} \otimes {\cal E} \otimes {\cal N}_{S/X})
\oplus H^1( {\cal E}^{\vee} \otimes {\cal E} )$.
In a few moments, we shall describe an example of this form.

\subsection{A spectral sequence and its physical realization}

A mathematician reading these notes might not understand our
concern.  In particular, it turns out that there is a spectral sequence
relating the sheaf cohomology we obtained to the Ext groups we desire:
\begin{equation}    \label{specseq}
E_2^{p,q} \: = \: H^p\left( S, {\cal E}^{\vee} \otimes {\cal F}
\otimes {\cal N}_{S/X} \right) \: \Longrightarrow \:
\mbox{Ext}^{p+q}_X\left( i_* {\cal E}, i_* {\cal F} \right).
\end{equation}
However, as physicists we know that the mere existence of some
abstract mathematical operation does not suffice -- to be relevant,
that spectral sequence must be realized physically.
If it is not realized physically at all, then the open string spectrum is
counted by sheaf cohomology, {\it not} Ext groups, contradicting
our earlier claims (and posing a severe problem for the derived categories
program).  If the spectral sequence is realized physically, but via
some sort of target-space operation, {\it e.g.} through some superpotential,
then again we have a problem -- the open string spectrum is
again counted by sheaf cohomology, and not Ext groups, and again the
derived categories program has run into trouble.
What we really want is for open string spectra to be counted directly
by Ext groups, not sheaf cohomology -- we want the BRST cohomology computation
to be modified, somehow.  On the other hand, the BRST cohomology computation
we have just outlined is both well-known and standard, making any errors,
unlikely.

To understand the role that the spectral sequence plays, we need
to re-examine some of our initial assumptions.  Recall we assumed that the
curvature on the gauge bundle vanishes, and that $TX|_S$ splits holomorphically
as $TS \oplus {\cal N}_{S/X}$, so as to simplify the description of
the boundary conditions on worldsheet fermions. 
It can be shown that in the special case that either of these
conditions holds, the spectral sequence~(\ref{specseq}) trivializes,
implying that\footnote{The isomorphism is not unique mathematically,
just as the splitting $TX|_S = TS \oplus {\cal N}_{S/X}$ need not
be unique.  However, the precise choice of open string B model
boundary conditions fixes a particular splitting $TX|_S = TS
\oplus {\cal N}_{S/X}$, and also fixes a particular
isomorphism between Ext's and sheaf cohomology.}
\begin{displaymath}
\mbox{Ext}^n_X\left( i_* {\cal E}, i_* {\cal F} \right) \: \cong \:
\sum_{p+q=n} H^p\left(S, {\cal E}^{\vee} \otimes {\cal F} \otimes 
\Lambda^q {\cal N}_{S/X} \right).
\end{displaymath}
So, we did actually get the correct answer, in light of our
assumptions, but for the purposes of seeing Ext groups nontrivially,
our simplifying assumptions were {\it too} simple.

We claim that when those assumptions are relaxed, the resulting
nontrivial spectral sequence is realized directly in physics via
subtle implicit modification of BRST cohomology.
So, let us suppose that $TX|_S$ does not split holomorphically
as $TS \oplus {\cal N}_{S/X}$, and that 
the curvature of the gauge bundle is nonzero.

In this case, we can no longer associate the $\theta_i$ with
${\cal N}_{S/X}$, as the condition ``$\theta_i=0''$ no longer is
well-defined globally.  Instead, we can only think of the $\theta_i$
as coupling to $TX|_S$, with the property that for those $\theta_i$ in the
$TS$ subbundle, $\theta_i = F_{i \overline{\jmath}} \eta^{ \overline{\jmath}}$.
(Assuming the gauge bundle is a line bundle, for simplicity.)

Consider, for example, a boundary R-sector state of the form
\begin{displaymath}
b^{\alpha \beta i} \theta_i
\end{displaymath}
Previously this would have corresponded to an element of
$H^0(S, {\cal E}^{\vee} \otimes {\cal E} \otimes {\cal N}_{S/X})$,
but now, there is no longer a direct relationship between such states
and sheaf cohomology, since the $\theta_i$ couple to $TX|_S$ instead
of ${\cal N}_{S/X}$.

Given an element of 
$H^0(S, {\cal E}^{\vee} \otimes {\cal E} \otimes {\cal N}_{S/X})$,
we can lift the coefficients from ${\cal N}_{S/X}$ to
$TX|_S$, which gives us something that {\it can} be described
as a boundary R-sector state of the form above.
However, if $TX|_S$ does not split holomorphically as $TS \oplus
{\cal N}_{S/X}$, then such a lift to $TX|_S$ need no longer be
holomorphic, {\it i.e.} will no longer be annihilated by $\overline{\partial}$.

Now, since the BRST operator has been identified with $\overline{\partial}$,
we appear to have a problem -- our state, lifted to $TX|_S$ from
${\cal N}_{S/X}$, is no longer BRST invariant.  However, we have not yet
taken into account the boundary conditions on the $\theta_i$.

Although the result of the lift to $TX|_S$ is not holomorphic,
the image under $\overline{\partial}$ has coefficients that lie entirely
in the $TS$ subbundle, making it appropriate to apply the boundary
condition on the $\theta_i$.  Thus, instead of demanding that 
$\overline{\partial}$ annihilate the state immediately,
we demand that $\overline{\partial}$ annihilate the state {\it after}
we apply the boundary conditions $\theta_i =
F_{i \overline{\jmath}} \eta^{\overline{\jmath}}$.

We can summarize this as follows.  The coefficient lift from ${\cal N}_{S/X}$
to $TX|_S$, followed by $\overline{\partial}$, is just the coboundary
map 
\begin{displaymath}
\delta: \:
H^0\left(S, {\cal E}^{\vee} \otimes {\cal E} \otimes {\cal N}_{S/X} \right)
\: \longrightarrow \:
H^1\left(S, {\cal E}^{\vee} \otimes {\cal E} \otimes TS \right)
\end{displaymath}
in the long exact sequence associated to the short exact sequence
\begin{displaymath}
0 \: \longrightarrow \: TS \: \longrightarrow \:
TX|_S \: \longrightarrow \:  {\cal N}_{S/X} \: \longrightarrow \: 0
\end{displaymath}
So, we can describe the process we followed as the composition
\begin{equation}  \label{comp}
H^0\left(S, {\cal E}^{\vee} \otimes {\cal E} \otimes {\cal N}_{S/X} \right)
\: \stackrel{ \delta }{ \longrightarrow } \:
H^1\left(S, {\cal E}^{\vee} \otimes {\cal E} \otimes TS \right)
\: \stackrel{ eval }{ \longrightarrow } \:
H^2\left(S, {\cal E}^{\vee} \otimes {\cal E} \right)
\end{equation}
where the map $eval$ is the evaluation map obtained by replacing
$\theta_i$ in the $TS$ subbundle with $F_{ i \overline{\jmath} }
\eta^{ \overline{\jmath} }$.

As it happens, the composition~(\ref{comp}) is one of the differentials
in the spectral sequence:
\begin{displaymath}
d_2: \:
H^0\left(S, {\cal E}^{\vee} \otimes {\cal E} \otimes {\cal N}_{S/X} \right)
\: \longrightarrow \:
H^2\left(S, {\cal E}^{\vee} \otimes {\cal E} \right)
\end{displaymath}
and so demanding that a boundary R-sector state be annihilated by this
composition is the same as taking the kernel of $d_2$.
Similarly, we want to mod out anything in the image of the
composition~(\ref{comp}), which just corresponds to taking cokernels
of $d_2$.  We shall not discuss the other differentials in this
spectral sequence here, but they have the same form -- a composition
of $\delta$'s and $eval$ maps.

A very careful reader might have the following minor technical
objection.  Back in section~\ref{twoanom}, we discussed the
Freed-Witten anomaly, and its manifestation as the statement
that the sheaf $i_* {\cal E}$ really corresponds to a D-brane
with worldvolume gauge bundle ${\cal E} \otimes \sqrt{ K_S^{\vee}}$.
Since the Chan-Paton factors couple to the bundle
${\cal E} \otimes \sqrt{ K_S^{\vee}}$ and not to ${\cal E}$,
the Chan-Paton curvature is the curvature of ${\cal E} \otimes
\sqrt{K_S^{\vee}}$, and so is not the same as the curvature
of ${\cal E}$.  Hence, the evaluation map above
(involving the curvature of ${\cal E}$) and the open string
boundary condition (involving the curvature of ${\cal E}
\otimes \sqrt{ K_S^{\vee}}$) are not quite the same.
However, although they are not the same, the extra contribution
to the Chan-Paton curvature from $\sqrt{ K_S^{\vee}}$
drops out of the composition above, and so is irrelevant.
Although the Chan-Paton curvature and the evaluation map are not
the same, one gets the same composition regardless of which one uses,
so the Freed-Witten anomaly is irrelevant for the physical realization
of this spectral sequence.

As mentioned earlier, the spectral sequence proceeds by taking
kernels and cokernels of differentials, chipping away at the initial
approximation to eventually yield the final result.
Thus, we see that the spectral sequence is realized physically.

\subsection{Example of a nontrivial spectral sequence}

An example of a situation in which the spectral sequence
is nontrivial, and the physical calculation of open string
spectra is correspondingly hard, is as follows.
Let $X$ be a K3-fibered Calabi-Yau threefold, and let $S$ be a smooth
K3 fiber.  Assume further that $S$ contains a $C \cong {\bf P}^1$
which is rigid in $X$, having normal bundle ${\cal N}_{C/X}
\cong {\cal O}_C(-1) \oplus {\cal O}_C(-1)$, and such that the
bundle ${\cal O}_S(C)$ does not deform to first order as $S$ moves
in the fibration.  (More explicitly still,
the Calabi-Yau ${\bf P}(1,1,2,2,2)[8]$ has the desired properties.)
In this case, 
\begin{displaymath}
\mbox{Ext}^1_X\left( i_* {\cal O}_S(C) ,  i_* {\cal O}_S(C) \right) \: = \:
0
\end{displaymath}
so the combination of submanifold and bundle are rigid,
yet the submanifold by itself can move:
\begin{displaymath}
H^0\left(S, {\cal O}_S(C)^{\vee} \otimes {\cal O}_S(C)
\otimes {\cal N}_{S/X} \right) \: = \: 
H^0(S, {\cal O}) \: = \:
{\bf C}
\end{displaymath}
(using the fact that ${\cal N}_{S/X} = {\cal O}_S$).
Clearly, in this case
\begin{displaymath}
\mbox{Ext}^1_X \left( i_* {\cal E}, i_* {\cal E} \right) \: \neq \:
H^0\left(S, {\cal E}^{\vee} \otimes {\cal E} \otimes {\cal N}_{S/X}
\right) \: \oplus \:
H^1\left(S, {\cal E}^{\vee} \otimes {\cal E} \right)
\end{displaymath}
and also, both the curvature of the gauge bundle is nonzero,
and $TX|_S \neq TS \oplus {\cal N}_{S/X}$ holomorphically.

Regarding the spectral sequence, also note
\begin{displaymath}
H^2\left( S, {\cal O}_S(C)^{\vee} \otimes {\cal O}_S(C) \right) \: = \:
H^2(S, {\cal O}) \: = \: {\bf C}
\end{displaymath}
so the differential 
\begin{displaymath}
d_2: \:
H^0\left(S, {\cal E}^{\vee} \otimes {\cal E} \otimes {\cal N}_{S/X} \right)
\: \longrightarrow \:
H^2\left(S, {\cal E}^{\vee} \otimes {\cal E} \right)
\end{displaymath}
is a nonzero map ${\bf C} \rightarrow {\bf C}$,
with only the zero element in the kernel.

\subsection{More general intersections}

\subsubsection{Derivation of Ext groups}

So far we have only spoken about D-branes on the same submanifold
$S$.  More general intersecting D-branes can be treated similarly.

One physical complication for more general intersections is that
the modings on the worldsheet fields are shifted by the different
boundary conditions\footnote{Technically, the same subtlety can
arise for D-branes on the same submanifold, if the gauge bundles
on either side of the open string have different nonzero curvatures.
We shall not speak about this complication here.} 
on either side, as in \cite{berkoozetal}.
Since we are only interested in massless states, we are only interested
in zero modes of fields, so only some of the worldsheet fermions
(those with the same boundary conditions on either side) can contribute
to the massless states.  This changes the forms of the spectral sequences,
as discussed in \cite{ks}, but these are merely technical complications.

Assume that the two D-branes lie on submanifold $S$ and $T$,
and correspond to sheaves $i_* {\cal E}$, $j_* {\cal F}$,
respectively. 
Proceeding as before, boundary Ramond sector states are of the same form
as before:
\begin{equation}    \label{genstates}
b^{\alpha \beta j_1 \cdots j_m}_{ \overline{\imath}_1 \cdots 
\overline{\imath}_n} \eta^{\overline{\imath}_1} \cdots
\eta^{\overline{\imath}_n} \theta_{j_1} \cdots
\theta_{j_m}
\end{equation}
The bosonic zero modes are confined to $S \cap T$, and
the $\eta$ zero modes also only reflect $T^*(S \cap T)$,
so that the corresponding sheaf cohomology groups are
computed on $S \cap T$.
Assuming that $TX|_S$ and $TX|_T$ split holomorphically, and that
the bundles ${\cal E}$, ${\cal F}$ have no curvature, the $\theta$ couple to
\begin{displaymath}
\tilde{N} \: = \: \frac{ TX|_{S \cap T} }{ TS|_{S \cap T} + TT|_{S \cap T}}
\end{displaymath}
because of the moding shift of \cite{berkoozetal} alluded to above.

Finally, the Chan-Paton factors must be treated carefully.
Recall from section~\ref{twoanom} that because of the Freed-Witten anomaly,
the D-brane corresponding to the sheaf $i_* {\cal E}$ has `bundle'
${\cal E} \otimes \sqrt{ K_S^{\vee}}$ on its worldvolume,
and not ${\cal E}$.
Thus, the Chan-Paton factors $\alpha$, $\beta$ couple to the bundle
\begin{displaymath}
\left. \left( {\cal E} \otimes \sqrt{ K_S^{\vee} } \right)^{\vee} 
\right|_{S \cap T}
\otimes
\left. \left( {\cal F} \otimes \sqrt{ K_T^{\vee} } \right) \right|_{S \cap T}
\end{displaymath}
so that there is an extra factor of $\sqrt{ K_S/K_T }$ that the
reader might not have expected.

Often the `bundle' $\sqrt{ K_S/K_T}$ is not an honest bundle at all,
which poses a problem for us -- sheaf cohomology with coefficients
in non-honest bundles, is not well-defined.
However, recall from section~\ref{twoanom} that there is another
anomaly in the open string B model, which says that the open string
B model is only well-defined when the line bundle
\begin{equation}    \label{cyanalogue}
\Lambda^{top} {\cal N}_{S \cap T / S } \otimes 
\Lambda^{top} {\cal N}_{S \cap T / T }
\end{equation}
is trivializable.
When the line bundle~(\ref{cyanalogue}) is trivializable, it is
straightforward to show that
\begin{eqnarray*}
\sqrt{ \frac{  K_S|_{S \cap T} }{ K_T|_{S \cap T} } } & = &
\Lambda^{top} {\cal N}_{S \cap T / T} \\
\sqrt{ \frac{ K_T|_{S \cap T} }{ K_T|_{S \cap T} } } & = &
\Lambda^{top} {\cal N}_{S \cap T / S }
\end{eqnarray*}

Thus, when $TX|_S$ and $TX|_T$ split holomorphically and
the bundles ${\cal E}$, ${\cal F}$ have trivial curvature,
we see that the states~(\ref{genstates}) are counted by
the sheaf cohomology groups
\begin{displaymath}
H^n\left( S \cap T, {\cal E}^{\vee}|_{S \cap T} \otimes {\cal F}|_{S \cap T}
\otimes \Lambda^m \tilde{N} \otimes \Lambda^{top} {\cal N}_{S \cap T/T}
\right)
\end{displaymath}
Moreover, there is a spectral sequence 
\begin{displaymath}
H^p\left( S \cap T, {\cal E}^{\vee}|_{S \cap T} \otimes {\cal F}|_{S \cap T}
\otimes \Lambda^q \tilde{N} \otimes \Lambda^{top} {\cal N}_{S \cap T/T}
\right)
\: \Longrightarrow \:
\mbox{Ext}^{p+q+m}_X\left( i_* {\cal E}, j_* {\cal F} \right)
\end{displaymath}
(where $m$ is the rank of ${\cal N}_{S \cap T/T}$)
that we believe (though have not checked) should be realized directly
in BRST cohomology when the boundary conditions on open string worldsheet
fields are nontrivial, just as we explicitly saw happens for parallel coincident
D-branes.

In passing, we should mention that the Freed-Witten anomaly plays an
essential role here.  For example, there is {\it no} spectral
sequence relating the sheaf cohomology groups
\begin{displaymath}
H^p\left( S \cap T, {\cal E}^{\vee}|_{S \cap T} \otimes {\cal F}|_{S \cap T} 
\otimes
\Lambda^q \tilde{N} \right)
\end{displaymath}
to Ext groups, as was discussed in \cite{ks}.
The factor $\Lambda^{top} {\cal N}_{S \cap T/T}$, arising from
the Freed-Witten anomaly, is absolutely essential.

\subsubsection{Observations on general intersections}

One complication, not usually present in topological field theories,
is that the $U(1)_R$ charges of the states does {\it not} seem to be the
same as the degrees of the Ext groups in general.
We saw previously in section~\ref{extrev} that the $U(1)_R$ charges
implied by the GSO projection are not the same as the mathematical
degrees of the Ext groups, in examples involving different boundary
conditions on either side of the open string.  Sometimes
Ext group degrees are the same as $U(1)_R$ charges; sometimes the
Ext group degree is the same as the $U(1)_R$ charge of the state
minus the charge of the vacuum; sometimes the relationship is even
less clear.  In order to properly
sort out this issue, one would need to systematically compute
fractional vacuum charges in open strings, following the analysis
of \cite{wenwitten} for closed strings.  Ultimately the
Ext group degree is very unlikely to coincide with the $U(1)_R$ charge
of the state except for mutually supersymmetric brane configurations,
as the vacuum can have fractional charge in general, and Ext group
degrees are integral.

A more interesting physical point concerns supersymmetry.
We mentioned earlier that one objection to claiming that open string
states are counted by Ext groups is that one usually only gets nice
connections to algebraic geometry in supersymmetric situations,
and generic boundary conditions will break supersymmetry.

For general intersections, we can see more clearly how this issue
is resolved.  Since we are interested in calculating R-sector states,
and the mode-shifting of \cite{berkoozetal} acts equally on the
worldsheet bosons and fermions, there is no net shift of the
vacuum zero-point energy in the R-sector, and hence massless
modes are still obtained merely by multiplying by zero modes.
However, in the boundary NS-sector the contributions to
the zero-point energy from the worldsheet bosons and fermions
do {\it not} cancel.  

For example, consider the old statement that two D-branes, one on a submanifold
of the other, will break supersymmetry unless the difference in 
(real) dimensions
is a multiple of four.  This is argued by showing that if that 
condition is not met, then there are no massless boundary NS-sector states
to be superpartners to the massless boundary R-sector states that 
one still has.  Similarly, open string tachyons arise from
the boundary NS-sector, not the boundary R-sector.

Thus, by counting boundary R-sector states instead of 
boundary NS-sector states, we avoid many issues involving supersymmetry
breaking, and so it is not unreasonable that we can count boundary
R-sector states with Ext groups even in non-supersymmetric brane configurations.

\subsection{Notes on the boundary chiral ring}

An entire paper should probably be written on boundary ${\cal N}=2$
algebras and their chiral primary rings for general boundary conditions.
We shall not try to do so here.  Instead, we will take a few moments
to compare some properties of boundary chiral rings with
bulk chiral rings, following \cite{lvw}.

In particular, let us only consider chiral rings of boundary
R-sector states in open strings with the same boundary conditions
on both sides, {\it i.e.} open strings between D-branes on the same
submanifold with the same gauge bundle.  The analysis here is somewhat
simpler, as for example the $U(1)_R$ charges coincide with the
Ext degrees, which as we have seen 
does not seem to be true for more general boundary
conditions.

Most of the analysis of \cite{lvw} seems to apply equally well
in this situation, with a few quirks.

First, whereas for bulk states \cite{lvw} there was a unique
charge zero state, corresponding geometrically to a zero-form
on the manifold, here the charge zero states are elements of
\begin{displaymath}
\mbox{Ext}^0_X\left( i_* {\cal E}, i_* {\cal E} \right)
\end{displaymath}
and there can be many charge zero states.  However, although charge zero
states need not be unique, there is a distinguished charge zero state,
specifically
\begin{displaymath}
1_{ i_* {\cal E} } \: \in \:
\mbox{Ext}^0_X\left( i_* {\cal E}, i_* {\cal E} \right)
\: = \:
\mbox{Hom}_X\left(  i_* {\cal E}, i_* {\cal E} \right)
\end{displaymath}
which corresponds to the identity map $i_* {\cal E} \rightarrow
i_* {\cal E}$.

Next, the appearance of the holomorphic top form is slightly subtle.
There are many top-degree states, coinciding with elements
of 
\begin{displaymath}
\mbox{Ext}^{\dim \: X}_X\left( i_* {\cal E}, i_* {\cal E} \right)
\end{displaymath}
just as there were many zero-degree states.
The Serre dual of $1_{ i_* {\cal E} }$ is an element of
the {\it dual} vector space
\begin{displaymath}
\mbox{Ext}^{\dim \: X}_X\left( i_* {\cal E}, i_* {\cal E} \right)^*
\end{displaymath}
and is canonically associated to the holomorphic top form.
However, there is no canonical way to associate an element of the
original vector space
\begin{displaymath}
\mbox{Ext}^{\dim \: X}_X\left( i_* {\cal E}, i_* {\cal E} \right)
\end{displaymath}
with the holomorphic top-form, using the data given so far.
Technically, we need to specify an inner product, in addition
to other data given so far, in order to pick a specific element
of the original vector space of top-degree states.
(There is a similar subtlety involving the B model bulk states.)

Also note that when the boundary conditions on the two sides of the open string
are not the same, there need not be any degree-zero states, or any
top-degree states.

The product structure on the fields above is believed to be the
same as the Yoneda pairing of Ext groups.  However, the Yoneda pairing is
{\it not} the same as a wedge product of differential forms in general,
and in fact, seems to be realized in physics in a rather nontrivial
way, even when the boundary conditions on the two sides of the
open string are the same.  
We shall discuss the Yoneda pairing in much greater detail in
section~\ref{yoncheck}.

\subsection{Dimensional reduction of holomorphic Chern-Simons}

For the special case of Calabi-Yau three-folds,
if one believes that the dimensional reduction of open string
field theory is the open string field theory of D-branes wrapped
on submanifolds, then as a consistency check one ought to be able to reproduce
computations of open string spectra (albeit only for open strings connecting
D-branes back to themselves) via dimensional reduction of holomorphic
Chern-Simons \cite{edcs}.

Let us first assume that $TX|_S$ splits holomorphically as
$TS \oplus {\cal N}_{S/X}$, and the gauge bundle has no curvature, 
so that spectral sequences trivialize.  After we have checked that
the advertised states appear as zero modes of the Chern-Simons gauge
field, we shall relax these conditions and outline how the
spectral sequences recur in this context.

With the simplifying assumptions just mentioned, the spectral sequences
trivialize and
\begin{displaymath}
\mbox{Ext}^n_X\left( i_* {\cal E}, i_* {\cal E} \right) \: = \:
\bigoplus_{p+q=n} H^p\left(S, {\cal E}^{\vee} \otimes {\cal E}
\otimes \Lambda^q {\cal N}_{S/X} \right).
\end{displaymath}
States with both $p \leq 1$ and $q \leq 1$, not both one, arise from 
dimensional reduction of the 
Chern-Simons gauge field in an obvious way, but states with
larger values of $p$ or $q$ appear more mysterious.
It turns out that states with larger values of $p$ or $q$
are related by Serre duality to states that clearly arise from
the gauge field, and so are merely their antiparticles.

Let us check that claim systematically.
When the dimension of $S$ is two, the available sheaf cohomology
groups are
\begin{displaymath}
\begin{array}{cc}
H^0\left(S, {\cal E}^{\vee} \otimes {\cal E} \right)
& H^2\left(S, {\cal E}^{\vee} \otimes {\cal E} \otimes {\cal N}_{S/X} \right)\\
H^1\left(S, {\cal E}^{\vee} \otimes {\cal E} \right) &
H^1\left(S, {\cal E}^{\vee} \otimes {\cal E} \otimes {\cal N}_{S/X} \right)\\
H^0\left(S, {\cal E}^{\vee} \otimes {\cal E} \otimes {\cal N}_{S/X} \right) &
H^2\left(S, {\cal E}^{\vee} \otimes {\cal E} \right)
\end{array}
\end{displaymath}
Groups in the right column are Serre duals of groups in the left column,
so we need only describe how the groups appearing in the left column arise
from the holomorphic Chern-Simons gauge field.

The first group, $H^0(S, {\cal E}^{\vee} \otimes {\cal E})$,
is the set of zero modes that correspond to automorphisms that
multiply the base gauge field by a scalar, as $\phi A_i$ for
some scalar automorphism $\phi \in H^0(S, {\cal E}^{\vee} \otimes {\cal E})$.

The second group, $H^1(S, {\cal E}^{\vee}\otimes {\cal E})$,
corresponds to deformations of the gauge field,
{\it i.e.}, $A_{\overline{\imath}} + \delta A_{\overline{\imath}}$ 
for $\delta A_{\overline{\imath}} \in H^1(S, {\cal E}^{\vee}\otimes {\cal E})$.

The third group, 
$H^0(S, {\cal E}^{\vee} \otimes {\cal E} \otimes {\cal N}_{S/X})$,
is the zero modes of the Higgs scalars arising from the dimensional reduction
of the Chern-Simons gauge field.  In passing, we should note that
it only makes sense to recover scalars from dimensional reduction
in the case that $TX|_S \cong TS \oplus {\cal N}_{S/X}$ holomorphically,
an issue we shall return to later.

Thus, for $S$ of dimension two inside the Calabi-Yau three-fold $X$,
in the case that the spectral sequence trivializes we recover all
of the states described by sheaf cohomology from the dimensional
reduction of the holomorphic Chern-Simons gauge field.

The analysis works similarly for $S$ of other dimension.
When the dimension of $S$ is one, for example, the available sheaf
cohomology groups are
\begin{displaymath}
\begin{array}{cc}
H^0\left(S, {\cal E}^{\vee} \otimes {\cal E} \right) &
H^1\left(S, {\cal E}^{\vee} \otimes {\cal E} \otimes \Lambda^2
{\cal N}_{S/X} \right) \\
H^1\left(S, {\cal E}^{\vee} \otimes {\cal E} \right) &
H^0\left(S, {\cal E}^{\vee} \otimes {\cal E} \otimes \Lambda^2
{\cal N}_{S/X} \right) \\
H^0\left(S, {\cal E}^{\vee} \otimes {\cal E} \otimes {\cal N}_{S/X} \right)
&
H^1\left(S, {\cal E}^{\vee} \otimes {\cal E} \otimes {\cal N}_{S/X} \right)
\end{array}
\end{displaymath}
Again, we have listed Serre duals in the right column,
and again, note that the available states modulo Serre duals are
the same sheaf cohomology groups as before, and so can be derived
from dimensional reduction of holomorphic Chern-Simons in the same
way as before.

Finally, consider $S$ of dimension zero.
In this case, the available sheaf cohomology groups are
\begin{displaymath}
\begin{array}{cc}
H^0\left(S, {\cal E}^{\vee} \otimes {\cal E} \right) &
H^0\left(S, {\cal E}^{\vee} \otimes {\cal E} \otimes \Lambda^3
{\cal N}_{S/X} \right) \\
H^0\left(S, {\cal E}^{\vee} \otimes {\cal E} \otimes {\cal N}_{S/X} \right) &
H^0\left(S, {\cal E}^{\vee} \otimes {\cal E} \otimes \Lambda^2
{\cal N}_{S/X} \right)
\end{array}
\end{displaymath}
Again, we have listed the Serre duals in the right column, and again,
the available states modulo Serre duals can be derived, just as before,
by dimensional reduction of holomorphic Chern-Simons.

Next, let us drop the assumptions that $TX|_S$ splits holomorphically
as $TS \oplus {\cal N}_{S/X}$ and that the curvature of the gauge
bundle vanishes.  In this more general situation,
one can see the same subtleties that drove us to spectral sequences
cropping up when one tries to perform the dimensional reduction.
Consider, for example, dimensional reduction from a three-fold to a surface $S$
in the three-fold, as in the example we discussed where the spectral sequence
is nontrivial.  Ideally, one would like to split the
holomorphic gauge field $A_i$ into a piece along $S$, also a gauge
field, plus a complex scalar, denoted $\Phi$, say, corresponding to
the component of $A_i$ normal to $S$.

However, such a description is already making a false assumption.
For $C^{\infty}$ bundles, $TX|_S$ always splits as $TS \oplus
{\cal N}_{S/X}$, so for most physical theories the process 
we have just described
would be sensible.  However, in the present case
the gauge field $A_i$ is parametrized
by a {\it holomorphic} index, and for holomorphic bundles,
$TX|_S$ does {\it not} split as $TS \oplus {\cal N}_{S/X}$,
one of the subtleties that made our spectral
sequence nontrivial.  Thus, splitting a holomorphic gauge field
on the three-fold into a piece along $S$ and a piece $\Phi$ normal
to $S$ is already not a sensible description in general, because of
one of the same subtleties that gave us nontrivial spectral sequences.

The other subtlety that made spectral sequences nontrivial
(and physical computations difficult) is the twisting of the
boundary conditions on worldsheet fermions due to curvature in the
Chan-Paton factors.  This also has a clear analogue in dimensional
reductions of holomorphic Chern-Simons:  even when $TX|_S$ splits,
zero modes of fields are solutions of equations such as $D_A \Phi=0$,
reflecting the fact that the fields originate as excitations of 
the holomorphic Chern-Simons gauge field, and mixing Higgs moduli
and the gauge field in a clear fashion.

We shall not pursue the dimensional reduction of holomorphic Chern-Simons
any farther, but it is clear that the same subtleties that give
rise to nontrivial spectral sequences also arise when trying to
dimensionally-reduce holomorphic Chern-Simons.
In general, it is mechanically easier to compute string spectra
in nontrivial compactifications by computing zero modes of 
spinors
than by computing zero modes of their superpartners,
one of the reasons why standard references such as \cite{dg,edtft}
compute spectra in Ramond sectors rather than Neveu-Schwarz sectors,
and a reason why we emphasize computing open string spectra
directly
in Ramond-sector states, 
rather than trying to 
slog through a messy dimensional reduction.

\section{Brane-antibrane annihilation vs Ext groups}   \label{tachyonext}

As mentioned in the introduction, one of the motivations for
modelling D-branes with sheaves is as a stepping-stone to
making physical sense of derived categories, where instead of
individual sheaves, one has complexes.  Very basic parts of the relationship
are clear:  a complex should correspond to a set of branes and antibranes,
and passing to the cohomology of the complex should correspond
to brane-antibrane annihilation.  However, understanding why
one has a complex, instead of just a random collection of maps;
understanding the grading of the complex physically; and understanding
the equivalence relations that lead to derived categories physically
are not trivial matters.  In this section, we shall describe
how giving vevs to tachyons leads to working with complexes of D-branes,
and how one can -- formally, modulo physical complications --
see Ext groups emerge in a different fashion than before.

\subsection{Basic analysis}

Let us take a moment to describe how one gives a vev to the tachyon
between top-dimensional branes and antibranes
in the open string B model.  In the process, we will learn how
it is natural to describe complexes of sheaves on the boundaries
of open strings, recreating a construction that first appeared in
\cite{paulalb}.

The boundary NS-sector state associated
with the tachyon 
corresponds (under spectral flow) to a degree zero Ext group element
labelling a boundary R-sector state in the TFT.
Let $\phi^{\alpha \beta}$ denote the element of $\mbox{Ext}^0({\cal E},
{\cal F})$ that we want to `turn on,' so to speak.
To give a vev to this boundary state, we insert the operator
\begin{equation}   \label{pvev}
P \exp \int_{ \partial \Sigma} \, [ G, \phi^{\alpha \beta} ]
\end{equation}
on the boundary.

%In the present case,
%\begin{displaymath}
%[ G, \phi^{\alpha \beta} ] \: = \:
%i \partial_i \phi^{\alpha \beta} \rho^i.
%\end{displaymath}

Using the fact that $\{ G, Q \} \propto d$,
adding this term has the effect of modifying BRST cohomology.
We can see that deformation easily, without having to explicitly
work out Noether charges, and so forth.
In particular, if we evaluate the correlation function
\begin{displaymath}
\langle \{ Q, V \} \rangle
\end{displaymath}
for some vertex operator $V$,
then commuting the BRST operator $Q$ past the insertions of 
equation~(\ref{pvev}) effectively deforms the BRST operator to the form
\begin{displaymath}
Q_{old} \: + \: \phi|_{\sigma = 0} \: - \: \phi|_{\sigma = \pi}.
\end{displaymath}
After all, for a vertex operator $V$ to be BRST-closed in the sense that
$< \{ Q, V \} >_{new} = 0$ now implies not that it is closed under
only $Q_{old}$, but commuting $Q$ past the insertions~(\ref{pvev}) generates
extra factors, and so $V$ must be closed under the combination of
$Q_{old}$ plus the new factors brought down by moving $Q_{old}$ past
the insertions.

Next, let us demand that these insertions preserve worldsheet
${\cal N}=2$ supersymmetry.  
One of the constraints one must impose is that $Q^2 = 0$,
for the new BRST operator $Q$.
For a single tachyon $\phi$, this merely implies that the
tachyon be holomorphic:  $\overline{\partial} \phi = 0$.
For multiple deformations of the form~(\ref{pvev}) to
preserve supersymmetry, 
the condition $Q^2 = 0$ factors into two pieces, one of which
is the holomorphicity constraint above, and the other of which is that
if we have vevs $\phi_n$ turned on between neighboring
branes, then $\phi_{n+1} \phi_n = 0$.
Notice that this second condition is the same condition as arising in
a complex of sheaves on the differentials of the complex.

Another condition we must impose in order to preserve 
boundary ${\cal N}=2$ supersymmetry is on the boundary conditions
on the $U(1)_R$ current.  Our insertion~(\ref{pvev}) is not
neutral, but rather has $U(1)_R$ charge $-1$, and so appears to
break the $U(1)_R$ symmetry, which would destroy the boundary
${\cal N}=2$ algebra.  We can fix this problem by imposing different
boundary conditions on the $U(1)_R$ charge operator, so that the two
sides of the open string, on either side of the insertion~(\ref{pvev}),
have relative $U(1)_R$ charge $+1$, so that the total charge of
insertion~(\ref{pvev}), taking into account the charge of the
underlying boundary-condition-changing operator, is now zero.

This charge constraint plays two important roles for us:
\begin{itemize}
\item Physically, this difference in charges enforces the intuitive
notion that neighboring elements of the complex should be antibranes
with respect to one another.  After all, on the worldsheet the distinction
between a brane and an antibrane lies essentially in the GSO projection.
By forcing the boundary-condition-changing vertex operator to have
charge $1$, the two sides of the open string effectively obey
different GSO projections, and hence if one is a brane in the physical
theory, the other must be an antibrane.
\item Mathematically this difference in charges provides a notion
of grading of a complex.  We can not meaningfully assign an absolute
grading to elements of a complex, but we can say that neighboring
elements have grading differing by one unit. 
\end{itemize}

From the conditions imposed by worldsheet ${\cal N}=2$ boundary
supersymmetry, we see that tachyon vevs are naturally related to
complexes:  the composition of neighboring tachyons must vanish,
just as the composition of neighboring maps in a complex must vanish,
and each tachyon shifts the $U(1)_R$ charge by one unit, reflecting
a grading on a complex.  These observations played an important
role in \cite{paulalb} in arguing that D-brane boundary states
are related to complexes of sheaves, which led the authors of
\cite{paulalb} to consider derived categories.

Previously we have seen how massless modes in open strings
are counted by Ext groups.  Let us try to repeat that analysis here.

In section~\ref{vertexext}, we saw how BRST cohomology can be very subtle
when describing D-branes wrapped on submanifolds.  We can simultaneously
avoid such technical problems, and also forge a link to
some relevant mathematics, by only considering D-branes wrapped
on the entire Calabi-Yau, {\it i.e.} locally-free sheaves,
and replacing D-branes on submanifolds with locally-free resolutions
of the corresponding sheaves.  Since we never have nontrivial boundary
conditions, the BRST cohomology is merely the 
the cohomology of the new
BRST operator
\begin{displaymath}
Q_{BRST} \: = \: \overline{\partial} \: + \: \sum_n \phi_n
\end{displaymath}
without any of the subtleties that previously led us to spectral
sequences.  We have replaced the spectral sequence computation,
with a more complicated BRST operator. 

Indeed, the cohomology of this operator, acting on differential forms on
$X$ valued in the bundles appearing in the locally-free resolutions,
also calculates Ext groups.  Put another way, it is a true fact that
if a complex $C_{\cdot}$
is the same as a locally-free resolution of a sheaf ${\cal S}$,
then
\begin{displaymath}
\mbox{Ext}^n_X\left( C_{\cdot}, {\cal T} \right) \: = \:
\mbox{Ext}^n_X\left( {\cal S}, {\cal T} \right).
\end{displaymath}
Our BRST cohomology calculation in this massive theory is a computation
of Ext groups between complexes (a statement proven in
appendix~\ref{pf}); but if the complexes are resolutions
of sheaves, then we are also calculating Ext groups between the sheaves.
For completeness, we describe an example in the next section,
where we see that this method is actually not very useful as a practical
computational tool, though it does give insight.
This method was used in \cite{paulalb} to check that open string
states should be counted by Ext groups, albeit in a massive theory
instead of directly in BCFT. 

Setting aside questions of usefulness for practical computations,
this method of studying open string spectra suffers from a physical
drawback:  in turning on tachyon vevs, we have broken conformal invariance.
The computation of Ext groups just described is akin to computing
spectra in a gauged linear sigma model, or a Landau-Ginzburg model,
and does not tell us the explicit form of the states in the
open string BCFT.  In order to write down states in the BCFT,
or even to directly check that BCFT states really are counted by
Ext groups, the methods of the section~\ref{vertexext} are required.

Another objection the reader might raise
is that we have only spoken about maps between neighboring elements
in a complex -- why can we not consider more general maps?
Our discussion was motivated by the desire to give a physical
realization of ordinary complexes, but that hardly implies that
complexes are the only such objects physically realizable.
We shall see in section~\ref{gencpx} that including more general vevs
leads to a notion of `generalized complex' first discussed mathematically
in \cite{bk}, and ultimately also leads back to derived categories.

In passing, we should mention that the massive theories described
by branes, antibranes, and tachyon vevs, corresponding to complexes
of sheaves, are precisely off-shell boundary states, and so in principle
can be efficiently handled using the methods of string field theory.
For the purposes of these introductory lectures, however,
we shall not work with string field theory, but content ourselves
merely to mention that in principle, this is a motivation.

\subsection{Example}

Now, let us see how to reproduce Ext group calculations in this language.
Let us calculate $\mbox{Ext}^*_{ {\bf C} } \left( {\cal O}_D, {\cal O}
\right)$ where $D$ is some divisor on ${\bf C}$.
The torsion sheaf ${\cal O}_D$ has a two-step resolution:
\begin{displaymath}
0 \: \longrightarrow \: {\cal O}(-D) \: \longrightarrow \:
{\cal O} \: \longrightarrow \: {\cal O}_D \:
\longrightarrow \: 0
\end{displaymath}
so it is easy to check that the answer should be given by
\begin{displaymath}
\mbox{dim }\mbox{Ext}^n_{ {\bf C} } \left( {\cal O}_D, {\cal O} \right) \: = \:
\left\{ \begin{array}{cl}
        0 & n=0 \\
        1 & n=1
        \end{array} \right.
\end{displaymath}
We shall check our physical picture by verifying, through a physical
calculation, the Ext groups just described.
(We will find that the physical calculation is too messy for practical
use, but working through the details should illuminate the ideas.)

On one end of the open string, put Chan-Paton factors corresponding
to the bundle ${\cal O}(-D) \oplus {\cal O}$,
and on the other end of the open string, put Chan-Paton factors corresponding
to ${\cal O}$.  We shall also give a vev to the tachyon corresponding
to the holomorphic map $\phi: {\cal O}(-D) \rightarrow {\cal O}$,
on the corresponding end of the open string.
As a result of this tachyon vev, the BRST operator is now deformed
to the form $Q_{BRST} = \overline{\partial} + \phi^{\vee}$.
(We write $\phi^{\vee}$ instead of $\phi$ to emphasize the fact that
$\phi$ acts solely within the dualized half of the Chan-Paton
factors.)

Proceeding as before, boundary R-sector states are of the general
form
\begin{displaymath}
b^{\alpha \beta}_{ \overline{\imath}_1 \cdots \overline{\imath}_n }
\eta^{\overline{\imath}_1} \cdots \eta^{\overline{\imath}_n}
\end{displaymath}
where $\alpha$, $\beta$ are Chan-Paton factors coupling to
${\cal O}(-D) \oplus {\cal O}$ and ${\cal O}$, respectively.
Since we will need to work with the detailed components,
let us write
\begin{displaymath}
( b^{\alpha \beta}_{ \overline{\imath}_1 \cdots \overline{\imath}_n } )
\: = \:
\left[ \begin{array}{c}
       b_{0 \overline{\imath}_1 \cdots \overline{\imath}_n }  \\
       b_{1 \overline{\imath}_1 \cdots \overline{\imath}_n }  
       \end{array} \right]
\end{displaymath}
where $b_0$ is associated with ${\cal O}(-D)^{\vee}\otimes {\cal O}$ and $b_1$ is associated
with ${\cal O}^{\vee} \otimes {\cal O}$.  

Now, let us compute BRST cohomology.  We need to be careful to keep
track of degrees properly -- for example, degree zero states are {\it not}
of the form
\begin{displaymath}
\left[ \begin{array}{c}
       b_0 \\
       b_1 
       \end{array} \right]
\end{displaymath}
since the tachyon $\phi$ forces $b_0$ and $b_1$ to have different
charges.  Instead, the only degree zero state is $b_1$, and the condition
for this state to be BRST closed is that $\overline{\partial} b_1 = 0$
and $\phi^{\vee} b_1 = 0$.  The only holomorphic function on ${\bf C}$
that is annihilated by multiplication by $x$ (assuming $D$ is the
divisor $\{ x=0\}$, without loss of generality) 
is the zero function, hence the space of degree zero
states is zero-dimensional, exactly as desired.

The degree one states are of the form
\begin{displaymath}
b_0 \: + \: b_{1 \overline{\imath} } \eta^{ \overline{\imath} }
\end{displaymath}
The condition for these states to be BRST-closed is that
\begin{eqnarray}
\overline{\partial} b_0  & = & - \phi^{\vee} \left( b_{1 \overline{\imath}} 
\eta^{\overline{\imath}}\right)  \label{clos1} \\
\overline{\partial} \left( b_{1 \overline{\imath}} d \overline{z}^{
\overline{\imath}} \right) & = & 0   \label{clos2}
\end{eqnarray}
and BRST-exact states are of the form
\begin{eqnarray*}
b_0 \: = \: \phi^{\vee} a \\
b_{1 \overline{\imath}} d \overline{z}^{\overline{\imath}} 
\: = \: \overline{\partial} a 
\end{eqnarray*}
for some $a$.
Condition~(\ref{clos2}) means that $b_{1 \overline{\imath}} 
\eta^{\overline{\imath}}$ is an element of $H^1({\cal O})$.
Condition~(\ref{clos1}) means that if we define $b_0' = b_0
\mbox{ mod } \mbox{im }\phi^{\vee}$, then
$\overline{\partial} b_0' = 0$, and more to the point,
\begin{displaymath}
b_0' \: \in \: H^0\left(D, {\cal O}(-D)^{\vee}|_D \otimes {\cal O}|_D \right)
\: = \: \mbox{Ext}^1\left({\cal O}_D, {\cal O} \right)
\end{displaymath}
(Technically $b_0'$ can be interpreted as defining a form on $D$ because
we are modding out the image of an element of $H^1({\cal O})$.)
Conversely, given an element of 
\begin{displaymath}
\mbox{Ext}^1\left( {\cal O}_D, {\cal O} \right) \: = \:
H^0\left(D, {\cal O}(-D)^{\vee}|_D \otimes {\cal O}|_D \right)
\end{displaymath}
we can define $b_0$ and $b_1$, using the long exact sequence
\begin{displaymath}
\cdots \: \longrightarrow \:
H^0({\cal O}) \: \longrightarrow \:
H^0({\cal O}(D) ) \: \longrightarrow \:
H^0(D, {\cal O}(D)|_D ) \: \stackrel{ \delta }{\longrightarrow} \:
H^1({\cal O}) \: \longrightarrow \: \cdots
\end{displaymath}
The element $b_1$ is the image under $\delta$, and $b_0$ is the lift
to an element of ${\cal A}^{0,0}( {\cal O}(D) )$.

Clearly, using tachyon vevs to compute Ext group elements directly
is not very efficient in general, but this example should help clarify
the relevant ideas.

\section{Derived categories}    \label{dercatoverview}

One of the motivations for describing D-branes with sheaves is to
understand the precise relationship between open string B model boundary
states and derived categories.  In this section, we shall attempt to
briefly outline derived categories.  Interested readers should consult
\cite{weibel,hred,thomasdc} for more information.  A more thorough
overview will appear in \cite{paultoappear}, so we will content ourselves
with merely outlining the highlights of derived categories.

The objects in a derived category are equivalence classes of
complexes.  We have seen in the previous section how complexes 
emerge naturally
on open string boundaries.  Very briefly,
there are two levels of equivalences of complexes that one mods out
to get a derived category:
\begin{enumerate}
\item First, one identifies any two complexes $C_{\cdot}$,
$D_{\cdot}$ related by a {\it chain
homotopy}, which is defined by a collection of maps
$f_n: C_n \rightarrow D_n$ where each $f_n = s \circ d + d \circ s$
for some set of maps $s_n: C_n \rightarrow D_{n-1}$.
\item Second, one identifies complexes related by quasi-isomorphisms,
which are maps between complexes that define isomorphisms between the
cohomologies of the complexes.  Intuitively, if the end-product of
renormalization-group flow in each collection of branes and antibranes
is described by the cohomology complexes, then two complexes that are
quasi-isomorphic should be in the same universality class physically.
\end{enumerate}
We shall describe how these two equivalences are realized physically.
The first set is closely related to BRST equivalence;
the second set is believed to correspond to universality classes
of renormalization-group flow.

\subsection{Chain maps and chain homotopies}

Consider an open string with boundaries on two complexes of locally-free
sheaves $C_{\cdot}$, $D_{\cdot}$, as described in the previous section.
We shall show here that, formally,
a zero-degree BRST-closed map between two such complexes
is an example of a chain map, and that BRST-exact chain maps
are formally chain homotopies, two notions more familiar to
mathematicians than physicists.

A {\it chain map} between two complexes $C_{\cdot}$, $D_{\cdot}$
is a collection of holomorphic maps $f_n: C_n \rightarrow D_n$, for each $n$,
such that each of the following diagrams commutes:
\begin{equation}   \label{chainmap}
\xymatrix{
C_n \ar[r]^{ \phi_C } \ar[d]^{f_n} & C_{n+1} \ar[d]^{f_{n+1}} \\
D_n \ar[r]^{ \phi_D } & D_{n+1}
}
\end{equation}
where $\phi$ denote the tachyons, or mathematically, the differentials
of the complex.
Given a set of maps $f_n: C_n \rightarrow D_n$, 
demanding BRST-closure implies that
\begin{displaymath}
\overline{\partial} f \: + \: \sum \phi_D f \: - \: \sum f \phi_C
\: = \: 0
\end{displaymath}
This condition factorizes into two pieces, namely that
\begin{displaymath}
\overline{\partial} f_n \: = \: 0
\end{displaymath}
for all $n$, and that
\begin{displaymath}
\phi_D f_n \: = \: f_{n+1} \phi_C
\end{displaymath}
for all $n$.
The first condition is just the statement that all the maps are
holomorphic, and the second condition is the statement that
diagram~(\ref{chainmap}) commutes, so we see that for a set of $f_n$ to
be annihilated by the BRST operator is equivalent to demanding that they
be chain maps.

A {\it chain homotopy} is a special kind of chain map,
namely a set of $f_n$ with the property that
$f = \phi_D s - s \phi_C$ for some set of maps $s_n: C_n \rightarrow
D_{n-1}$.  However, it should be clear that if $f$ can be written
in this form, then $f = Q s$, and so is BRST-exact.

Thus, modding out chain homotopies, the first mathematical step
in defining a derived category, formally appears to be
realized in BRST cohomology in
the massive worldsheet theory.

In looking for a relation to BRST cohomology, we have been
a little too quick -- as we shall see in more detail in section~\ref{gencpx},
preserving ${\cal N}=2$ supersymmetry constrains the relative
degrees of maps, and it is not completely clear that all of the
maps bandied about above can actually be realized physically.

There is an alternative way of thinking about this computation
which both reinforces the relationship to BRST cohomology, as well
as points out shortcomings of the attempted physical realization.
We have already argued that Ext groups count BRST cohomology in
the massive theory.  However, Ext groups have an alternative description:
\begin{displaymath}
\mbox{Ext}^n_X\left( C_{\cdot}, D_{\cdot} \right) \: = \:
\mbox{Hom}_X\left( C_{\cdot}, D_{\cdot}[n] \right)
\end{displaymath}
where the $[n]$ denotes a complex shifted by $n$ units, and
Hom between complexes counts chain maps {\it between quasi-isomorphic
projective/injective complexes}, modulo chain homotopies.
{\it If} $C_{\cdot}$ and $D_{\cdot}$ were complexes of projectives and
injectives, then our naive description of BRST cohomology in this section
would
be exactly right.  Unfortunately, the complexes ${\bf C}_{\cdot}$ and
$D_{\cdot}$ will almost certainly not\footnote{There is a potential confusion 
here that can trap the unwary.  Locally-free sheaves are examples
of projectives for local $\underline{\mbox{Ext}}$,
but {\it not} for global Ext.  What is needed here are injectives and
projectives for global Ext.}
have such properties.
Thus, 
in plain english, sometimes we can count Ext groups by counting chain maps
(modulo chain homotopies) between one complex and the other complex,
shifted by $n$ units, as well as by differential forms valued
in the locally-free sheaves in the complex (our strategy 
in section~\ref{tachyonext}).  
Thus, we should not be at all surprised to see 
chain maps and chain homotopies being realized in a fashion that
closely resembles BRST cohomology,
since this is yet another calculation of Ext groups.

\subsection{Quasi-isomorphisms}

Quasi-isomorphisms are chain maps that induce isomorphisms between
cohomology sheaves.  The second level of equivalence classes that
one must take to get a derived category are obtained by 
``localizing on quasi-isomorphisms,'' which for our purposes
essentially means identifying complexes related by quasi-isomorphisms.

Although the physical realization of
the first level of equivalences (chain homotopies) is easy to check,
properly checking the second level of equivalences requires a more
detailed knowledge of universality classes of D-branes than we
seem to possess at the moment.  For example, it is possible for two
complexes to have the same cohomology sheaves, and yet not be
quasi-isomorphic.  Physically we would need the two collections of
branes and antibranes to not be in the same universality class,
despite having the same cohomology sheaves.  Such assertions are,
at present, difficult to check physically. 

Reference \cite{paulalb} attempts to work around this difficulty
by defining a notion of ``physical equivalence'' of complexes:
two complexes are said to be physically equivalent if, for any other
fixed complex, all open strings between the fixed complex and
either of the two original complexes always have the same spectrum.
Then, \cite{paulalb} argues that the corresponding category
whose objects are equivalence classes of physically-equivalent complexes
is closely related to the derived category.  However, this method
by itself still leaves open questions of how to count Ext groups
directly in BCFT (which were separately answered in
\cite{ks}), as well as the issue of directly checking that
quasi-isomorphism classes really are the same as universality classes
of RG flow.

In any event, properly checking that renormalization group flow
realizes localization of quasi-isomorphisms may be moot.
If two complexes are quasi-isomorphic, then the corresponding
massive theories will have the same BRST cohomology and product structures,
and so as topological field theories, are essentially equivalent.
Furthermore, our direct computations in BCFT \cite{ks} 
tell us that massive theories
describing complexes which allegedly flow in the IR to BCFT boundaries
describing quasi-isomorphic sheaves do indeed have the same spectrum,
verifying to some extent the assumptions about behavior of RG flow.
Now that these statements have also been checked explicitly in
conformal field theories that are the end-points of the RG flows,
there seems little remaining ambiguity.

\subsection{Naturality of derived categories}   \label{cones}

Sometimes even physicists naively argue that, if a part of mathematics
is sufficiently `nice,' it should have some role in string theory.
The author has even heard this argument applied to derived categories.

However, there is a technical sense in which derived categories are
not precisely natural:  one of the standard constructions involving
derived categories, namely the cone construction, does not behave
functorially.

The cone construction alluded to above is a mechanism for creating
a new complex from two other complexes.
Let $C_{\cdot}$ and $D_{\cdot}$ be two complexes, and
$f: C_{\cdot} \rightarrow D_{\cdot}$ be a chain map between them.
The ``mapping cone'' over $f$, denoted
$\mbox{Cone}(f: C_{\cdot} \rightarrow D_{\cdot})$, is a complex of the
form
\begin{displaymath}
\cdots \: \longrightarrow \: C_1 \oplus D_0 \: \longrightarrow \:
C_2 \oplus D_1 \: \longrightarrow \: \cdots
\end{displaymath}
with differentials of the form
\begin{displaymath}
\left[ \begin{array}{cc}
       \phi_C & 0 \\
        f & \phi_D 
        \end{array} \right]
\end{displaymath}
Intuitively, if $C_{\cdot}$ and $D_{\cdot}$ represent collections
of branes and antibranes, with $f$ a set of tachyons between them,
then the cone construction morally should give something that flows
to the result of mutual brane/antibrane annihilation between
$C_{\cdot}$ and $D_{\cdot}$.
For example, if $C_{\cdot}$ is the one-element complex with
${\cal O}(-D)$ in the zero position, $D_{\cdot}$ is the 
one-element complex with ${\cal O}$ in the zero position,
and $f$ is a nonzero map ${\cal O}(-D) \rightarrow {\cal O}$,
then $\mbox{Cone}(f)$ is the complex
\begin{displaymath}
0 \: \longrightarrow \: {\cal O}(-D) \:
\stackrel{ f }{ \longrightarrow } \: {\cal O} \:
\longrightarrow \: 0
\end{displaymath}
which is quasi-isomorphic to the sheaf ${\cal O}_D$, and one expects
the D-brane corresponding to ${\cal O}_D$
should be obtained by brane/antibrane annhilation between the
branes corresponding to ${\cal O}(-D)$ and ${\cal O}$.

Technically the cone construction above, commonly used in derived
categories, is not ``functorial.''  In plain english, although as
elements of a derived category the complexes are only defined up to
isomorphism, in order to define the cone one must choose a particular
complex representing any given equivalence class,
and the resulting cone depends upon that choice.  If one then tries
to compose morphisms in the derived category, applied to the 
cone construction, one finds that in order for the composition
to be well-defined, one might have to change the choices of representative
complexes one made to define the cones in the first place.

There are workarounds for this lack of functoriality of the cone
construction, involving more complicated theories
(such as \cite{bk}) that encode the same information as derived categories.
Some of those more complicated theories may, strictly speaking,
be somewhat more directly related to physics, as we shall see in
the next section.

\section{Generalized complexes}   \label{gencpx}

Previously we argued how one could associate a complex
of bundles to an open string boundary.  However, the fact that we only
gave vevs to (off-shell) tachyons between neighboring elements in the
complex should have seemed somewhat ad-hoc.  
More generally, one can add additional vertex operators, corresponding
to extra fields mapping between non-neighboring elements of the complex,
as explained in \cite{bk,calin1,calin2,diac} in a string field theory context.
We shall now outline the same ideas using purely worldsheet manipulations.

So, we would like to add boundary vertex operators corresponding to
more general maps between elements of a complex than we considered
in the previous section.  However, worldsheet supersymmetry imposes
certain restrictions.  In the previous section we saw how demanding
boundary worldsheet supersymmetry in the B model not only forces us
to consider holomorphic tachyons, forming a complex, but preserving
the $U(1)_R$ charge of the boundary ${\cal N}=2$ algebra also forces
us to impose boundary conditions on the bulk Noether charge, 
reproducing the grading of a complex.

\begin{figure}
\centerline{\psfig{file=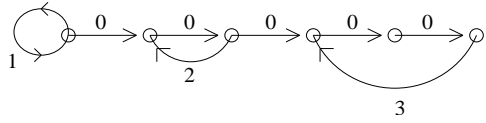,width=2.5in}}
\caption{\label{calin} Example of generalized complex.  
Each arrow is labelled by the
degree of the corresponding vertex operator.}
\end{figure}

We saw in the previous section that for a degree zero operator $\phi$,
since $[ G, \phi ]$ has $U(1)_R$ charge $-1$, the boundary conditions
on the bulk $U(1)_R$ Noether charge must be such that
the boundary-condition-changing operator underlying $\phi$ has
charge $+1$, and the neighboring complexes have degree differing
by one, otherwise the $U(1)_R$ is explicitly broken.
Suppose by contrast we have a degree $n$ operator $\phi$.
Then, $[ G, \phi ]$ has charge $n-1$, and so the two boundaries
it lies between must have relative $U(1)_R$ charge $1-n$.
If the $U(1)_R$ Noether charge boundary conditions were fixed
by a set of tachyons, as in the previous section, then we immediately
read off that
\begin{itemize}
\item A degree one operator $\phi$ must map an element of the complex
back into itself.  In other words, charge one vertex operators define
D-brane endomorphisms, exactly as one would expect.
\item A degree two operator $\phi$ must map an element of the complex
to the {\it previous} element -- it acts between neighboring elements
of the complex, but runs in the opposite direction.
\item A degree three operator $\phi$ must map between next-nearest-neighboring
elements of the complex.
\end{itemize}
We have sketched an example of such a generalized complex, with the degrees
of each map labelled, in figure~\ref{calin}.

Now, not any set of vertex operators is compatible with ${\cal N}=2$ 
supersymmetry, just as not any set defines a generalized complex in
the sense of \cite{bk}.  We have already discussed the constraint imposed
by charge conservation.  In addition, we must demand that the
BRST operator square to zero.  The fields above deform the BRST operator,
just as in our previous analysis, and the result can be expressed
schematically as
\begin{displaymath}
Q \: = \: \overline{\partial} \: + \: \sum_a \phi_a
\end{displaymath}
Demanding that the BRST operator square to zero implies that
\begin{displaymath}
\sum_a \overline{\partial} \phi_a \: + \: \sum_{a,b} \phi_b \cdot \phi_a
\: = \: 0
\end{displaymath}
which is the same constraint imposed in \cite{bk} to recover
a ``generalized complex.''

In our previous analysis, involving ordinary complexes, the condition
above factorized, as
\begin{eqnarray*}
\overline{\partial} \phi_n & = & 0 \\
\phi_{n+1} \cdot \phi_n & = & 0
\end{eqnarray*}
telling us that the maps are all holomorphic, and enforcing the constraint
that the image of any one map lie in the kernel of the next, {\it i.e.}
enforcing the condition that we have a complex, as opposed to merely
a random collection of maps.
Here, by contrast, since maps of all degrees are allowed, no such factorization
is possible in general. 
(The lack of factorization yields a counterintuitive non-holomorphicity in
allowed deformations, as noted in \cite{diac}.)

Mathematically, one motivation \cite{tonypriv} for \cite{bk} 
was to find something
closely related to derived categories, that did not suffer from certain
technical issues of ordinary derived categories, such as non-functoriality
of the cone construction.  Intuitively, these generalized complex constructions
seem to contain more information than ordinary derived categories.
These properties may well play an important role
in more detailed future discussions of derived categories in physics,
as these generalized complex constructions seem more directly applicable
to general off-shell states than mere ordinary derived categories.
On the other hand, for some purposes these generalized complexes
encode essentially the same information as ordinary derived categories
\cite{tonypriv}, so for some purposes, they may be less
useful.  In any event, we shall not discuss them further here.

\section{Orbifolds}  \label{orb}

Having now studied spectra of open strings on large-radius Calabi-Yau's,
let us now turn to orbifolds, another limit where calculations are
possible.  

Recently, quiver methods have become fashionable for manipulating
D-brane gauge theories near singularities; however, in order to
write down a quiver in the first place, one must know the open string
spectrum, as that determines the arrows in the quiver diagram.
For relatively simple cases, one can make educated guesses about quiver
diagrams and corresponding gauge theories, 
but for nontrivial geometries, one needs more powerful methods
in order to calculate the open string spectra, to make quivers useful.
Our results in this section provide the more powerful methods needed.

\subsection{General result}

Our approach \cite{kps} to solving this problem and calculating open 
string spectra in orbifolds for more general cases closely
follows \cite{dm}.  There, open string spectra in orbifolds $[X/G]$
were calculated by first calculating the open string spectrum
on the covering space $X$, and then taking $G$-invariants,
where $G$ is the orbifold group. 
However, they were only able to perform calculations in topologically
trivial cases, because methods for calculating open string spectra
in general cases were not known.  Having solved that problem,
we can now calculate open string spectra in topologically nontrivial
cases in orbifolds, and we can also relate the results mathematically to an
intellectually-simpler picture of orbifolds.

The basic result has the following form.
Let $X$ be a Calabi-Yau with a $G$-action, and let $S$, $T$ 
be submanifolds of $X$ that are both closed under $G$
({\it i.e.}, $G$ maps each submanifold back into itself).
Wrap D-branes on $S$ and $T$, and let ${\cal E}$ denote the
corresponding holomorphic vector bundle on $S$,
and ${\cal F}$ denote the holomorphic vector bundle on $T$.
We must define a group action on the Chan-Paton factors;
this is known more formally as choosing a $G$-equivariant structure
on ${\cal E}$ and on ${\cal F}$.

As discussed earlier, the (R-sector) spectrum of open strings connecting
the D-branes above on $X$ is given by $\mbox{Ext}^*_X\left( i_* {\cal E},
j_* {\cal F} \right)$.  Thus, following the same procedure as \cite{dm},
the obvious conjecture is that the open string states in the orbifold
should be counted by the $G$-invariant part of the Ext groups just named,
denoted $\mbox{Ext}^*_X\left( i_* {\cal E}, j_* {\cal F} \right)^G$.

We do need to be slightly more careful than that, however.
The Ext groups were not quite obtained directly, but rather as
the end-product of a physically-realized spectral sequence.

We need to check that our spectral sequence makes sense when applied
to $G$-invariants.  Let us examine this matter explicitly in the
case that $T=S$ and ${\cal F} = {\cal E}$ are line bundles, 
as discussed in section~\ref{vertexext}.
Recall from that discussion that the differential:
\begin{displaymath}
d_2: \: H^0\left(S, {\cal E}^{\vee} \otimes {\cal E} \otimes
{\cal N}_{S/X} \right) \: \longrightarrow \:
H^2\left(S, {\cal E}^{\vee} \otimes {\cal E} \right)
\end{displaymath}
factorizes as
\begin{displaymath}
 H^0\left(S, {\cal E}^{\vee} \otimes {\cal E} \otimes
{\cal N}_{S/X} \right)
\: \stackrel{ \delta }{ \longrightarrow } \:
H^1\left(S, {\cal E}^{\vee} \otimes {\cal E} \otimes TS \right)
\: \stackrel{ eval }{ \longrightarrow } \:
H^2\left(S, {\cal E}^{\vee} \otimes {\cal E} \right)
\end{displaymath}
where $\delta$ is the coboundary map in the long exact sequence
associated to the short exact sequence
\begin{displaymath}
0 \: \longrightarrow \: TS \: \longrightarrow \:
TX|_S \: \longrightarrow \: {\cal N}_{S/X} \: \longrightarrow \: 0
\end{displaymath}
and $eval$ is the evaluation map that replaces $\theta_i 
\mapsto F_{i \overline{\jmath} } \eta^{\overline{\jmath}}$.
It turns out that both of these maps map the subset of $G$-invariant
elements to other $G$-invariant elements.  This should be clear
for the evaluation map, since the curvature $F_{i \overline{\jmath}}$
is necessarily a $G$-invariant two-form if the bundle is $G$-equivariant.

Thus, the maps $\delta$ and $eval$ map $G$-invariants to $G$-invariants,
and so it turns out that the differential $d_2$ does indeed descend
to $G$-invariants, {\it i.e.}, 
\begin{displaymath}
d_2: \:
 H^0\left(S, {\cal E}^{\vee} \otimes {\cal E} \otimes
{\cal N}_{S/X} \right)^G
\: \stackrel{ \delta }{ \longrightarrow } \:
H^1\left(S, {\cal E}^{\vee} \otimes {\cal E} \otimes TS \right)^G
\: \stackrel{ eval }{ \longrightarrow } \:
H^2\left(S, {\cal E}^{\vee} \otimes {\cal E} \right)^G
\end{displaymath}
is well-defined.  More generally, it can be shown that
all of the differentials in the
original spectral sequence $H^p\left(S, {\cal E}^{\vee} \otimes
{\cal F} \otimes \Lambda^q {\cal N}_{S/X} \right) \: \Longrightarrow \:
\mbox{Ext}^{p+q}_X\left( i_* {\cal E}, i_* {\cal F} \right)$
descend to $G$-invariants, and the resulting spectral sequence
on $G$-invariants converges to the $G$-invariant part of Ext groups.

In other words, there is a spectral sequence
\begin{displaymath}
E_2^{p,q} \: = \: H^p\left( S, {\cal E}^{\vee} \otimes {\cal F}
\otimes {\cal N}_{S/X} \right)^G \: \Longrightarrow \:
\mbox{Ext}^{p+q}_X\left( i_* {\cal E}, i_* {\cal F} \right)^G
\end{displaymath}
(as well as analogues for more general intersections), and it
is clear that it
is also realized physically, in exactly the same form as
in section~\ref{vertexext}.

Thus, we now see in detail that open string states in orbifolds
are counted by $G$-invariant parts of Ext groups, a result certainly
in line with the general methods of \cite{dm}.

\subsection{More elegant phrasing}

There is a much more elegant way to phrase this result,
in terms of mathematical objects called ``stacks.''
Stacks are generalizations of spaces, and have most of the properties
that one requires of a space -- one can do differential geometry on stacks,
one can define metrics, spinors, gauge fields, {\it etc} on stacks.

In particular, quotient stacks (denoted with brackets as
$[X/G]$ to distinguish them from quotient spaces $X/G$)
are easy examples of stacks, and are closely related to orbifolds.
For example, functions on a quotient stack $[X/G]$ turn out to
be $G$-invariant functions on $X$; metrics on $[X/G]$ are
$G$-invariant metrics on $X$; bundles on $[X/G]$ are 
$G$-equivariant bundles on $X$, and so forth.
By contrast, $G$-equivariant bundles on $X$ are {\it not} the
same as bundles on quotient spaces $X/G$ -- stacks encode information
that is lost when passing to quotient spaces.

More generally, quotient stacks give a geometric realization
of properties of orbifolds.  Given that stacks appear to have all
of the basic properties that one requires to make sense of target spaces
of nonlinear sigma models, it was conjectured in \cite{qs} that
nonlinear sigma models on (Calabi-Yau) stacks are a well-defined notion,
and moreover that string orbifolds are the same as nonlinear sigma models
on quotient stacks.  In \cite{qs} it was shown how the twisted sector
sum that defines string orbifold CFT's emerges immediately as a property
of a sigma model on a stack \cite[section 4]{qs}, 
as part of the sum over all maps into the stack.
Actions for sigma models on stacks were defined \cite[section 9]{qs}, 
and closed string massless
spectra computed, as cohomology of the zero-momentum part of the loop
space of the stack, which turns out to precisely duplicate the
massless spectrum of a string orbifold CFT. 

With this in mind, it is natural to conjecture that the spectrum
of open strings in orbifolds should be counted by Ext groups on 
the quotient stack $[X/G]$, and in fact, this turns out to be true.
In particular, when $G$ is a finite group,
Ext groups on a quotient stack are the same\footnote{
This equality follows from the Grothendieck spectral sequence
\cite[section 5.6]{groth} $H^p\left(G, \mbox{Ext}^q_X\left( {\cal S}, {\cal T}
\right) \right) \Longrightarrow \mbox{Ext}^{p+q}_{ [X/G] } \left( {\cal S},
{\cal T} \right)$ and the observation that since $G$ is finite and the 
coefficients are vector spaces, $G$ has no cohomology with finite-dimensional
coefficients of degree $p > 0$.}
as the $G$-invariant
part of Ext groups on the covering space:
\begin{displaymath}
\mbox{Ext}^n_{ [X/G] }\left( i_* {\cal E}, j_* {\cal F} \right) 
\: = \:
\mbox{Ext}^n_X\left( i_* {\cal E}, j_* {\cal F} \right)^G
\end{displaymath}
so in arguing that the open string spectrum in orbifolds is the
same as the $G$-invariant part of Ext groups on the covering space,
we have also implicitly been arguing that the spectrum is counted
by Ext groups on quotient stacks.

In passing, we should observe that fractional branes can be understood
very simply in this language.  Fractional branes were defined in
\cite{dm} as D-branes on the covering space with $G$-actions,
{\it i.e.} $G$-equivariant sheaves on the cover, which means they can be
described as
sheaves on the stack $[X/G]$.  However, for reasons explained in \cite{kps},
fractional branes often cannot be reasonably modelled as sheaves
on quotient spaces -- the stack perspective is essential.

\subsection{Example:  ADHM/ALE construction}

As a quick check of our results, let us derive the setup ADHM construction
on ALE spaces from D-branes on orbifolds, as in \cite{dm}.
In particular, consider the moduli space of $k$ instantons of $U(N)$,
or at least its ADE version, built from $k$ D5-branes on $N$ D9-branes, say.
To match the literature, define $V = {\bf C}^k$ and $W = {\bf C}^N$.
Let ${\cal E}$ denote the trivial rank $N$ bundle on ${\bf C}^2$,
and ${\cal F}$ the triviail rank $k$ bundle over the origin of ${\bf C}^2$,
so that the sheaf $i_* {\cal E}$ corresponds to the D9-branes,
and $j_* {\cal F}$ corresponds to the D5-branes.
Then, on the covering space, the spectrum of boundary R-sector states
connecting D9-branes to D5-branes is given by
\begin{displaymath}
\mbox{Ext}^n_{ {\bf C}^2 }\left( i_* {\cal E}, j_* {\cal F} \right)
\: = \: \left\{ \begin{array}{cl}
                 \mbox{Hom}(W,V) & n=0 \\
                 0 & n \neq 0
                 \end{array} \right.
\end{displaymath}
and the spectrum of states connecting the D5-branes to the D9-branes is given
by 
\begin{displaymath}
\mbox{Ext}^n_{ {\bf C}^2 } \left( j_* {\cal F}, i_* {\cal E} \right)
\: = \: \left\{ \begin{array}{cl}
                \mbox{Hom}(V,W) & n=2 \\
                0 & n \neq 2
                \end{array} \right.
\end{displaymath}
The spectrum of states in the orbifold theory is given by the $G$-invariants,
{\it i.e.},
\begin{eqnarray*}
\mbox{Ext}^n_{ [ {\bf C}^2/G ] }\left( i_* {\cal E}, j_* {\cal F} \right)
& = & \left\{ \begin{array}{cl}
               \mbox{Hom}_G(W,V) & n=0 \\
               0 & n \neq 0
               \end{array} \right. \\
\mbox{Ext}^n_{ [ {\bf C}^2/G ] }\left( j_* {\cal F}, i_* {\cal E} \right)
& = & \left\{ \begin{array}{cl}
              \mbox{Hom}_G(V,W) & n=2 \\
              0 & n \neq 2 
              \end{array} \right.
\end{eqnarray*}
exactly reproducing the needed states from D5-D9 strings appearing
in the ADHM/ALE construction \cite{kronnak,nak}.

\section{$B$ field twistings}   \label{btwist}

Physicists often think of the $B$ field as a globally-defined two-form
field on spacetime.  However, the fact that the curvature $H = d B$
can be nontrivial in cohomology tells us that that such a description
is somewhat inexact.  Just like $U(1)$ gauge fields on a nontrivial bundle,
the $B$ field should really be thought of as two-form fields on patches,
related on overlaps by gauge transformations, as described in 
\cite{hitchin}.  Such a collection of two-form fields on patches
can be thought of as a ``connection on a gerbe,'' where a gerbe is a
mathematical object that plays
the same role for two-form fields that line bundles do for 
one-form gauge fields.

Another careful analysis reveals that $B$ fields and D-brane gauge
bundles are also intertwined.  On a closed-string worldsheet $\Sigma$, 
the $B$ field coupling
\begin{displaymath}
\int_{\Sigma} B
\end{displaymath}
is invariant under the gauge transformation $B \mapsto B + d \Lambda$,
but on an open string worldsheet, this term is not invariant:
\begin{displaymath}
\int_{\Sigma} B \: \mapsto \: \int_{\Sigma} B \: + \:
\int_{\partial \Sigma} \Lambda
\end{displaymath}
The solution to this difficulty is well-known -- one must combine
gauge transformations of the $B$ field with affine translations
of the Chan-Paton gauge field:
\begin{eqnarray*}
B & \mapsto & B \: + \: d \Lambda \\
A & \mapsto & A \: - \: \Lambda
\end{eqnarray*}
so that the combination
\begin{displaymath}
\int_{\Sigma} B \: + \: \int_{\partial \Sigma} A
\end{displaymath}
is gauge-invariant.

One effect of this relationship between the $B$ field and the
Chan-Paton gauge field involves the
projectivization of group actions on D-brane gauge bundles
due to discrete torsion conjectured in \cite{mikedt}.
Because of the coupling above between $B$ field gauge transformations
and Chan-Paton affine translations, group actions on Chan-Paton
gauge fields are interrelated to the group actions on $B$ fields. 
Once one properly understands discrete torsion in terms of
group actions on $B$ fields \cite{dt3,dtrev,dtshift},
it is straightforward to check \cite{dt3,dtrev,dtshift}
that consistency of a group action
on a D-brane gauge bundle with the group action on the $B$ requires
the projectivization conjectured in \cite{mikedt}.

In any event, in the presence of a $B$ field, one no longer has
an honest bundle on the worldvolume of a D-brane, but rather
a twisted bundle, so our description of D-branes in terms of sheaves
must be re-examined.  This matter is examined in detail in
\cite{cks}; we will outline the results here.

\subsection{Twisted bundles and sheaves}

Ordinarily on triple overlaps the transition functions $g_{i j}$
of a bundle obey the condition
\begin{displaymath}
g_{i j} g_{j k} g_{k i}
\: = \: 1
\end{displaymath}
However, as mentioned above,
$B$ field gauge transformations combine nontrivially
with Chan-Paton translations, 
so if the $B$ field has nonzero transition functions,
then the connection on our `bundle' picks up affine translations
between coordinate charts.
The effect is that the transition functions $g_{i j}$
of the `bundle' obey the condition \cite{freeded}
\begin{equation}
\label{cochain}
g_{i j} g_{j k} g_{k i}
\: = \: \alpha_{i j k} 
\end{equation} 
for a 2-cocycle $(\alpha_{i j k})$ representing a cohomology class
in $H^2(X, C^{\infty}(U(1))) \cong H^3(X, {\bf Z})$.  
These cohomology classes classify gerbes \cite{hitchin},
and the $\{g_{i j}\}$ satisfying~(\ref{cochain}) define twisted bundles.
From the exact sequence
\begin{displaymath}
0 \: \longrightarrow \:
{\bf Z} \: \longrightarrow \:
C^{\infty}( {\bf R} ) \: \longrightarrow \:
C^{\infty}( U(1) ) \: \longrightarrow \: 0
\end{displaymath}
we get a map $H^2(X,C^{\infty}(U(1)))\to H^3(X,\mathbf{Z})$ which 
takes $[\alpha]$ to a class in $H^3(X,\mathbf{Z})$, the characteristic
class of the gerbe.  Note that
$H^2(X, C^{\infty}(U(1))) \cong H^3(X, {\bf Z})$.
See \cite{hitchin} for more details.

Note that modulo scalars, equation~(\ref{cochain})
becomes
\begin{equation}
\label{azumaya}
\bar{g}_{i j} \bar{g}_{j k} \bar{g}_{k i}
\: = \: 1
\end{equation}
where $\bar{g}$ is the $\mathrm{PGL}(n)$ image of $g \in \mathrm{GL}(n)$.
If the twisted bundle has rank $r$, then (\ref{azumaya}) says that we
get a cohomology class $[\bar{g}]\in H^1(X, C^{\infty}(\mathrm{PGL}(r)))$.
{}From the exact sequence
\[
0\to {\bf Z}_r \to C^{\infty}(\mathrm{GL}(r)) \to
C^{\infty}(\mathrm{PGL}(r)) \to 0
\]
we get a coboundary map $H^1(X, C^{\infty}(\mathrm{PGL}(r)))\to 
H^2(X, {\bf Z}_r )$, 
taking $[\overline{g}]$ to an element of $H^2(X, {\bf Z}_r)$.
It is 
straightforward to check that the image of this class in 
$H^2(X, C^{\infty}(U(1)))$ induced by the inclusion 
${\bf Z}_r \hookrightarrow U(1)$
is precisely the class of the gerbe defined by
the original $\alpha$.  Since $H^2(X, {\bf Z}_r )$ is torsion, we see that
rank $r$ twisted bundles can only exist if the class of the underlying gerbe
is $r$ torsion. In particular, if the gerbe is not a torsion class, then there
are no finite-rank twisted bundles at all.

Such twisted bundles are understood in mathematics
(see for example \cite{andreithesis,bouwmathai,mathaisteven} and
references therein).
One can define not only twisted bundles, but also twisted
coherent sheaves on a space.
In fact, sheaves can also be defined in terms of local data
and transition functions obeying a cocycle condition
(see for example \cite[cor. I-14]{eh}), so we can define twisted
sheaves in exactly the same form as twisted bundles.
On the other hand, although twisted bundles are immediately
realized in D-brane physics, the physics of twisted sheaves
is more obscure.  Part of what we are doing in this paper
amounts to giving concrete evidence that D-branes in flat
$B$ field backgrounds can be accurately modeled by twisted sheaves
(see also \cite{kap,kaporlov} for previous work in this vein).

In this holomorphic context, the twisting is defined by
an element $\alpha\in H^2(X, {\cal O}_X^*)$ associated
to the $B$ field.  Moreover, one can also define Ext groups
between twisted sheaves.  If ${\cal S}$, ${\cal T}$ are two
coherent sheaves on $X$, both twisted by the same
element $\omega \in H^2(X, {\cal O}_X^*)$, then one can
define $\mbox{Ext}^*_{X, \omega}\left( {\cal S}, {\cal T} \right)$.
We will say more about such Ext groups and how to arrive at
elements of $H^2(X, {\cal O}_X^*)$ from a $B$ field momentarily.
First we will review twisted sheaves in more detail.

The most convenient definition of a twisted sheaf is closely
analogous to a definition of an ordinary sheaf given in
\cite[cor. I-14]{eh}. 
If $\alpha\in H^2(X,
{\cal O}_X^*)$ is represented by a \v{C}ech 2-cocycle 
$(\alpha_{ijk})$ on an
open cover $\{U_i\}_{i\in I}$, with the $\alpha_{ijk}$ holomorphic,
an $\alpha$-twisted sheaf $\cF$ consists of a
pair 
\[ (\{\cF_i\}_{i\in I}, \{\phi_{ij}\}_{i,j\in I}), \]
where $\cF_i$ is an ordinary sheaf on $U_i$ for all $i\in I$ and
\[ \phi_{ij}: \cF_j|_{U_i\cap U_j} \ra \cF_i|_{U_i\cap U_j} \]
is an isomorphism for all $i,j\in I$, subject to the conditions:
\begin{enumerate}
\item $\phi_{ii} = \mathrm{Id}$;
\item $\phi_{ij} = \phi_{ji}^{-1}$;
\item $\phi_{ij}\circ \phi_{jk}\circ \phi_{ki} = \alpha_{ijk} 
\cdot \mathrm{Id}$.
\end{enumerate}
One can show that the category of twisted sheaves, defined in
this fashion, is independent of the choice of the covering
$\{U_i\}$~(\cite[1.2.3]{andreithesis}) and of the particular cocycle
$\{\alpha_{ijk}\}$~(\cite[1.2.8]{andreithesis}).

Twisted sheaves are not coherent in the usual sense,
essentially because multiplication by an algebraic function is not
well-defined across overlaps because of the twisting.
However, there is an applicable notion of twisted coherence.
If a sheaf if twisted by $\alpha \in H^2(X, {\cal O}_X^*)$ as above,
then we say it is $\alpha$-coherent if each of the sheaves
${\cal F}_i$ in the definition above is coherent.

The class of twisted sheaves together with the obvious notion of
homomorphism is an abelian category,
the category of $\alpha$-twisted sheaves, leading to the stated
notion of Ext groups by general homological algebra. 
It can be shown that the category of $\alpha$-coherent sheaves
has enough injectives.  However, defining projective or locally-free
resolutions is somewhat more tricky.

For completeness, we describe the
relationship between elements of $H^2(X, {\cal O}_X^*)$
and gerbes, at least in the case of a Calabi-Yau threefold.  Extensions
are straightforward.
There is an exact sequence
\begin{displaymath}
H^2(X, {\cal O}_X ) \: \longrightarrow \:
H^2(X, {\cal O}_X^* ) \: \longrightarrow \:
H^3(X, {\bf Z} ) \: \longrightarrow \: H^3(X, {\cal O}_X )
\end{displaymath}
and on a Calabi-Yau threefold, $H^2(X, {\cal O}_X) = 0$.  Thus $H^2(X,
{\cal O}_X^* )$ is the kernel of 
\begin{displaymath}
H^3(X, {\bf Z} ) \to H^3(X, {\cal
O}_X )\simeq H^{0,3}(X).
\end{displaymath}
This map can be described by first mapping
$H^3(X, {\bf Z} )\to H^3(X, {\bf R} )$, then following with the
Hodge-theoretic projection of a de Rham third cohomology class onto
its $(0,3)$ component.  Since an element of $H^3(X, {\bf Z} )$ is
real, the $(3,0)$ component vanishes whenever its $(0,3)$ component
vanishes.  Thus we see that given an element of $H^3(X,{\bf Z})\cap
(H^{2,1}(X)\oplus H^{1,2}(X))$,\footnote{We are not being quite precise in
notation here,
as $H^3(X, {\bf Z})$ does not embed in $H^3(X, {\bf R})$ in the presence of
torsion.} we can uniquely reconstruct a
corresponding element of $H^2(X, {\cal O}_X^*)$.  Note that
$H^3(X,{\bf Z})\cap (H^{2,1}(X)\oplus H^{1,2}(X))$ plays an important role
in the generalized Hodge conjecture in dimension 3.  For more details on
this circle of ideas, see \cite{cg}.

In these lectures, we are only interested in flat $B$ fields,
which means that the curvature (the de Rham image of the
characteristic class) vanishes, so the corresponding
element of  $H^3(X, {\bf Z})$ is purely torsion.  When this happens, its
image in $H^3(X, {\cal O}_X )$ necessarily vanishes for the above reasons.

In summary, the torsion subgroup of $H^2(X,{\cal O}_X^*)$ is isomorphic to the
torsion subgroup of $H^3(X,{\bf Z})$.  So the elements 
of $H^2(X,{\cal O}_X^*)$ relevant to our study of D-branes correspond to 
purely topological information.

As we have seen above, one can define holomorphic {\it locally free}
twisted sheaves of finite rank only when the twisting
is by an element of $H^2(X, {\cal O}_X^*)$ corresponding
to a torsion element of $H^3(X,{\bf Z})$.  The elements arising this way
form the {\em Brauer group}.
(A widely-believed conjecture of Grothendieck says that the elements 
of the Brauer group
are the same as the torsion elements of $H^2(X, {\cal O}_X^*)$.)
For twistings by non-torsion elements, no locally free twisted
sheaves of finite rank exist. 
So our physical constraint of flat $B$ fields translates
directly into mathematics.

In passing, we will note that 
there is a closely analogous phenomenon in the study of WZW models.
There, one has strings propagating on a group manifold with a nontrivial
$B$ field background, in which the curvature $H$ of the $B$ field
is nonzero (this is the point of the Wess-Zumino term, after all).
D-branes cannot wrap the entire group manifold.  Instead,
D-branes can only wrap submanifolds $S$ such that the restriction
of the characteristic class to $S$ is torsion \cite{wzw1,wzw2}, mirroring the
statement that finite-rank locally-free twisted sheaves only exist when
the characteristic class of the twisting is torsion.

Another way to think about twisted holomorphic bundles is that they
are associated to projective space bundles that cannot be obtained
by projectivizing an ordinary holomorphic vector bundle.
In detail, given a projective space bundle, imagine lifting 
the fibers locally from ${\bf P}^n$ to ${\bf C}^{n+1}$.
The original projective space bundle was a {\it bundle},
so its transition functions close on triple overlaps.
However, if we lift those same transition functions from
${\bf P}^n$ to ${\bf C}^{n+1}$, then they need only close
up to the action of ${\bf C}^{\times}$ on triple overlaps.
In other words, given a projective space bundle, it need not
lift to an honest vector bundle, but in general will only
lift to a twisted vector bundle, twisted by an element of
$H^2\left(X, {\cal O}_X^*\right)$ defined by the failure of the
transition functions to close on triple overlaps.

Given two twisted sheaves ${\cal E}$, ${\cal F}$, one can
define a twisted sheaf of local homomorphisms $\underline{
\mbox{Hom}}({\cal E}, {\cal F} )$.
If ${\cal E}$ is twisted by $\omega \in H^2(X, {\cal O}_X^*)$
and ${\cal F}$ is twisted by $\omega' \in H^2(X, {\cal O}_X^*)$,
then $\underline{\mbox{Hom}}({\cal E}, {\cal F})$ is
a twisted sheaf, twisted by $\omega^{-1} \cdot \omega' \in
H^2(X, {\cal O}_X^*)$.

In particular, if two bundles ${\cal E}$, ${\cal F}$ are twisted
by the {\it same} element of $H^2(X, {\cal O}_X^*)$,
then $\underline{\mbox{Hom}}({\cal E}, {\cal F})$ is an
ordinary, untwisted sheaf.

We can use that fact to set up homological algebra in the
usual form.  If ${\cal S}$, ${\cal T}$ are two twisted sheaves,
but twisted by the {\it same} element of $H^2(X, {\cal O}_X^*)$,
then we can define local $\underline{\mbox{Ext}}^n_{ {\cal O}_X }
\left( {\cal S}, {\cal T}\right)$ and global $\mbox{Ext}^n_X
\left( {\cal S}, {\cal T} \right)$.

This last fact is directly relevant for this paper,
as in any given theory there is only a single $B$ field that
twists the D-brane bundles.  Thus, open string spectra are
always computed between twisted sheaves that have been
twisted by the same element of $H^2(X, {\cal O}_X^*)$,
which means that for all our string spectrum calculations,
there exists a relevant notion of Ext groups
in mathematics, and the associated Ext sheaves are ordinary sheaves.

Serre duality functions for twisted sheaves much as for 
ordinary untwisted sheaves.
For $\omega \in H^2(X, {\cal O}_X^*)$ an element of the Brauer
group ({\it i.e.} corresponding to a flat $B$ field),
for any two $\omega$-twisted coherent sheaves ${\cal S}$,
${\cal T}$ on a Calabi-Yau $n$-fold $X$, we have
\begin{displaymath}
\mbox{Ext}^i_{X, \omega}\left( {\cal S}, {\cal T} \right) \: \cong \:
\mbox{Ext}^{n-i}_{X, \omega}\left( {\cal T}, {\cal S} \right)^*.
\end{displaymath}

We have spoken a great deal so far about twisted sheaves
in general terms.  Although our remarks apply in principle,
in practice we unfortunately do not know of any examples
of smooth Calabi-Yau threefolds with non-zero torsion in
$H^2(X, {\cal O}_X^*)$.  Thus, our analysis is necessarily somewhat
formal.  In \cite{cks} we illustrated our calculations with some
examples involving spaces that are almost but not quite Calabi-Yau,
as in \cite{agm}.

\subsection{Twisted Ext groups from open strings}

From our preceding discussion, it should be clear that in the presence
of flat nontrivial $B$ fields, D-branes should be modelled by
twisted sheaves, in the sense just described.  This hypothesis has also
been investigated in \cite{kap,kaporlov}.  We will check this hypothesis
by comparing open string spectra in flat $B$ field backgrounds to
twisted Ext groups between corresponding twisted sheaves.

Since we can locally gauge the $B$ field to zero,
the analysis is very similar to that in section~\ref{vertexext}.
Again, boundary R-sector states are constructed from BRST-invariant
combinations of zero modes, and have the general form
\begin{displaymath}
b^{\alpha \beta j_1 \cdots j_m}_{ \overline{\imath}_1 \cdots \overline{\imath}_n }(\phi)
\eta^{\overline{\imath}_1} \cdots \eta^{\overline{\imath}_n }
\theta_{j_1} \cdots \theta_{j_m}
\end{displaymath}
just as in section~\ref{vertexext}.
The only significant difference between the present case and that
discussed in section~\ref{vertexext} is that the Chan-Paton factors
no longer couple to honest bundles on D-brane
worldvolumes, but rather to twisted bundles.

Let us specialize to the case of two D-branes wrapped on the
same submanifold $S$ in $X$.
Just as in section~\ref{vertexext}, if we assume that $TX|_S$ splits
holomorphically as $TS \oplus {\cal N}_{S/X}$,  
then the boundary R-sector
states above are counted by the sheaf cohomology groups
\begin{displaymath}
H^n\left(S, {\cal E}^{\vee} \otimes {\cal F} \otimes \Lambda^m {\cal N}_{S/X}
\right)
\end{displaymath}
where ${\cal E}$, ${\cal F}$ are the $i^* \omega$-twisted 
bundles on the D-brane 
worldvolume.  Also just as in section~\ref{vertexext}, there is a spectral
sequence
\begin{displaymath}
H^n\left(S, {\cal E}^{\vee} \otimes {\cal F} \otimes \Lambda^m {\cal N}_{S/X}
\right)
\: \Longrightarrow \:
\mbox{Ext}^{n+m}_{X, \omega}\left( i_* {\cal E}, i_* {\cal F} \right).
\end{displaymath}
Just as in section~\ref{vertexext}, this spectral sequence can be
realized physically, by lifting ${\cal N}_{S/X}$-valued sheaf cohomology
to $TX|_S$-valued sheaf cohomology, and demanding that the 
$\overline{\partial}$-image of such $TX|_S$-valued sheaf cohomology be
annihilated after applying the boundary conditions
$\theta_i = \left( F_{i \overline{\jmath}} + B_{i \overline{\jmath}} \right)
\eta^{\overline{\jmath}}$ \cite[(3.3)]{aboo2}.
In all important respects, the physical analysis and construction
of boundary R-sector states is identical to section~\ref{vertexext}.
Reference \cite{cks} works through these details extensively, and also
proves that the needed spectral sequences (giving twisted Ext groups
rather than ordinary Ext groups) do exist and have the form needed to
match physics.

Thus, we see that open string spectra between D-branes in flat
nontrivial $B$ field backgrounds are counted by
twisted Ext groups between twisted sheaves corresponding to the
D-branes.

\section{Higgs vevs, nonreduced schemes, and other sheaves} \label{nonred}

\subsection{Basics}

In this paper we have really only discussed sheaves that are of the
form $i_* {\cal E}$, {\it i.e.}, pushforwards of vector bundles along
inclusion maps.  However, there are many sheaves that are not of this form,
begging the question whether any of them also have physical interpretations.

We shall see momentarily that
some of those other sheaves do have physical interpretations,
in terms of nilpotent Higgs fields.
(Rather, we shall outline the relevant results; see \cite{dks} for a more
complete exposition.)
Now, typically nilpotent Higgs fields are excluded by D-terms in physical
theories.  However, even in physical theories that is not always the case
 -- D-branes in string orbifolds are an example where nilpotent 
Higgs vevs are not
only allowed, but play an important role, as mapping out the exceptional
divisors of resolutions of quotient spaces \cite{dgm}, as we shall see
more explicitly in subsection~\ref{nilorb}.  Also, in topological
theories there are no such D-term constraints, only the analogues of F-term
constraints, so nilpotent Higgs vevs are certainly allowed in 
the B model.

In order to talk sensibly about Higgs vevs, we must already make
an assumption about the D-branes.  In section~\ref{vertexext} we saw
how holomorphic sections of ${\cal E}^{\vee} \otimes {\cal E} \otimes
{\cal N}_{S/X}$, {\it i.e.} Higgs fields in the usual sense,
can be inextricably intertwined with deformations of the gauge bundle:
not only can one type of deformation not be distinguished from another,
but also
not all holomorphic sections of 
${\cal E}^{\vee} \otimes {\cal E} \otimes
{\cal N}_{S/X}$ are even states in the low-energy spectrum.
So, the notion of Higgs field from algebraic geometry does not
apply in general.  

In order for Higgs fields in the sense
of algebraic geometry ({\it i.e.} holomorphic sections of 
${\cal E}^{\vee} \otimes {\cal E} \otimes
{\cal N}_{S/X}$) to appear in the low-energy spectrum, we must
assume that $TX|_S$ splits holomorphically into $TS \oplus {\cal N}_{S/X}$.
Although such a splitting always happens as $C^{\infty}$ bundles,
we have seen that holomorphically $TX|_S$ need not split, 
a fact which played an important role in understanding how Ext groups
solve the problem of calculating the open string spectrum with nontrivial
boundary conditions.  Once we assume that $TX|_S$ splits holomorphically,
we are able to distinguish gauge bundle moduli from Higgs fields.
In fact, if $TX|_S$ splits holomorphically then the spectral sequences
that played an important role in section~\ref{vertexext} all
trivialize.

In this section, for simplicity we shall not only assume
that $TX|_S$ splits, but also that the gauge bundle on the worldvolume
of each D-brane is trivial.  In fact, almost all of the examples we will
discuss here will involve coincident D-branes supported at points
on ${\bf C}^n$ (see \cite{dks} for a much more general and detailed
analysis).

We can see the effect of giving a Higgs field a vev by adding a boundary
term $\{ G, V \}$ to the boundary action, where 
$V = \Phi^i \theta_i$ is the vertex operator
for the Higgs field.  As before, the effect is to deform the BRST operator,
by adding a term $V$ to $\overline{\partial}$.
If we add Higgs vevs $\Phi_1^i \theta_i$, $\Phi_2^i \theta_i$ to either
side of an open string, then the BRST operator is deformed to the form
\begin{displaymath}
Q \: = \: \overline{\partial} \: + \:
\Phi_1^i \theta_i \: - \: \Phi_2^i \theta_i
\end{displaymath}
and demanding that $Q^2 = 0$, to preserve boundary ${\cal N}=2$ supersymmetry,
gives the constraints that the Higgs fields be holomorphic, and that
the Higgs fields on either side of the open string commute\footnote{Note that
this is a worldsheet realization of what are usually F-term constraints
in a physical target space theory.} with one another:
$[\Phi_1^i, \Phi_1^j] = 0$, $[\Phi_2^i, \Phi_2^j] = 0$.
As mentioned above, since we assume $TX|_S$ splits holomorphically,
all spectral sequences trivialize, so the physical spectrum of
boundary R-sector states is merely the cohomology of $Q$, in the obvious
way, without the complications described in section~\ref{vertexext}.

Mathematically, we can associate Higgs vevs to sheaves by
interpreting (commuting) Higgs vevs as defining a deformation of the
action of the ring of algebraic functions on the module describing
the D-branes (without Higgs vevs) as sheaves.  Technically, the resulting
sheaf is a sheaf in the normal bundle ${\cal N}_{S/X}$, not $X$,
as is discussed
in more detail in \cite{dks}, but this distinction will only
be relevant in the last example we discuss.

It can be proven \cite{dks} 
that Ext groups between sheaves obtained from Higgs vevs in this fashion
are the same as open string spectra, computed as cohomology of the
BRST operator described above, giving us a highly nontrivial check that this
mathematical prescription has physical content.
See \cite{dks} for a general proof; we shall verify the result in 
the examples below.

Our methods apply to any set of commuting Higgs vevs.  For `typical' cases,
the effect on the sheaves of the operation above is to move the support,
{\it i.e.} to move the D-brane, as one would expect naively.
Nilpotent Higgs vevs often are interpreted as giving rise to
nonreduced schemes, as originally suggested in \cite{tomasme}.
We mentioned above that nilpotent Higgs fields play an important role
in discussions of D-branes on orbifolds \cite{dgm}, where they correspond to the
exceptional divisors seen on classical moduli spaces 
(see subsection~\ref{nilorb}).
The interpretation as nonreduced schemes could have been anticipated
from consistency with the McKay correspondence \cite{bkr}, 
where the McKay image of
a D0 brane on the exceptional divisor of a resolution of a quotient space
is typically a $G$-equivariant nonreduced scheme on the covering space
\cite[section 7.2]{kps}.

To illustrate these methods, we shall consider a few easy examples.

\subsection{First example:  separable D0 branes}

Our methods apply equally well to both nilpotent and non-nilpotent
Higgs vevs, so let us begin by considering a simple case involving
non-nilpotent Higgs vevs, that should be interpretable as moving
the D-branes.

Consider two D0 branes sitting at the origin of ${\bf C}$, the
complex line.  Let us give the Higgs field on one of the D0 branes
a nonzero vev.  That should be equivalent to moving the D0 brane
away from the origin, which should be reflected both in the sheaf
description, and in the open string spectrum.

First, let us verify that in the sheaf description, giving the
Higgs field a vev corresponds to moving the support of the sheaf.
The original sheaf describing a single D0 brane, with no Higgs vev,
is the skyscraper sheaf supported at the origin of ${\bf C}$,
and corresponding to the ${\bf C}[x]$-module ${\bf C}[x]/(x)$
 -- a module with one generator $\alpha$ and the relation 
$x \cdot \alpha = 0$.  Now, giving a nonzero vev $\phi$ to the Higgs field
corresponds to creating a new module from the old one, with a new ring action.
The new module is also defined by a single generator $\alpha$,
but instead of the relation $x \cdot \alpha = 0$,
the Higgs vev defines a new action of $x$:  $x \cdot \alpha = \phi \alpha$,
or $(x-\phi) \alpha = 0$.  This new module is also a skyscraper sheaf,
but supported at the point $x=\phi$ instead of $x=0$,
consistent with our expectation that giving a nonzero (and non-nilpotent)
Higgs vev should move the D-brane.

Next, let us verify that the open string spectrum is consistent with
the picture we have just described.  Since the two D-branes no longer
intersect, there cannot be any massless states between them,
or more mathematically, all the Ext groups between the two sheaves
(the skyscraper sheaf at the origin, and the skyscraper sheaf at
$x = \phi$) should vanish.

The open string spectrum is calculated by starting with states of the
same form as for two D0-branes both at the origin, and then taking
cohomology of the deformed BRST operator.
In this simple case, the open string spectrum reduces to the cohomology
of the operator $Q = \phi \theta$.  (There is no longer an 
$\overline{\partial}$ contribution to the BRST operator, since the D-branes
are supported at {\it points}.)

Degree zero open string states connecting the two original D0-branes
at the origin
have the form $V = b$ for a constant $b$ (a $1 \times 1$ matrix), 
following \cite{ks}.
The BRST operator maps $b \mapsto b \phi \theta$, and so if $\phi \neq 0$,
then for the state $V = b$ to be in the kernel of the BRST operator
implies that $b=0$, hence there are no degree zero states.

Degree one states are of the form $V = b \theta$.
These are all annihilated by the BRST operator, but they are also
all in the image of the BRST operator, hence the space of physical
states has dimension zero.

Thus, we see explicitly from the spectrum of the deformed BRST operator
that there are no physical states, precisely as expected on both
mathematical and physical grounds in this simple case.

\subsection{Second example:  nilpotent Higgs vevs}

Here we shall consider open strings between two pairs of D0 branes,
all at the origin of ${\bf C}^2$,
with nonzero Higgs vevs on only one side of the open string.
We shall take the Higgs fields to be
\begin{displaymath}
\Phi_x \: = \: \left[ \begin{array}{cc}
                       0 & 1 \\
                       0 & 0
                       \end{array} \right],
\: \:
\Phi_y \: = \: \left[ \begin{array}{cc}
                      0 & 0 \\
                      0 & 0
                      \end{array} \right]
\end{displaymath}
If we start with the (trivial) rank 2 vector bundle on a point
in ${\bf C}^2$, (equivalently a pair of skyscraper sheaves ${\cal O}_0^2$),
and deform the action of the ambient ring by the Higgs fields above,
then our bundle becomes the structure sheaf of the nonreduced scheme
of order 2 at the origin, defined by the ideal $(x^2,y)$,
and denoted $D_x$.  Thus, our open string computation should
be correctly reproduced by $\mbox{Ext}^*_{ {\bf C}^2} ( {\cal O}_0^2, D_x )$.

First, consider degree zero states.
These are just matrices
\begin{displaymath}
V \: = \: \left[ \begin{array}{cc}
       a & b \\
       c & d 
       \end{array} \right]
\end{displaymath}
Demanding that the state above be in the kernel of $Q = \overline{\partial}
+ \Phi^i_1 \theta_i - \Phi^i_2 \theta_i $ means that
$V \Phi_x = 0$, so the $V$ in the kernel of $Q$ have the form
\begin{displaymath}
\left[ \begin{array}{cc}
       0 & b \\
       0 & d 
       \end{array} \right]
\end{displaymath}
Since there is no image to mod out, we see that the space of degree zero
states has dimension two.

Next, consider degree one states.
These can be written in the form
\begin{displaymath}
V \: = \: \left[ \begin{array}{cc}
                 a_x & b_x \\
                 c_x & d_x 
                 \end{array} \right] \theta_1 \: + \:
\left[ \begin{array}{cc}
       a_y & b_y \\
       c_y & d_y 
       \end{array} \right] \theta_2 
\end{displaymath}
States in the kernel of the BRST operator $Q$ can be written
\begin{displaymath}
V \: = \: \left[ \begin{array}{cc}
                 a_x & b_x \\
                 c_x & d_x 
                 \end{array} \right] \theta_1 \: + \:
\left[ \begin{array}{cc}
       0 & b_y \\
       0 & d_y 
       \end{array} \right] \theta_2 
\end{displaymath}
and states in the image of $Q$ can be written in the form
\begin{displaymath}
\left[ \begin{array}{cc}
       0 & a \\
       0 & c 
       \end{array} \right] \theta_1
\end{displaymath}
for some $a$, $c$.  Since the kernel is six-dimensional,
and the image is two-dimensional, we see that the space of
BRST-closed degree one states, modulo BRST exact states,
has dimension four.

Next, consider degree two states.
These can be written in the form
\begin{displaymath}
V \: = \: \left[ \begin{array}{cc}
                 a & b \\
                 c & d 
                 \end{array} \right] \theta_1 \theta_2
\end{displaymath}
The BRST operator $Q$ annihilates all of the degree two states,
and it is straightforward to check that the image of $Q$ in
degree two states has the form
\begin{displaymath}
\left[ \begin{array}{cc}
       0 & \alpha \\
       0 & \beta 
       \end{array} \right] \theta_1 \theta_2
\end{displaymath}
for some $\alpha$, $\beta$.
Hence the kernel of $Q$ has dimension four, and the image of $Q$
has dimension two, so the space of BRST-closed states, modulo
BRST-exact states, has dimension two.

Now, in order to interpret the results of these calculations
as coming from Ext groups between sheaves, we must find a sheaf-theoretic
interpretation of the pair of D0-branes with nontrivial Higgs fields.
Since we start with two D0 branes, described by a pair of
skyscraper sheaves ${\cal O}_0^2$,
we have a module with two generators,
say, $\alpha$ and $\beta$, which in the undeformed sheaf are
both annihilated by both $x$ and $y$.  The Higgs field associated to $x$
maps 
\begin{displaymath}
\left[ \begin{array}{c}
       \alpha \\ \beta \end{array} \right] \: \mapsto \:
\left[ \begin{array}{c}
       \beta \\ 0 \end{array} \right]
\end{displaymath}
so we deform the ring action as
$x \cdot \alpha = \beta$ and $x \cdot \beta = 0$.
The Higgs field associated to $y$ annihilates both generators:
$y \cdot \alpha = x \cdot \beta = 0$.
Such a module is precisely ${\bf C}[x,y]/(x^2,y)$,
where we identify $\alpha$ with the image of $1 \in {\bf C}[x,y]$
in the quotient, and $\beta$ with the image of $x$ in the quotient.

This module defines an example of a nonreduced scheme.
It is supported at the origin, and has length\footnote{Length is a property
of ideals that we will not try to define here.} two, reflecting the fact
that ultimately it is describing a pair of D-branes, but is not the
same as two copies of the skyscraper sheaf.
In fact, there is a ${\bf P}^1$'s worth of ideals of length two
by which we could quotient -- $D_x$ is not the only example of
a nonreduced scheme of length two supported at the origin of ${\bf C}^2$.

Thus, our computations predict
\begin{displaymath}
\mbox{dim }\mbox{Ext}^n_{ {\bf C}^2 } \left( {\cal O}_0^2, D_x \right) \: = \:
\left\{ \begin{array}{cl}
         2 & n=0 \\
         4 & n=1 \\
         2 & n=2
         \end{array} \right.
\end{displaymath}

This result is easy to check.
First, Serre dualize to $\mbox{Ext}^*\left( D_x, {\cal O}_0^2\right)$,
so that we can use the locally-free resolution of $D_x$,
given by
\begin{displaymath}
0 \: \longrightarrow \: {\cal O}_{ {\bf C}^2 }
\: \stackrel{{\scriptsize \left[ \begin{array}{c}
                      -y \\ x^2 \end{array} \right]} }{\longrightarrow} \:
{\cal O}^2_{ {\bf C}^2 } \:
\stackrel{ [x^2,y] }{\longrightarrow} \: {\cal O}_{ {\bf C}^2 } \:
\longrightarrow \: D_x \: \longrightarrow \: 0
\end{displaymath}
to calculate local $\underline{\mbox{Ext}}^n_{ {\bf C}^2 }(D_x,{\cal O}_0^2)$.
Since the maps in the resolution vanish on the support of
${\cal O}_0^2$, calculating local Ext is now trivial:
\begin{displaymath}
\underline{\mbox{Ext}}^n_{ {\cal O}_{ {\bf C}^2} }\left(
D_x, {\cal O}_0^2 \right) \: = \:
\left\{ \begin{array}{cl}
        \underline{\mbox{Hom}}( {\cal O}, {\cal O}_0^2 ) \: = \: {\cal O}_0^2 & n=0 \\
        \underline{\mbox{Hom}}( {\cal O}^2, {\cal O}_0^2 ) \: = \: {\cal O}_0^4 & n=1 \\
        \underline{\mbox{Hom}}( {\cal O}, {\cal O}_0^2 )  \: = \: {\cal O}_0^2 & n=2
        \end{array} \right.
\end{displaymath}
Applying the local-to-global spectral sequence, which is trivial
since the supports are on a point, 
we immediately recover the Serre dual of the Ext groups listed above,
as expected.

\subsection{Aside:  nilpotent Higgs vevs and orbifolds}  \label{nilorb}

In the previous example we saw explicitly how nilpotent Higgs vevs
correspond to structure sheaves of nonreduced schemes.
One question the reader might ask is, where do such nilpotent Higgs
vevs pop up physically?  After all, typically nilpotent Higgs vevs are
excluded by D-terms.
One answer to this question is that since we are working in a topological
field theory, target-space D-terms are irrelevant.
A better answer is that such nilpotent Higgs vevs play an important
role when describing D-branes on orbifolds \cite{dgm}, 
where they map out exceptional
divisors of resolutions, implicitly realizing a version of the
McKay correspondence.

Let us consider a specific example, to make explicit how nilpotent Higgs
vevs arise in orbifolds.  Consider a pair of D0-branes at the origin
of ${\bf C}^2$.  Consider the usual ${\bf Z}_2$ action on ${\bf C}^2$,
and put the D0-branes in the regular representation of ${\bf Z}_2$.
On the covering space, there are two Higgs fields, which we shall
label $X$ and $Y$.  These two Higgs fields must be invariant under
the orbifold group, which restricts them to the form
\begin{displaymath}
X \: = \: \left[ \begin{array}{cc}
                 0 & b_x \\
                 c_x & 0 \end{array} \right], \: \: \:
Y \: = \: \left[ \begin{array}{cc}
                 0 & b_y \\
                 c_y & 0 \end{array} \right]
\end{displaymath}
There is an F-term condition $[X,Y]=0$, which implies the constraint
\begin{displaymath}
b_x c_y \: - \: b_y c_x \: = \: 0
\end{displaymath}
The original $U(2)$ gauge symmetry is broken to $U(1)^2$,
of which one of the $U(1)$'s is trivial.  The $b$'s and $c$'s have
equal and opposite charges under the remaining $U(1)$,
so we have a D-term condition 
\begin{displaymath}
| b_x |^2 \: + \: | b_y |^2 \: - \: |c_x|^2 \: - \: |c_y|^2 \: = \: r
\end{displaymath}

Thus, the classical Higgs moduli space has the form
\begin{displaymath}
\left\{ \{ b_x c_y - b_y c_x = 0 \} \subset {\bf C}^4 \right\} // {\bf C}^{
\times}
\end{displaymath}
which turns out to be the same as the minimal resolution of 
${\bf C}^2/{\bf Z}_2$.

Suppose (without loss of generality) that $r > 0$.
Then the exceptional divisor of the resolution is given by
$c_x = c_y = 0$, so that $b_x$ and $b_y$ are homogeneous coordinates
on a ${\bf P}^1$, namely the exceptional divisor.
Notice that the Higgs vevs on the covering space can be written
in the form
\begin{displaymath}
X \: = \: \left[ \begin{array}{cc}
                 0 & b_x \\
                 0 & 0 \end{array} \right], \: \: \:
Y \: = \: \left[ \begin{array}{cc}
                 0 & b_y \\
                 0 & 0 \end{array} \right]
\end{displaymath}
which are manifestly nilpotent.

More generally, given a set of Higgs fields invariant under a $G$-action,
the eigenvalues and eigenvectors\footnote{F-term conditions will always
force the Higgs vevs to commute with one another, hence they are
simultaneously diagonalizable.} will form a representation of $G$.
There is an obvious map to the quotient space, determined by the
eigenvalues.  This map fails to be one-to-one when the Higgs fields
are nilpotent, corresponding to the case that the Higgs vevs are
mapping out the exceptional divisor of a resolution.

Thus, nilpotent Higgs vevs do occur in physics, and we see that they
are responsible for the exceptional divisors seen by D-branes on
orbifolds in \cite{dgm}.

The appearance of nilpotent Higgs vevs in this context
can also be viewed as a consequence of the McKay correspondence,
as formulated by \cite{bkr}.  If we start with a skyscraper sheaf
on the exceptional divisor of a resolution, corresponding to a D0-brane
on the exceptional divisor, and apply the McKay correspondence,
we get a $G$-equivariant nonreduced scheme on the covering space,
where the scheme structure encodes the location on the exceptional divisor.
To be consistent with the physics of \cite{dgm}, ideally one would like
for nonreduced schemes and nilpotent Higgs vevs to be closely related,
and that is exactly what we have described earlier in this section.

\subsection{Third example:  obstructed ${\bf P}^1$}

Another example involves D-branes wrapped on an obstructed
${\bf P}^1$ in a Calabi-Yau threefold.  Turning on a Higgs vev
corresponds to moving the sheaf inside the normal bundle,
but in this case, the curve cannot be moved inside the Calabi-Yau.
Since people typically identify Higgs vevs with finite deformations
inside the ambient space, rather than infinitesmial deformations /
deformations only inside the normal bundle, this example is both
interesting and important.

We will consider a single D-branes wrapped
on an obstructed ${\bf P}^1$, whose normal bundle
in the ambient Calabi-Yau threefold is ${\cal O} \oplus {\cal O}(-2)$.
The restriction of the tangent bundle $TX|_S$ does split holomorphically
as $TS \oplus {\cal N}_{S/X}$, so the usual notion of Higgs field
is applicable, and furthermore we shall assume the D-brane gauge bundle
is trivial, so as to simplify boundary conditions on worldsheet fermions.

Since the normal bundle contains an ${\cal O}$ factor, there is an
infinitesimal deformation of the D-brane -- a Higgs field, in the
sense of algebraic geometry -- corresponding to the holomorphic section
of the normal bundle.  But, how should that Higgs field be interpreted?

A better question is perhaps, if the curve is obstructed,
then how can the normal bundle be ${\cal O} \oplus {\cal O}(-2)$?
The total space of that bundle admits a one-parameter family of
${\bf P}^1$'s, including the zero section of the bundle (the original
${\bf P}^1$), yet we described the ${\bf P}^1$ as obstructed,
not admitting any finite deformations.

The mathematical resolution of this puzzle lies in the fact that in algebraic
geometry, normal bundles do {\it not} give a good local description
of the holomorphic geometry of the ambient space, unlike topology
and differential geometry, where the notion of a `tubular neighborhood'
is commonly used.  Instead, the holomorphic structure on the normal
bundle is only a linearization of the holomorphic structure on the
ambient space, and in this case, that linearization fails to capture
information about the obstruction.  We will return to this issue
in section~\ref{yoncheck}.
Thus, although the ${\bf P}^1$
can be deformed inside the normal bundle, it cannot be deformed
inside the ambient space.

Physically, in this case although the operator corresponding to the
Higgs field is marginal, it is not truly marginal, as some higher
correlation functions do not vanish,
but rather encode the
mathematical obstruction data.  Giving a vev to this Higgs field
breaks conformal invariance, albeit in a subtle fashion.
Again, see section~\ref{yoncheck}.

The reader should now be better equipped to understand remarks
made earlier, that giving a vev to a Higgs field should be
interpreted as moving the D-brane inside its normal bundle,
but not the ambient space.  Here, giving a vev to the Higgs field
creates a new sheaf inside the total space of the normal bundle
whose support has been shifted away from the original ${\bf P}^1$,
completely consistent with the fact that
the ${\bf P}^1$ cannot be moved inside the ambient space.

\subsection{Commutivity versus obstructedness of moduli}

Earlier we mentioned that in order to consistently give a Higgs
field a vev, one constraint (arising from demanding $Q^2=0$) is that
the Higgs vevs must commute with one another:  $[\Phi^i, \Phi^j] = 0$,
a condition that in the target space theory is typically an F-term
condition.

Mathematically this condition is a necessary, but not sufficient,
condition for a modulus to be obstructed.  We will take a moment
to review this fact.

In general, given an infinitesimal modulus of a sheaf $i_* {\cal E}$,
corresponding to an element of
\begin{displaymath}
\mbox{Ext}^1_X\left( i_* {\cal E}, i_* {\cal E} \right)
\end{displaymath}
the first obstruction to deforming a finite distance in that direction
is given by the Yoneda pairing:
\begin{displaymath}
\mbox{Ext}^1_X\left( i_* {\cal E}, i_* {\cal E} \right)
\: \times \:
\mbox{Ext}^1_X\left( i_* {\cal E}, i_* {\cal E} \right)
\: \longrightarrow \:
\mbox{Ext}^2_X\left( i_* {\cal E}, i_* {\cal E} \right).
\end{displaymath}

The Yoneda pairing fits into a commutative diagram:
\begin{displaymath}
\xymatrix{
\mbox{Ext}^1_X\left( i_* {\cal E}, i_* {\cal E} \right) \ar[d]
& \times &
\mbox{Ext}^1_X\left( i_* {\cal E}, i_* {\cal E} \right)
 \ar[r] \ar[d] & 
\mbox{Ext}^2_X\left( i_* {\cal E}, i_* {\cal E} \right) \ar[d] \\
H^0\left(S, {\cal E}^{\vee} \otimes {\cal E} \otimes {\cal N}_{S/X} \right)
& \times &
H^0\left(S, {\cal E}^{\vee} \otimes {\cal E} \otimes {\cal N}_{S/X} \right)
\ar[r] &
H^0\left(S, {\cal E}^{\vee} \otimes {\cal E} \otimes \Lambda^2{\cal N}_{S/X}
\right)
}
\end{displaymath}
and the commutivity statement is just the statement that
the image of the bottom product in 
$H^0\left(S, {\cal E}^{\vee} \otimes {\cal E} \otimes \Lambda^2{\cal N}_{S/X}
\right)$ vanishes.
(The necessary condition for the modulus to be unobstructed is that
the image in $H^0\left(S, {\cal E}^{\vee} \otimes {\cal E} \otimes \Lambda^2{\cal N}_{S/X}
\right)$ vanishes in cohomology, but since this is a degree {\it zero}
cohomology class, for this to vanish in cohomology means that
the wedge product must vanish identically.)

\section{Algebraic properties}   \label{algebraic}

In this section we shall consider algebraic properties of the open
string B model:  the boundary-boundary and bulk-boundary OPE's, and
Cardy's condition.  We shall discuss probable mathematical 
descriptions of these quantities, and outline the complications
involved in checking those assertions directly in BCFT.

\subsection{Boundary-boundary OPE's}   \label{yoncheck}

Boundary-boundary OPE's should give, in the B model, a pairing
\begin{displaymath}
\mbox{Ext}^p\left( {\cal E}, {\cal F} \right) \times
\mbox{Ext}^q\left( {\cal F}, {\cal G} \right) \: \longrightarrow \:
\mbox{Ext}^{p+q}\left( {\cal E}, {\cal G} \right).
\end{displaymath}
A well-known example of a pairing of this form is the Yoneda pairing,
and it is believed (though as yet not checked in detail) that the
B model boundary-boundary OPE algebra should be determined by the
Yoneda pairing.

In the special case that the
sheaves ${\cal E}$, ${\cal F}$, ${\cal G}$ are all
actually bundles, so that the Ext groups are sheaf cohomology groups,
representable by differential forms, the Yoneda pairing can be
explained much more simply:  it is merely the wedge product of the
corresponding differential forms, with a trace taken over the 
internal (${\cal F}$) indices, closely analogous to the
bulk-bulk OPE reviewed in section~\ref{bulkstates}.

However, more generally the Yoneda pairing is much more complicated.
Let us work through one particularly nasty example, to illustrate
some of the technical complications that arise when trying to check
that general boundary-boundary OPE's really are given by the Yoneda pairing.

We shall consider a single D-brane wrapped on an obstructed ${\bf P}^1$,
which is to say, a ${\bf P}^1$ which admits an infinitesimal deformation,
but whose deformation is obstructed at some order.
Such a setup was one of the motivations given in \cite{dv} for 
having an adjoint-valued field $\phi$ with a $\phi^n$-type 
superpotential\footnote{The superpotential was checked indirectly in \cite[section 2.2]{shamitkatz}
using a dimensionally-reduced holomorphic Chern-Simons theory
(implicitly assuming the dimensional reduction of the open string field
theory on the total space coincides with open string field theory of
D-branes on a submanifold).  However, a direct derivation in BCFT was
not given in that paper, and that is our interest here.}.

We shall assume that the gauge bundle on the D-brane is trivial,
so the boundary conditions on worldsheet fields take a simple form.
Furthermore, the restriction of the tangent bundle of the Calabi-Yau
to the ${\bf P}^1$ does split holomorphically.  Thus, neither of the usual
subtleties associated with open string computations is relevant here.

The normal bundle to the ${\bf P}^1$ is ${\cal O} \oplus {\cal O}(-2)$.
Since the normal bundle admits a holomorphic section, 
$\mbox{Ext}^1\left( {\cal O}_{ {\bf P}^1 }, {\cal O}_{ {\bf P}_1 } \right)$
is one-dimensional.  
For `generic' obstructions ({\it i.e.} of order 3),
the Yoneda pairing
\begin{displaymath}
\mbox{Ext}^1\left( {\cal O}_{ {\bf P}^1 }, {\cal O}_{ {\bf P}_1 } \right)
\times
\mbox{Ext}^1\left( {\cal O}_{ {\bf P}^1 }, {\cal O}_{ {\bf P}_1 } \right)
 \: \longrightarrow \:
\mbox{Ext}^2\left( {\cal O}_{ {\bf P}^1 }, {\cal O}_{ {\bf P}_1 } \right)
\end{displaymath}
is nonzero, and the obstruction is characterized by the image in Ext$^2$.
For (nongeneric) obstructions of higher order,
the Yoneda pairing will vanish, but a higher-order computation will
be nonvanishing.

Already at the level of vertex operators we can begin to see
some of the complications involved in realizing the Yoneda pairing.
In the present example, both Ext$^1$ and Ext$^2$ above are one-dimensional.
In fact,
\begin{eqnarray*}
\mbox{Ext}^1\left( {\cal O}_{ {\bf P}^1 }, {\cal O}_{ {\bf P}_1 } \right)
& = & H^0\left( {\cal N}_{ {\bf P}^1/X } \right) \: = \: {\bf C} \\
\mbox{Ext}^2\left( {\cal O}_{ {\bf P}^1 }, {\cal O}_{ {\bf P}_1 } \right)
& = & H^1\left( {\cal N}_{ {\bf P}^1/X } \right) \: = \: {\bf C} 
\end{eqnarray*}
From our earlier description of vertex operators, and the fact that
the only holomorphic section of ${\cal O}$ is the constant section,
we see that the elements of Ext$^1$ are described by the vertex operator
$\theta$ (associated to the ${\cal O}$ factor in the normal bundle),
and elements of Ext$^2$ are described by the vertex operator $\eta \theta$.
If the Yoneda pairing in this case were as trivial as just a wedge product,
then the image in Ext$^2$ would just be a product of $\theta$'s
 -- but by the Grassman property, such a product vanishes.
Instead, in a case in which the Yoneda pairing is nontrivial,
the image in Ext$^2$ is $\eta \theta$ instead of $\theta \theta$
 -- so the operator product must necessarily involve some sort of
interaction term that has the effect of changing a $\theta$ into
an $\eta$.

The fact that the normal bundle has this form might confuse the reader
 -- after all, the ${\bf P}^1$ is supposed to be obstructed,
and yet there is a one-parameter-family of rational curves inside the
normal bundle containing the ${\bf P}^1$.
The solution to this puzzle gives another reason why the Yoneda pairing
computation in this case is extremely difficult.
Unlike differential geometry, where normal bundles capture
local geometry, in algebraic geometry the normal bundle need {\it not}
encode the local holomorphic structure, only the local smooth structure.
In the present case, local coordinates in a neighborhood of the
obstructed ${\bf P}^1$ can be described as follows.
Let one coordinate patch on a holomorphic neighborhood have coordinates
$(x, y_1, y_2)$, and the other coordinate patch on a holomorphic neighborhood
have coordinates $(w, z_1, z_2)$, where
\begin{eqnarray*}
w & = & x^{-1} \\
z_1 & = & x^2 y_1 \: + \: x y_2^n \\
z_2 & = & y_2
\end{eqnarray*}
The integer $n$ is the degree of the obstruction,
the coordinates $x$, $w$ are coordinates on the ${\bf P}^1$, 
$z_2 = y_2$ is a coordinate on the ${\cal O}$ factor on the normal bundle,
and $z_1$, $y_1$ morally would be coordinates on the ${\cal O}(-2)$ factor,
except that the coordinate transformation is {\it not} that of
${\cal O}(-2)$ -- it's complicated by the $x y_2^n$ term, which means that
this local holomorphic neighborhood is not equivalent to the normal bundle.
The normal bundle is only a linearized approximation to local holomorphic
coordinates.  Unfortunately, data concerning the degree of the obstruction
({\it i.e.} the `extra' term in the expression for $z_1$)
is omitted by the linearization that gives rise to the normal bundle.

Thus, in order to see the obstruction, we need more data than the normal
bundle itself provides.  In order to recover the obstruction, the
BCFT calculation corresponding to the Yoneda pairing must have some
nonlocal component.

So, already before trying to set up the physics calculation, we see
two features that the result must have:
\begin{itemize}
\item The calculation must take advantage of some interaction term
in the worldsheet action -- the result is not just a wedge product,
unlike the closed string B model bulk-bulk OPE's.
\item The calculation must give a result that is somehow nonlocal.
\end{itemize}

Now, let us actually perform the calculation.
In principle, the Yoneda pairing should be encoded in the three-point
function
\begin{displaymath}
< \theta(\tau_1) \theta(\tau_2) \theta(\tau_3) >
\end{displaymath}
involving vertex operators for three copies of the element of Ext$^1$
inserted at various places along the boundary.
Now, in topological field theories, correlation functions should
reduce to zero modes.  In the present case, there is one $\eta$
zero mode and two $\theta$ zero modes, yet here we have three $\theta$'s.
The only way to get a nonzero result is to use some interaction terms.

Put another way, this correlation function should encode three
copies of the Yoneda pairing -- one for each pair of $\theta$'s.
In principle, each boundary-boundary OPE should take two $\theta$'s
and generate a $\eta \theta$ term, so that the result is a correlation
function involving one $\eta$ and two $\theta$'s, perfect to match
the available zero modes.  However, in order for the OPE to operate
in this fashion, we shall need some sort of interaction term.

Ordinarily one available interaction term would be the boundary interaction
\begin{displaymath}
\int_{\partial \Sigma} F_{i \overline{\jmath}} \rho^i \eta^{\overline{\jmath}}
\end{displaymath}
We could contract the $\rho$ on one of the $\theta$'s,
leaving us with two $\theta$'s and one $\eta$, perfect to match the
available zero modes.  The $\rho-\theta$ contraction
would generate a propagator factor proportional to $1/z$,
and the boundary integral would give a scale-invariant result.
The obvious log divergence can be handled by regularizing the propagator,
as discussed in \cite{edcs}, leaving a factor of an inverted laplacian.

In the present case, the curvature of the Chan-Paton factors can
be assumed to be trivial, so there is no such available interaction term,
but the general idea is on the right track.

The only available interaction term is the bulk four-fermi term:
\begin{displaymath}
\int_{\Sigma} R_{i \overline{\imath} j \overline{\jmath}}
\rho^i \rho^j \eta^{ \overline{\imath} } g^{\overline{\jmath} k}
\theta_k 
\end{displaymath}
We could contract the two $\rho$'s on two of the three $\theta$'s,
leaving us with a total of two $\theta$'s (one from the interaction term,
plus one of the original correlators) and one $\eta$, exactly as needed
to match the available zero modes.  Each $\rho-\theta$ contraction
would generate a propagator factor proportional to $1/z$, 
which would be cancelled by the integral over the bulk of the disk.
Boundary divergences can be handled by regularizing the propagators,
leaving us with factors of inverted laplacians.

Thus, we see the structure that we predicted earlier -- the correlation
function is nonvanishing thanks to an interaction term, and we have
nonlocal effects due to the presence of inverted laplacians.

What remains is to check that the resulting expression really does
correctly calculate the Yoneda pairing, which has not yet been
completed \cite{klswip}.

Assuming that this calculation is correct, note that we have found
a counterexample to the spirit of the 
open-closed decoupling conjecture described
in \cite[section 5.1]{edcs}.

\subsection{Bulk-boundary OPE's}

On general principles one expects to have bulk-boundary OPE's in
open string theory \cite{cardyope}. 
In the B model, a bulk-boundary OPE would be a pairing
\begin{displaymath}
H^p\left(X, \Lambda^q TX \right) \times
\mbox{Ext}^n_X\left( {\cal E}, {\cal F} \right) \: \longrightarrow \:
\mbox{Ext}^{n+p+q}_X\left( {\cal E}, {\cal F} \right).
\end{displaymath}

Such a mathematical pairing exists, and it is natural to conjecture that
it realizes the bulk-boundary OPE above, although no work has been done
to check that conjecture in nontrivial cases.  In the remainder of this
section, we shall outline the mathematics of that mathematical
pairing.

In the special case that the sheaves are bundles on $X$,
so that the Ext groups reduce to sheaf cohomology on $X$, represented
by differential forms, the mathematical pairing just mentioned reduces
to a wedge product of differential forms \cite{andreipriv}, 
just like the Yoneda pairing
in such circumstances, and the bulk-bulk OPE in all circumstances.

More generally, this bulk-boundary pairing is defined as follows:
\begin{enumerate}
\item First, we must define a map 
\begin{displaymath}
H^p(X, \Lambda^q TX ) \: \mapsto \:
\mbox{Ext}_X^{p+q}( {\cal E}, {\cal E})
\end{displaymath}
that maps bulk states
to states defined on the boundary.
\item Then, we use the Yoneda pairing
\begin{displaymath}
\mbox{Ext}_X^{p+q}\left( {\cal E}, {\cal E} \right) \times
\mbox{Ext}^n_X\left( {\cal E}, {\cal F} \right) \: \longrightarrow \:
\mbox{Ext}^{n+p+q}_X\left( {\cal E}, {\cal F} \right).
\end{displaymath}
\end{enumerate}

The mathematical explanation of the first map is somewhat technical,
and we are indebted to A.~C\u ald\u araru for explaining it to us.
Further details will appear in \cite{andreicardy1,andreicardy2}.
The idea is as follows.
First, we need to identify
\begin{displaymath}
\bigoplus_{p+q=n} H^p\left(X, \Lambda^q TX \right)
\end{displaymath}
with the groups
\begin{displaymath}
\mbox{Ext}^n_{X \times X} \left( {\cal O}_{\Delta}, {\cal O}_{\Delta} \right)
\end{displaymath}
where $\Delta$ is the diagonal in $X \times X$.
There is a well-known isomorphism between these two groups
(the Hochschild-Kostant-Rosenberg isomorphism).  Technically, for reasons
we shall not try to explain here, it turns out that to match physics
one needs a slight modification of the usual isomorphism,
involving the square root of the Todd class of $X$.

In any event, given the isomorphism above, we can now define
the desired bulk-boundary map.  A Fourier-Mukai transform with kernel
${\cal O}_{\Delta}$ maps ${\cal E}$ to ${\cal E}$, and with kernel
${\cal O}_{\Delta}[n]$ maps ${\cal E}$ to ${\cal E}[n]$.
A bulk state, now identified via the isomorphism above as an element
of 
\begin{displaymath}
\mbox{Ext}^n_{X \times X} \left( {\cal O}_{\Delta}, {\cal O}_{\Delta} \right)
\: = \: \mbox{Hom}_{X \times X}\left( {\cal O}_{\Delta},
{\cal O}_{\Delta}[n] \right)
\end{displaymath}
is a map $f: {\cal O}_{\Delta} \rightarrow {\cal O}_{\Delta}[n]$.
A map between the kernels of two Fourier-Mukai transforms
defines a map ${\cal E} \rightarrow {\cal E}[i]$ between the images
of a given object (${\cal E}$), as
\begin{displaymath}
\xymatrix{
Rp_2^*\left( {\cal O}_{\Delta} \stackrel{L}{\otimes} p_1^* {\cal E} \right)
\ar@{=}[r] \ar[d]^{f} & {\cal E} \ar[d]^{R p_2^*(f)} \\
Rp_2^*\left( {\cal O}_{\Delta}[n] \stackrel{L}{\otimes} p_1^* {\cal E} \right)
\ar@{=}[r] & {\cal E}[n]
}
\end{displaymath}
hence our bulk state $f$ defines
an element of
\begin{displaymath}
\mbox{Hom}_X\left( {\cal E}, {\cal E}[n] \right) \: = \:
\mbox{Ext}^n_X\left( {\cal E}, {\cal E} \right)
\end{displaymath}
Thus, given an element of $H^p(X, \Lambda^q TX)$, {\it i.e.}, a bulk state,
we have defined an element of $\mbox{Ext}^{p+q}_X\left( {\cal E},
{\cal E} \right)$, as desired.  

As one trivial example, an element of $H^0(X, \Lambda^0 TX)$
maps to $1_{\cal E} \in \mbox{Ext}^0({\cal E}, {\cal E})$,
up to a scale factor.  In particular, $1 \in H^0(X, \Lambda^0 TX)$
maps to the identity element of
$\mbox{Ext}^0_{X \times X} \left( {\cal O}_{\Delta}, {\cal O}_{\Delta}
\right)$, which is mapped to the identity element under the map above,
and all maps are linear.

Given this map from bulk states to boundary states,
the
bulk-boundary OPE is very plausibly the Yoneda product of the output
of this map with the given boundary state.
At present, we have not even completely checked that the Yoneda
product correctly duplicates boundary-boundary OPE's, so a direct
BCFT-based check that the mathematical operation discussed here does
indeed correctly describe bulk-boundary OPE's will have to wait to
a later date.

\subsection{Cardy's condition}

There is an interesting constraint on open string theories known
as Cardy's condition.  This constraint can be viewed as the statement
that the physical results of interpreting an annulus diagram as a one-loop
open string diagram, should be identical to physical results obtained
from thinking of annulus diagrams as describing closed strings propagating
between two boundary states.  

Some work of A.~C\u ald\u araru \cite{andreicardy1,andreicardy2} 
implies that
in the B model, Cardy's condition boils down to the
Hirzebruch-Riemann-Roch index theorem:
\begin{displaymath}
\sum_n (-)^n \mbox{dim } \mbox{Ext}^n_X\left( {\cal E}, {\cal F} \right)
\: = \:
\int_X \mbox{ch}\left( {\cal E} \right)^* \wedge
\mbox{ch}\left( {\cal F} \right) \wedge
\mbox{td}\left( TX \right)
\end{displaymath}
as we shall now outline.
See \cite{andreicardy1,andreicardy2} for a much more detailed explanation.

In general terms, the idea is that if we interpret the annulus
diagram in the open string B model as describing open strings at one-loop,
then it is calculating
\begin{displaymath}
\sum_n (-)^n \mbox{dim } \mbox{Ext}^n_X\left( {\cal E}, {\cal F} \right)
\end{displaymath}
where ${\cal E}$ and ${\cal F}$ are the sheaves corresponding to the
boundaries of the annulus.
If we interpret the annulus diagram as describing a closed string
propagating between two boundary states, then it should be calculating
\begin{displaymath}
\int_X \mbox{ch}({\cal E})^* \wedge \mbox{ch}({\cal F}) \wedge
\mbox{td}(TX)
\end{displaymath}

In order to make sense of the annulus diagram as a closed string
diagram, we must introduce some mathematics.
In the previous subsection, we defined a map
\begin{displaymath}
H^p\left(X, \Lambda^q TX \right) \: \mapsto \:
\mbox{Ext}^{p+q}_X\left( {\cal E}, {\cal E} \right)
\end{displaymath}
for any sheaf ${\cal E}$ on $X$.
Using very similar methods, and generalizing to $X$ not necessarily
Calabi-Yau, one can define a map
\begin{displaymath}
\mbox{Hom}_{X \times X}\left( {\cal O}_{\Delta}, S_{\Delta} \right)
\: \mapsto \:
\mbox{Hom}_X\left( {\cal E}, {\cal E} \otimes S_X \right)
\end{displaymath}
where $S_X = \omega_X[\mbox{dim }X]$, $\omega_X$ is the canonical line
bundle on $X$,
and $S_{\Delta} = i_{\Delta *} S_X$, which on a Calabi-Yau
essentially maps
\begin{displaymath}
\bigoplus_{p+q=dim\:X} H^p\left(X, \Lambda^q TX \right) \: \mapsto \:
\mbox{Hom}_X \left( {\cal E}, {\cal E}[\mbox{dim }X] \right)
\end{displaymath}
This map has a dual given by 
\begin{displaymath}
\mbox{Hom}_X\left( {\cal E}, {\cal E} \right) \: \mapsto \:
\mbox{Hom}_{X \times X}\left( S_{\Delta}, {\cal O}_{\Delta} \otimes
S_{X \times X} \right)
\end{displaymath}
The right-hand-side is isomorphic to its Serre dual, 
so we can define a pairing.
In particular, if we interpret the annulus diagram as a closed string
diagram, then in the present language, that closed string diagram should
be interpreted as taking the inner product of the images of
$1_{ {\cal E} } \in \mbox{Hom}_X\left( {\cal E},
{\cal E} \right)$ and $1_{ {\cal F} } \in \mbox{Hom}_X\left( {\cal F},
{\cal F} \right)$ under the map above.
It can be shown (see \cite{andreicardy1,andreicardy2} for details) that the
result of that pairing is given by
\begin{displaymath}
\int_X \mbox{ch}({\cal E})^* \wedge \mbox{ch}({\cal F}) \wedge
\mbox{td}(TX)
\end{displaymath}
(at least, after modifying the usual isomorphism 
\begin{displaymath}
\mbox{Ext}^n_{X \times X}\left( {\cal O}_{\Delta}, {\cal O}_{\Delta} \right)
\: \cong \:
\bigoplus_{p+q=n} H^p\left( X, \Lambda^q TX \right)
\end{displaymath}
by the square
root of the Todd class, as alluded to in the previous subsection).

Thus, in this very formal context, the Cardy condition seems to reduce
to the statement of Hirzebruch-Riemann-Roch.

\section{Stability}    \label{stability}

One of the motivations for using derived categories to 
model D-branes is to help make sense of stability of D-branes
in regimes where quantum corrections are nontrivial.

Now, in addition to physical notions of stability,
there are also mathematical notions of stability.
Curiously, these notions often coincide.
For example, for D-branes on large-radius Calabi-Yau's,
the holomorphic gauge field must satisfy essentially the
Donaldson-Uhlenbeck-Yau partial differential equation, which is
equivalent to a mathematical stability condition on the corresponding
holomorphic vector bundle. 

Both mathematical and physical stability are functions of K\"ahler moduli 
(and hence, are properties
of a physical theory, not a topological field theory).
As one varies K\"ahler moduli, both mathematical and physical stability
of a D-brane configuration can change.

We begin by discussing the correspondence between mathematical and
physical stability on large-radius Calabi-Yau's, where quantum corrections
can be ignored.  Then, we shall discuss pi-stability, an attempt
to create a mathematical notion of stability that encodes quantum corrections
to physical stability.

A much more thorough overview of pi-stability and related notions
will appear in \cite{paultoappear}, so we shall content ourselves
with merely outlining the highlights.

\subsection{Classical stability for bundles}

In \cite[equ'n (4.5b)]{hm} it was argued that, for a D-brane
wrapped on a submanifold of a large-radius Calabi-Yau,
the curvature $F$
of the gauge field on a D-brane must satisfy
\begin{equation}  \label{mtpde}
J^{d-1}\wedge F \: = \: \lambda J^d
\end{equation}
as a necessary condition for supersymmetry,
where $d$ is the (complex) dimension of the D-brane worldvolume,
$J$ is the K\"ahler form\footnote{Mathematicians will object to using
$J$ to denote the K\"ahler form, but this notation is standard in
the physics community.},
and $\lambda$ is some fixed constant.  
This partial differential equation 
closely generalizes the Donaldson-Uhlenbeck-Yau equation
that appears in heterotic strings, namely $J^{d-1} \wedge F = 0$.

Now, solving these partial differential equations is extremely difficult
in general.  However, as is often the case, this difficult problem
can be translated into an equivalent problem in algebraic geometry.
The gauge field associated to a holomorphic vector bundle ${\cal E}$
satisfies the partial differential equation above 
if and only if the holomorphic vector bundle ${\cal E}$ is
{\it Mumford-Takemoto semistable}, which means that if we define
\begin{displaymath}
\mu({\cal F}) \: = \: \frac{ \int_X J^{d-1} \wedge c_1({\cal F}) }{
\mbox{rank } {\cal F} }
\end{displaymath}
then for all
(well-behaved) subsheaves ${\cal F}$ of ${\cal E}$,
$\mu({\cal F}) \leq \mu({\cal E})$.
The function $\mu({\cal E})$ is known as the {\it slope} of ${\cal E}$,
and so Mumford-Takemoto stability is sometimes known as $\mu$-stability.

Up until now, we have ignored stability, as it does not enter into
the topological field theory (the open string B model) that we have
been studying.  However, if we untwist the open string B model to
a physical theory, then stability becomes very important.
This mathematical notion of stability, a necessary condition for a 
supersymmetric configuration, coincides with the physical notion of
stability, as well as interesting effects in heterotic compactifications,
as we shall now outline.

Let us consider an example.
Consider two D-branes wrapped on a K3 that is elliptically-fibered
and has a section.  Let $S$ denote the section, and $F$ a fiber,
so $F^2=0$ (since the fiber is elliptic), $S^2=-2$ (since the base
is a ${\bf P}^1$), and $S \cdot F = 1$.

The K\"ahler cone of this example is defined by K\"ahler forms
\begin{displaymath}
J \: = \: a S \: + \: b F
\end{displaymath}
where $a > 0$ (from demanding that $J \cdot F > 0$) and
$b > 2a$ (from demanding $J \cdot S > 0$).

To see how classical stability works in this context, first consider
the rank two bundle ${\cal O}(S-F) \oplus {\cal O}(F-S)$,
which has $c_1=0$ (and hence vanishing slope) and $c_2=4$.
Most of the time, this bundle is unstable -- for most points in the K\"ahler
cone, one of the two factors is a destabilizing subsheaf.
For example, when $b > 3a$, then $J \cdot (S-F) > 0$,
hence ${\cal O}(S-F)$ has positive slope, hence ${\cal O}(S-F)$ is a
destabilizing subsheaf.  The exception to this rule is along the line
$b=3a$ inside the K\"ahler cone, where this rank two bundle turns out
to be properly semistable.

For another example, consider the rank two bundle ${\cal E}$,
also of $c_1=0$ and $c_2=4$, defined as a nontrivial extension
\begin{displaymath}
0 \: \longrightarrow \: {\cal O}(S-F) \: \longrightarrow \:
{\cal E} \: \longrightarrow \: {\cal O}(F-S) \: \longrightarrow \: 0
\end{displaymath}
When $b > 3a$, ${\cal O}(S-F)$ is a destabilizing subsheaf,
and so ${\cal E}$ is unstable.  However, unlike the previous split
example, ${\cal O}(F-S)$ is not a subsheaf, and so does not destabilize
${\cal E}$ when $b < 3a$.  In fact, ${\cal E}$ turns out to be stable
for $b < 3a$.
Thus, we see explicitly that some bundles are only stable on
part of the K\"ahler cone.  More generally, the K\"ahler cone breaks
into subcones, with a different moduli space of bundles in each subcone
\cite{qin1,qin2,qin3,qin3}.

Along a chamber wall,
a close reading of the equivalence between solutions of the partial
differential equation~(\ref{mtpde})
and stability conditions tells us that
the solution to~(\ref{mtpde}) is the unique split bundle associated to a given
properly semistable bundle.  Thus, if a bundle becomes semistable,
then for physics purposes, it effectively splits.

In a heterotic compactification, such a split bundle generates an
enhanced $U(1)$ gauge symmetry in the low-energy effective theory.
Some of the (formerly neutral) bundle moduli become charged under this
enhanced $U(1)$ gauge symmetry, and the corresponding D-term equations
explicitly realize the change in moduli space alluded to above.
For example, if the bundle ${\cal E}$ splits into ${\cal F} \oplus {\cal G}$,
then the formerly neutral moduli $H^1(\mbox{End } {\cal E})$ split into
\begin{displaymath}
\begin{array}{c}
H^1( {\cal F}^{\vee} \otimes {\cal F} ) \\
H^1( {\cal G}^{\vee} \otimes {\cal G} ) \\
H^1( {\cal F}^{\vee} \otimes {\cal G} ) \\
H^1( {\cal G}^{\vee} \otimes {\cal F} )
\end{array}
\end{displaymath}
The first two components give moduli that are neutral under the
enhanced $U(1)$, but the second two components have equal and
opposite charges under the enhanced $U(1)$.  If we schematically
label the fields corresponding to the charged moduli by $\alpha_i$,
$\beta_i$, then we have a D-term constraint
\begin{displaymath}
\sum_i | \alpha_i |^2 \: - \: \sum_i | \beta_i |^2 \: = \: r
\end{displaymath}
where the Fayet-Iliopoulos term $r$ is determined by the K\"ahler metric.
Varying $r$ changes solutions to the D-terms, and realizes the
change in moduli spaces.
See \cite{kcsub} for a more detailed discussion.

For D-branes, a split bundle means that the D-branes are no longer bound,
and can be separated -- mathematical stability corresponds very precisely
to physical stability.

An analogous phenomenon involving special Lagrangian submanifolds
and changes of complex structure -- the mirror to the situation 
just described --
also is known to take place.  Mathematically \cite{joycespeclag}, as the complex
structure of the Calabi-Yau is varied, the special Lagrangian cycle
degenerates to a singular union, and then stops being special Lagrangian,
exactly mirror to the phenomenon just described, in which a stable
bundle becomes unstable as K\"ahler moduli are varied.
Physically \cite{shamitjohn}, D-branes wrapped on such special
Lagrangian submanifolds change from BPS configurations to non-BPS 
configurations as the complex structure of the Calabi-Yau is varied,
which is seen in a low-energy effective theory through D-terms,
in a manner closely analogous to the physical realization of
bundle stability discussed above.

\subsection{Pi-stability}

In the previous section, to make sense of stability of D-branes in
a physical theory we had to make the technical assumption that we
were working at large-radius on the Calabi-Yau, where worldsheet
instanton corrections could be neglected.  However, one would like
to extend stability to somehow take into account those worldsheet
instanton corrections.

A proposal for a notion of stability in D-branes that is sensible
away from the large-radius limit, that implicitly takes into account
worldsheet instanton corrections, was made by Douglas \cite{pistab},
who used some physical ideas to motivate a conjecture that reproduces
stability in two known limits (large-radius and orbifold points).

This proposed notion of stability is known as pi-stability.
In order to define pi-stability, we must first introduce a notion
of grading $\varphi$ of a D-brane.
Specifically, for a D-brane wrapped on the entire Calabi-Yau $X$
with holomorphic vector bundle ${\cal E}$,
the grading is defined as the mirror to the expression
$\int_X \mbox{ch}({\cal E}) \wedge \Pi$, where $\Pi$ encodes the
periods.  Close to the large-radius limit,
this has the form \cite[equ'n~(7)]{paulalb}:
\begin{displaymath}
\varphi({\cal E}) \: = \: \frac{1}{\pi} \mbox{Im } \log \int_X 
\exp\left( B + i J \right)
\wedge \mbox{ch}\left( {\cal E} \right) \wedge
\sqrt{ \mbox{td} \left( TX \right) }  \: + \: \cdots
\end{displaymath}
As defined $\varphi$ is clearly $S^1$-valued; however, we must choose
a particular sheet of the log Riemann surface, to obtain a ${\bf R}$-valued
function.  

This notion of grading of D-branes is an ansatz, introduced as part
of the definition of pi-stability.  Physically, it is believed that
the difference in grading between two D-branes corresponds to the
fractional charge of the boundary-condition-changing vacuum between
the two D-branes, though we know of no convincing first-principles
derivation of that statement.  Recall from earlier discussions that,
unlike closed string computations, the degree of the Ext group element
corresponding to a particular boundary R-sector state is not always
the same as the $U(1)_R$ charge -- for example, it is often
determined by the
$U(1)_R$ charge {\it minus} the charge of the vacuum.
The grading gives us 
the mathematical significance of that vacuum charge.
This mismatch between Ext degrees and $U(1)_R$ charges is
necesary for the grading to make sense:
Ext groups degrees are integral, after all, yet we want the
grading to be able to vary continuously, so the grading had better
not be the same as an Ext group degree.
See \cite[section 2.2]{paulalb} for further discussions of the
grading and comparisons to vacuum charges.

Given a ${\bf R}$-valued function from a particular definition of
log in the definition of $\varphi$ above,
the statement of pi-stability is then that for all subsheaves
${\cal F}$, as in the statement of Mumford-Takemoto stability,
\begin{displaymath}
\varphi({\cal F}) \: \leq \: \varphi({\cal E})
\end{displaymath}

Before trying to understand the physical meaning of $\varphi$, or the
extension of these ideas to derived categories, let us try to 
confirm that Mumford-Takemoto stability emerges as a limit of pi-stability.

For simplicity, suppose that $X$ is a Calabi-Yau threefold.
Then, for large K\"ahler form $J$, we can expand $\varphi({\cal E})$
as,
\begin{displaymath}
\varphi({\cal E}) \: \approx \:
\frac{1}{\pi} \mbox{Im } \log \left[ - \frac{i}{3!} \int_X J^3 (\mbox{rk }
{\cal E}) \right] \: + \: \frac{3}{\pi} \frac{
\int_X J^2 \wedge c_1({\cal E}) }{ \int_X J^3 \left( \mbox{rk } {\cal E} 
\right) }
\: + \: \cdots
\end{displaymath}
Thus, we see that to leading order in the K\"ahler form $J$,
$\varphi({\cal F}) \leq \varphi({\cal E})$ if and only if
\begin{displaymath}
\frac{ \int_X J^2 \wedge c_1({\cal F} ) }{ \mbox{rk } {\cal F} }
\: \leq \:
\frac{ \int_X J^2 \wedge c_1({\cal E} ) }{ \mbox{rk } {\cal E} }
\end{displaymath}
which is precisely the statement of Mumford-Takemoto stability
on a threefold $X$.

One can define a notion of (classical) stability for more general
sheaves (see for example \cite{hl}), but what we really want
is to apply pi-stability to derived categories.

However, there is a technical problem that limits such an extension.
Specifically, in a derived category there is no meaningful notion
of ``suboject.''  Thus, a notion of stability formulated in terms
of subobjects cannot be immediately applied to derived categories.

There are two workarounds to this issue that have been
discussed in the literature:
\begin{enumerate}
\item One workaround involves picking a subcategory of the derived
category that {\it does} allow you to make sense of subobjects.
Such a structure is known as a ``T-structure,'' and so one can
imagine formulating stability by first picking a T-structure,
then specifying a slope function on the elements of the subcategory
picked out by the subcategory.  This is the strategy followed
in \cite{bridgeland,bridge2}, which also argue that the set of T-structures
plus slope functions has the structure of a manifold.
\item Another (surely equivalent) workaround is to work with a 
notion of ``relative stability.'' Instead of speaking about whether
something is stable against decay into all other objects,
you only speak about whether it is stable against decay into two specified
objects.  One can then relate this to stability in the usual sense by
going to a weak-coupling limit, where classical stability is relevant.
This approach has been followed in {\it e.g.} \cite{paulmike,ahk,ak}.
\end{enumerate}

Let us very briefly outline the second workaround,
following \cite{paulmike,ahk,ak}.
Suppose we have a complex that can be put in the form
$\mbox{Cone}(f: C_{\cdot} \rightarrow D_{\cdot})$ for
some complexes $C_{\cdot}$, $D_{\cdot}$, and a chain map $f$.
(See section~\ref{cones} for the definition of the cone
construction.)  We can then ask, whether this state is
stable to decay into $C_{\cdot}$, $D_{\cdot}$.

Define $Q$ to be the difference of the grades $\varphi(C_{\cdot})$,
$\varphi(D_{\cdot})$, suitably normalized\footnote{There is a numerical factor
proportional to $3/\pi$ that we will normalize out.}.
It is believed (though not yet checked in detail) that
$Q$ is the $U(1)_R$ charge of the boundary R-sector state in the open string
morally corresponding to a tachyon.  (As noted earlier, correlating 
$U(1)_R$ charges to geometric quantities for general intersections
is tricky, but so long as we
are dealing with complexes of locally-free sheaves, there should not be
any problems.)  The mass of the physical NS-sector state obtained from
this R-sector state after spectral flow is given by 
\begin{displaymath}
m^2 \: = \: \frac{1}{2} \left( Q \: - \: 1 \right)
\end{displaymath}
Whether the D-branes corresponding to $C_{\cdot}$, $D_{\cdot}$
are stable or not is a function of $m^2$.
If $m=o$, we are on a marginal-stability wall; if $m^2<0$, then the
open string becomes tachyonic, and we are in a bound state \cite{sen};
if $m^2>0$, then we are in an unbound state.
(Note that in effect we are considering whether the difference in 
grades is greater or less than one, whereas previously the marginal-stability
line was where the grades coincide.  This is because of the fact that
in order to
preserve boundary ${\cal N}=2$ supersymmetry, the $U(1)_R$ charge is
forced to shift by one unit across tachyonic vevs; taking that distinction
into account, at large-radius we recover the same stability condition
as before.)

\section{Conclusions}    \label{concl}

In these lecture notes we have reviewed how sheaves can be
used to model D-branes.  We have outlined how to check such models
by comparing massless spectra and operator products to mathematical
properties of sheaves, and described how the sheaf picture can be
used to simplify physics calculations that are extremely difficult.

We began by reviewing the mathematics of sheaves and the physics
of the open string B model, then directly calculated the boundary
R-sector states in the open string B model to see that those states
are counted by Ext groups.  In the process, not only did we gain a powerful
new computational tool, but we also saw how nontrivial boundary
conditions on worldsheet fermions are closely interrelated with the
mathematics of sheaves.  We discussed analogues of that calculation
in flat nontrivial $B$ field backgrounds, and in orbifolds, each time
gaining powerful new computational methods for physics.  We also
saw how Higgs vevs can be translated into sheaves, and how
many more sheaves than previously considered also have direct physical
interpretations, a prediction checked by comparing direct BCFT-based
calculations of open string spectra with Ext groups between the
proposed sheaf-theoretic models.  Algebraic properties of the open
string B model, {\it i.e.} the boundary-boundary OPE's, bulk-boundary
OPE's, and Cardy's condition, were also discussed, and we saw how
verifying basic mathematical predictions for these is physically
highly nontrivial in general.  

In addition to giving us powerful new methods to calculate
properties of open strings, such sheaf-based descriptions of D-branes
are also motivated by attempts to relate open string boundary states
to objects in derived categories.  Physically such objects are
massive theories, describing branes and antibranes with nonzero
tachyon vevs.
We discussed these massive models, corresponding to complexes of branes
and antibranes on the boundaries of the open strings.  We discussed
how properties of complexes (as well as derived categories)
are realized directly in physics, how massless spectra can
be calculated in these massive theories (and give another realization
of Ext groups),
and how generalizations of complexes can also appear.  

Finally, we outlined recent calculations of D-brane stability,
and how physical notions of stability are related to mathematical notions
of stability.

There are many remaining open problems that need to be solved before
the relationship between derived categories and B model boundary states
in string theory can be described as fully understood.
\begin{itemize}
\item First, we do not yet understand which subcategory of the category
of coherent sheaves on a Calabi-Yau corresponds to D-branes, because
there is no known direct map between open string boundary conditions
and sheaves.
Some sheaves (pushforwards of vector bundles on submanifolds) have
an obvious interpretation, which has been checked physically through
computations of open string spectra, and we have also seen in
section~\ref{nonred} that other
more obscure-looking sheaves can also have a physical interpretation.
But, not all sheaves are of one of these forms -- do any of the rest admit
a physical interpretation, and if so, what is it?
\item We have only been able to check explicitly that open string
spectra coincide with Ext groups.  However, more assumptions are usually
made, concerning product structures (OPE algebras) on open string states.
For example, boundary-boundary OPE's are believed to be described
by the Yoneda pairing.  Unfortunately, none of these product structures
have ever been checked in nontrivial cases, and as we have seen in
section~\ref{yoncheck}, the physical computations required to check
such claims can be extremely nontrivial.
\item The homological algebra methods that form an important part
of the mathematics of derived categories can not be immediately applied
to D-branes:
\begin{enumerate}
\item Complexes of branes-antibranes yield {\it massive} theories in
which conformal invariance is broken by boundary interactions, 
and so the resulting theories
are not BCFT's.  One can use gauged-linear-sigma-model-style techniques
in connection with such complexes to gain information about the 
boundary CFT obtained after RG flow, but such methods are usually limited
to statements about RG invariants, and necessarily make assumptions about
endpoints of RG flow which sound natural but can sometimes be misleading
\cite{distleranom}.
\item To be able to physically apply injective resolutions, 
each sheaf in the physically-relevant subcategory of the category
of coherent sheaves would have to admit an injective resolution
such that each injective in the resolution can also be physically realized.
At the moment, we do not even know how to describe the subcategory
that contains sheaves that correspond to physical D-branes,
so we cannot even begin to understand whether the physically-relevant
subcategory has enough injectives.
\item Localization of quasi-isomorphisms is accomplished through
boundary RG flow -- the universality classes of RG flow are 
assumed to coincide with the equivalence classes defined by the
localization procedure.  Unfortunately, checking this explicitly
is practically impossible at the moment.  On the other hand,
this particular issue may be moot, for two reasons.
First, two complexes that are
quasi-isomorphic define massive theories with the same BRST cohomology ring,
hence essentially isomorphic\footnote{But see
\cite{pauldave} for a cautionary note.} as topological field theories.
Second it has now been checked \cite{ks} that corresponding
BCFT's do indeed have the open string spectrum that one would predict
based on the massive theories and the usual assumptions about the behavior
of RG flow.  So, we do have some evidence that localization of
quasi-isomorphisms is accomplished through boundary RG flow, 
though clearly some more nearly direct computation would be desirable.
\end{enumerate}
\item It would be extremely desirable to give a first-principles
physical derivation of Douglas's proposed pi-stability.
A mathematically-sensible understanding of notions of stability
in derived categories,
including pi-stability, has been proposed
\cite{bridgeland,bridge2}, but it remains to derive pi-stability 
in physics in detail.
\end{itemize}

\section{Acknowledgements}

We would like to thank S.~Katz for many useful collaborations,
as well as our other collaborators A.~C\u ald\u araru, R.~Donagi, 
and T.~Pantev.  We would also like to thank P.~Aspinwall, S.~Dean,
E.~Diaconescu, A.~Kapustin,
R.~Karp, A.~Lawrence, C.~Lazaroiu, D.~Morrison, R.~Plesser, 
E.~Scheidegger, J.~Stasheff,
and last but not least, E.~Witten, for useful conversations.

\appendix

\section{Spectral sequences and filtrations}    \label{filt}

A {\it spectral sequence} is a mechanism for computing a graded series of
groups via a series
of successive approximations, much like a sculptor chipping away at
a block of marble to find the statue within.

At each given stage (indexed by an integer $r$), 
we have a double complex $E_r^{p,q}$, and a set of differentials
$d_r: E_r^{p,q} \rightarrow E_r^{p+r,q-r+1}$, that square to zero.
We form the next approximation by taking kernels and cokernels of
the differentials, as
\begin{displaymath}
E_{r+1}^{p,q} \: = \: \frac{  \{ \mbox{ker } d_r: \: 
E_r^{p,q} \: \longrightarrow \: E_r^{p+r,q-r+1} \} }{
\{ \mbox{im } d_r: \: E_r^{p-r,q+r-1} \: \longrightarrow \:
E_r^{p,q} \} }.
\end{displaymath}
After a finite number of steps, the approximations converge
to a double complex labelled $E_{\infty}^{p,q}$,
with which the result of the spectral sequence is derived
via filtrations.

In these lectures, we have ignored the filtrations in spectral
sequences, for reasons we shall explain shortly.
What it means for the result to be derived via filtrations is that
if the spectral sequence converges to some graded set of groups
$F^n$, say, then for each $n$, $F^n$ has a filtration
\begin{displaymath}
F^n \: = \: K^n_0 \: \supseteq \: K^n_1 \: \supseteq \: 
K^n_2 \: \supseteq \: \cdots
\end{displaymath}
which is related to the $E_{\infty}^{p,q}$ by
\begin{displaymath}
E_{\infty}^{p,q} \: = \: \frac{ K^{p+q}_p }{ K^{p+q}_{p+1} }.
\end{displaymath}
So, to be precise, the $E_{\infty}^{p,q}$ are not quite the same
thing as the objects $F^{p+q}$ which the spectral sequence is said to
be computing, but are closely related.

For our purposes in this paper, these filtrations are mostly irrelevant.
The groups we are interested in are always vector spaces -- there are no
torsion components to worry about.  Since the vector spaces are described
up to isomorphism by their dimensions, and 
the dimensions of the final groups are the same as the sum of the
dimensions of the $E_{\infty}^{p,q}$, we can ignore filtrations.
This is more problematic for computing product structures, however.

Let us illustrate these ideas in an example.
The most common example of a spectral sequence in these lectures
is the spectral sequence that computes $\mbox{Ext}^{p+q}_X\left(
i_* {\cal E}, i_* {\cal F} \right)$ by starting with an approximation
at level two
defined by sheaf cohomology.  This spectral sequence is usually
denoted
\begin{displaymath}
E_2^{p,q} \: = \: H^p\left(S, {\cal E}^{\vee} \otimes {\cal F}
\otimes \Lambda^q {\cal N}_{S/X} \right) \: \Longrightarrow \:
\mbox{Ext}^{p+q}_X\left( i_* {\cal E}, i_* {\cal F} \right).
\end{displaymath}
This spectral sequence is studied in considerable detail in
section~\ref{vertexext}, where, for example, the reader can find
explicit expressions for the differentials.  However, we did not discuss
filtrations in section~\ref{vertexext}, so we shall take a few moments
to repair that omission.

Suppose, for simplicity, that the spectral sequence trivializes at
level two, {\it i.e.}, all $d_r$ vanish for $r \geq 2$.
Then $E_{\infty}^{p,q} = E_2^{p,q}$.

Now, in this case, what is the {\it precise} relationship between the
sheaf cohomology groups and the Ext groups?
Consider for example $\mbox{Ext}^1_X\left( i_* {\cal E},
i_* {\cal F} \right)$.  From our general discussion,
there is implicitly a filtration
\begin{displaymath}
\mbox{Ext}^1_X\left( i_* {\cal E},
i_* {\cal F} \right) \: = \: K^1_0 \:
\supseteq \: K^1_1 \: \supseteq \: 0
\end{displaymath}
and the corresponding sheaf cohomology groups are related to the $K$'s as
\begin{eqnarray*}
H^1\left( S, {\cal E}^{\vee} \otimes {\cal F} \right)
\: = \: E_{2}^{1,0} & = & E_{\infty}^{1,0} \: = \:
\frac{ K^1_1 }{ K^1_2 } \: = \: K^1_1 \\
H^0\left(S, {\cal E}^{\vee} \otimes {\cal F} \otimes
{\cal N}_{S/X} \right) \: = \: E_2^{0,1}
& = & E_{\infty}^{0,1} \: = \: \frac{ K^1_0 }{ K^1_1 }
\end{eqnarray*}
The filtration defines a short exact sequence
\begin{displaymath}
0 \: \longrightarrow \:
K^1_1 \: \longrightarrow \: K^1_0 \: \longrightarrow \:
\frac{ K^1_0 }{ K^1_1 } \: \longrightarrow \: 0
\end{displaymath}
from which we immediately derive
\begin{equation}   \label{filtex}
0 \: \longrightarrow \:
H^1\left(S, {\cal E}^{\vee} \otimes {\cal F} \right) \: \longrightarrow \:
\mbox{Ext}^1_X\left( i_* {\cal E}, i_* {\cal F} \right) \:
\longrightarrow \:
H^0\left(S, {\cal E}^{\vee} \otimes {\cal F} \otimes {\cal N}_{S/X}
\right) \: \longrightarrow \: 0
\end{equation}
which tells us the {\it precise} relationship between Ext$^1$
and the corresponding sheaf cohomology groups,
in the special case that the spectral sequence degenerates at level
two.  (More generally, if the $d_2$'s do not all vanish, then 
the sheaf cohomology groups above would be replaced by $E_{\infty}^{p,q}$
which would no longer be sheaf cohomology.)

Physicists should note that the existence of a short exact sequence
does {\it not} always imply that the middle factor is the sum of the outer
two factors.  For example, there is a short exact sequence
\begin{displaymath}
0 \: \longrightarrow \: {\bf Z}_2 \: \longrightarrow \:
{\bf Z}_4 \: \longrightarrow \: {\bf Z}_2 \: \longrightarrow \: 0
\end{displaymath}
despite the fact that ${\bf Z}_4 \neq {\bf Z}_2 \oplus {\bf Z}_2$.
However, this is not an issue in
this paper, where all of the groups appearing are vector spaces,
since short exact sequences of vector spaces do always split.
For example, because sheaf cohomology and Ext groups are vector spaces,
the short exact sequence~(\ref{filtex}) implies that
\begin{displaymath}
\mbox{Ext}^1_X\left( i_* {\cal E}, i_* {\cal F} \right)
\: \cong \:
H^1\left(S, {\cal E}^{\vee} \otimes {\cal F} \right) \oplus
H^0\left(S, {\cal E}^{\vee} \otimes {\cal F} \otimes {\cal N}_{S/X}
\right)
\end{displaymath}
which is the reason we have ignored filtrations in this paper.
However, the isomorphism above is not uniquely determined, so to
compute products properly, something we have not described how to
do in these lectures, one would need to be much more careful
about filtrations.

\section{Proof of a calculation of Ext groups}  \label{pf}

(This appendix written by Sheldon Katz.)

In section~\ref{tachyonext} we described a calculation of
Ext groups as BRST cohomology in a massive theory obtained by
turning on tachyon vevs, so as to realize locally-free resolutions
of sheaves.  In this section we shall outline how to prove that
that physical calculation does yield Ext groups.
(Mathematicians should note that this is not intended to be a new
result, but rather is a review, intended for a physics audience.)

Let $X$ be a scheme (we are thinking of a CY threefold) 
and let $F$ and $G$ be coherent sheaves on $X$ (we are thinking of vector 
bundles on submanifolds $S\subset X$).  We pick finite resolutions
$F^\bullet\to F\to 0$ and $G^\bullet\to G\to 0$ by vector bundles.  Thus
$F^\bullet$ and $F$ describe the same object of $D^b(X)$; similarly for $G$.

We let $(\mathrm{Sh}_X)$ denote the category of sheaves of $\mathcal{O}_X$
modules and $(\mathrm{Ab})$ denote the category of abelian groups.  We
consider the usual bifunctors
$$\underline{\mathrm{Hom}}(\cdot,\cdot):
(\mathrm{Sh}_X)^{\mathrm{op}}
\times (\mathrm{Sh}_X)\to(\mathrm{Sh}_X)$$
$${\mathrm{Hom}}(\cdot,\cdot):
(\mathrm{Sh}_X)^{\mathrm{op}}\times (\mathrm{Sh}_X)\to(\mathrm{Ab})$$
$$\Gamma(X,\cdot):(\mathrm{Sh}_X)\to (\mathrm{Ab})$$
and denote their derived functors as usual by preceding them by an $R$.

By \cite[Prop 5.3 P102]{hred}

$$R\mathrm{Hom}^\bullet(F^\bullet,G^\bullet)\simeq R\Gamma(X,R
\underline{\mathrm{Hom}}^\bullet(F^\bullet, G^\bullet)).$$

This is an isomorphism in $D^b((\mathrm{Ab}))$.
If $F^\bullet$ and $G^\bullet$ are ordinary sheaves $F,G$, then the 
cohomology sheaves of the left hand side are the $Ext^i(F,G)$.

We now turn to the computation of the right hand side.

Since $F^\bullet$ and $G^\bullet$ are complexes of vector bundles, then
$\underline{\mathrm{Hom}}^\bullet(F^\bullet, G^\bullet)$ is just the
total complex of the natural double complex.  This is just
$\oplus_iG^{k+i}\otimes(F^{-i})^*$ in degree
$k$, with the natural differential induces from the differentials of 
$(F^\bullet)^*$ and $G^\bullet$ and including a sign.  
(Exercise/Problem on P91 of \cite{hred}).

More formally,

\bigskip\noindent
{\bf Lemma} Suppose that $F^\bullet$ is represented by a finite complex of 
locally free sheaves, and let $G^\bullet\in \mathrm{Ob}(D^b(X))$.  Then
$R\underline{\mathrm{Hom}}^\bullet
(F^\bullet,G^\bullet)\in \mathrm{Ob}(D^b(X))$ 
is represented by the complex 
$\underline{\mathrm{Hom}}^\bullet(F^\bullet, G^\bullet))$.

\bigskip
{\em Proof} (sketch): Since $F^\bullet$ is composed of locally frees,
the functor $\underline{\mathrm{Hom}}(F^\bullet,\cdot)$ is exact, so passes
to the derived functor in the first variable.  It is a general property
of the $\underline{\mathrm{Hom}}$ bifunctor that deriving in the second
variable is automatic \cite{hred}.

\bigskip
Now let $E^\bullet$ be a finite complex of locally free sheaves (in
our example, $E^\bullet$ will be
$\underline{\mathrm{Hom}}^\bullet(F^\bullet, G^\bullet))$.  Pass to the
analytic category.  Let $(A^{0,\bullet},\bar\partial)$ 
denote the Dolbeault complex on $X$.

\bigskip\noindent
{\bf Lemma} $R\Gamma(X,
\underline{\mathrm{Hom}}^\bullet(E^\bullet)$
is represented by the complex associated to the double complex
$E^\bullet\otimes A^{0,\bullet}$.
\bigskip

In the above, the differential is formed from the differentials of
$E^\bullet$ and $\bar\partial$, with the usual sign.

\bigskip\noindent
{\em Proof} (sketch) Consider one of the two spectral sequences associated to 
the double complex $E^\bullet\otimes A^{0,\bullet}$, the one which has
$E_1$ terms $E^p$ (since $E^p\otimes A^{0,\bullet}$ is a resolution of
$E^p$, the cohomology sheaves of $E^p\otimes A^{0,\bullet}$ are just
$E^p$ in degree 0 and everything else is 0).  
Then the $E_2$ terms are the cohomology sheaves $H^p(E^\bullet)$,
and the spectral sequence degenerates.  But the $E_\infty$ terms are the
cohomology sheaves of the complex associated to the double complex, proving
that its cohomology sheaves are also $H^p(E^\bullet)$,
hence the complex is quasi-isomorphic to $E^\bullet$ as claimed.  

\bigskip\noindent
{\bf Lemma} $R\Gamma(E^\bullet\otimes A^{0,\bullet})$ can be represented by
$\Gamma(E^\bullet\otimes A^{0,\bullet})$.

\bigskip\noindent
{\rm Proof} (sketch) The
complex we now have consists of fine sheaves
$E^\bullet\otimes A^{0,\bullet}$.
Compare with \cite[P 448 2]{gh}.

Putting this all together:

\bigskip\noindent
{\bf Proposition}
$\mathrm{Ext}^i(F,G)$ 
is the cohomology of the complex of $C^\infty$ sections of
$F^\bullet\otimes G^\bullet\otimes A^{0,\bullet}$ with the natural differential
constructed from $d_F,\ d_G$, and $\bar\partial$ with signs.

\end{document}